\documentclass[fleqn,usenatbib]{mnras}
\usepackage{url}
\usepackage{natbib}
\usepackage[normalem]{ulem}
\usepackage[T1]{fontenc}
\usepackage{txfonts}
\usepackage{graphicx}
\usepackage{savesym}
\savesymbol{leftroot}
\savesymbol{uproot}
\savesymbol{iint}
\savesymbol{iiint}
\savesymbol{iiiint}
\savesymbol{idotsint}
\savesymbol{Bbbk}
\usepackage{amsmath,amssymb,physics,bm,esint}
\usepackage{xlmacros}
\usepackage{cancel}
\usepackage[pdfpagelabels=false]{hyperref}
\hypersetup{colorlinks=True,linkcolor={RoyalBlue},citecolor={RoyalBlue},urlcolor={BrickRed},breaklinks=true}
\renewcommand{\vec}{\bm}

\DeclareRobustCommand{\VAN}[3]{#2}
\let\VANthebibliography\thebibliography
\def\thebibliography{\DeclareRobustCommand{\VAN}[3]{##3}\VANthebibliography}

\title[analytic shear estimator]{Analytical Weak-lensing Shear Responses of
Galaxy Properties and Galaxy Detection}
\author[]{
    Xiangchong Li $^{1}$\thanks{xiangchl@andrew.cmu.edu},
    Rachel Mandelbaum$^{1}$ \\
$^{1}$Department of Physics, McWilliams Center for Cosmology, Carnegie Mellon
University, Pittsburgh, PA 15213, USA
}
\date{Received Month XX, YYYY; accepted Month XX, YYYY}
\pubyear{2022}

\begin{document}
\label{firstpage}
\pagerange{\pageref{firstpage}--\pageref{lastpage}}
\maketitle

\begin{abstract}
Shear estimation bias from galaxy detection and blending identification is now
recognized as an issue for ongoing and future weak-lensing surveys.
Currently, the empirical approach to correcting for this bias involves
numerically shearing every observed galaxy and rerunning the detection and
selection process. In this work, we provide an analytical correction for this
bias that is accurate to subpercent level and far simpler to use. With the
interpretation that smoothed image pixel values and galaxy properties are
projections of the image signal onto a set of basis functions, we analytically
derive the linear shear responses of both the pixel values and the galaxy
properties (i.e., magnitude, size and shape) using the shear responses of the
basis functions. With these derived shear responses, we correct for biases from
shear-dependent galaxy detection and galaxy sample selection. With the
analytical covariance matrix of measurement errors caused by image noise on
pixel values and galaxy properties, we correct for the noise biases in galaxy
shape measurement and the detection/selection process to the second-order in
noise. The code used for this paper can carry out the detection, selection, and
shear measurement for $\sim$$1000$ galaxies per CPU~second. The algorithm is
tested with realistic image simulations, and we find, after the analytical
correction (without relying on external image calibration) for the
detection/selection bias of about $-4\%$, the multiplicative shear bias is
$-0.12 \pm 0.10\%$ for isolated galaxies; and about $-0.3 \pm 0.1\%$ for
blended galaxies with Hyper Suprime-Cam observational condition.
\end{abstract}

\begin{keywords}
gravitational lensing: weak; cosmology: observations; techniques: image
processing.
\end{keywords}

\section{INTRODUCTION}
\label{sec:Intro}

The observed shape (or shear) distortions of distant galaxy light profiles due
to deflection of light by the foreground mass distribution on its path to
observers can be used to study the distribution of matter, including both
baryonic and dark matter, in the Universe (see \citealt{dmrev,
cosmicShear-rev-Kilbinger15, revRachel17} for recent reviews). This deflection
of light is known as weak gravitational lensing, which is one of the main
science targets of the `Stage-IV' imaging surveys: the Vera C.\ Rubin
Observatory Legacy Survey of Space and Time\footnote{\url{http://www.lsst.org/}
} \citep[LSST,][]{LSSTOverviwe2019}, Euclid\footnote{ Euclid satellite mission:
\url{http://sci.esa.int/euclid/}} \citep{Euclid2011}, and the Nancy Grace Roman
Space Telescope High Latitude
Survey\footnote{\url{http://roman.gsfc.nasa.gov/}} \citep{WFIRST15}. These
surveys are designed to constrain the fundamental physics of the dark Universe
with unprecedented precision with weak lensing. A key challenge in the
scientific exploitation of these upcoming datasets is that the noisy images of
galaxies have been convolved by the point-spread function (PSF), which modifies
the apparent galaxy shapes in a spatially coherent way. To ensure that
systematic biases in cosmological weak-lensing analyses are within the
statistical uncertainties, these surveys require that the systematic bias in
measurement of the shear distortions from the noisy, PSF-convolved galaxy
images to be no worse than one part per thousand
\citep{euclidrequirement1,LSSTRequirement2018}.

A few methods developed by the community, namely \metacal{}
\citep[see][]{metacal-Huff2017,metacal-Sheldon2017,metaDet-Sheldon2020},
Bayesian Fourier Domain \citep[\BFD{},
see][]{BFD-Bernstein2014,BFD-Bernstein2016}, Fourier Quad \citep[\FQ{},
see][]{Z17,LZ2021} and Fourier power function shapelets \citep[\FPFS{},
see][]{FPFS-Li2018,FPFS_Li2022} are able to reach sub-percent level accuracy
for isolated galaxies without relying on calibrations using external image
simulation; these methods have other ways to correct for noise bias
\citep{noiseBiasRefregier2012}, model bias \citep{modelBias-Bernstein10} and
selection bias \citep{KaiserFlow2000} in the shear estimation.
\citet{metaDet-Sheldon2020} found that the shear-dependent detection and
blending identification causes a few percent-level shear estimation bias. They
demonstrated \metadet{} to correct for shear biases from source detection by
creating counterfactual images with different input shears before detecting
sources from pixels. The shear response of the detection process is estimated
from the difference in the average of galaxy shapes detected from the
counterfactual images. They demonstrated that \metadet{} is able to reduce
biases from shear-dependent detection below the stage-IV requirements on the
control of systematics even for blended galaxies.

\begin{figure*}
\centering
\begin{minipage}[t]{0.45\textwidth}
\begin{center}
    \includegraphics[width=1.\textwidth]{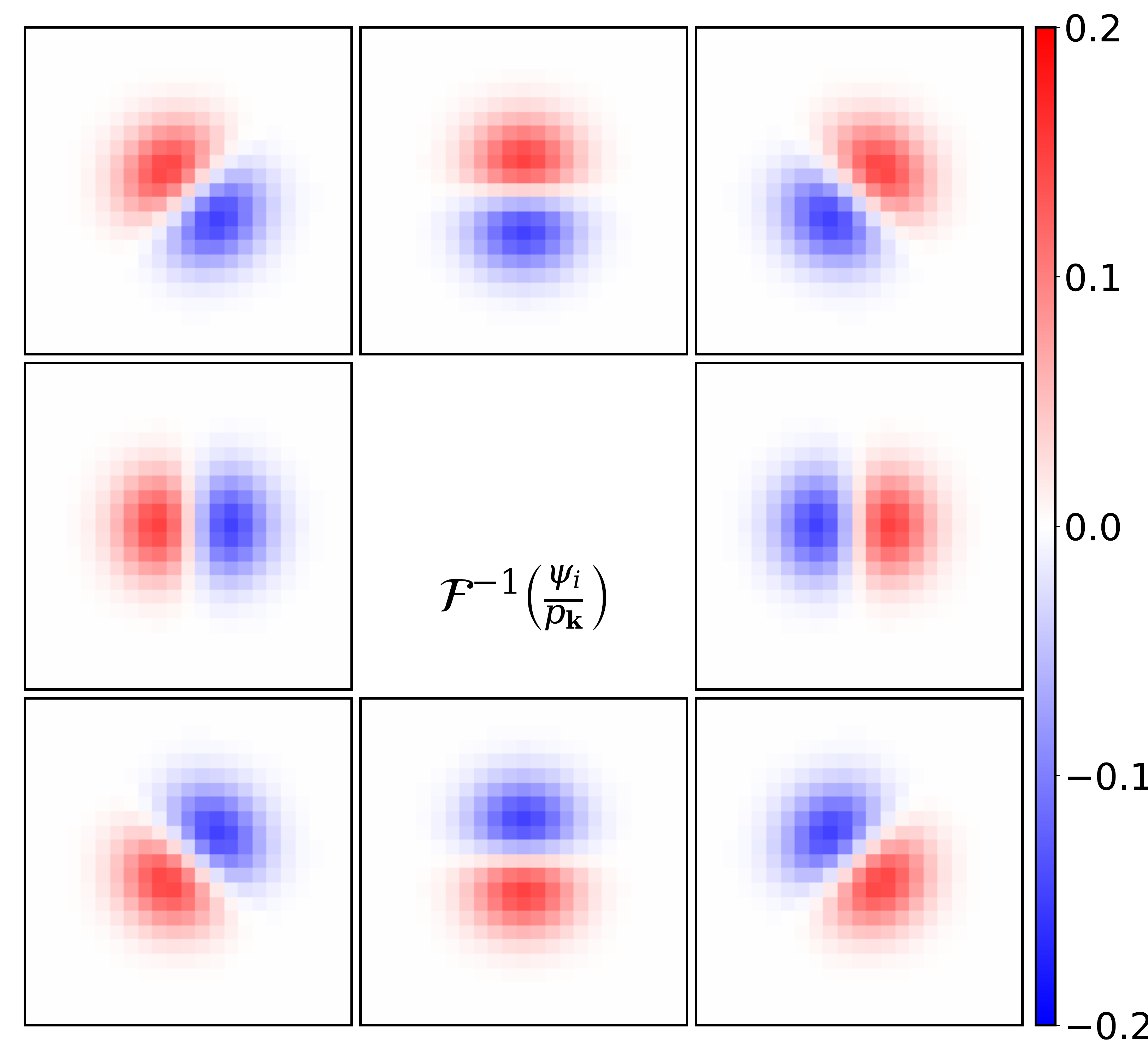}
\end{center}
\caption{
    Peak basis functions defined in configuration space with respect to
    postPSF images, which are used to define whether there is a detected peak
    at the location of each pixel. They are the inverse Fourier transform of
    $\psi_{i}$ deconvolved by the PSF, $p_\vk$\,, where $\mathcal{F}^{-1}$ in
    the figure refers to the inverse Fourier transform operator.
    }
    \label{fig:psi}
\end{minipage}
\hfill
\begin{minipage}[t]{0.45\textwidth}
\begin{center}
    \includegraphics[width=1.\textwidth]{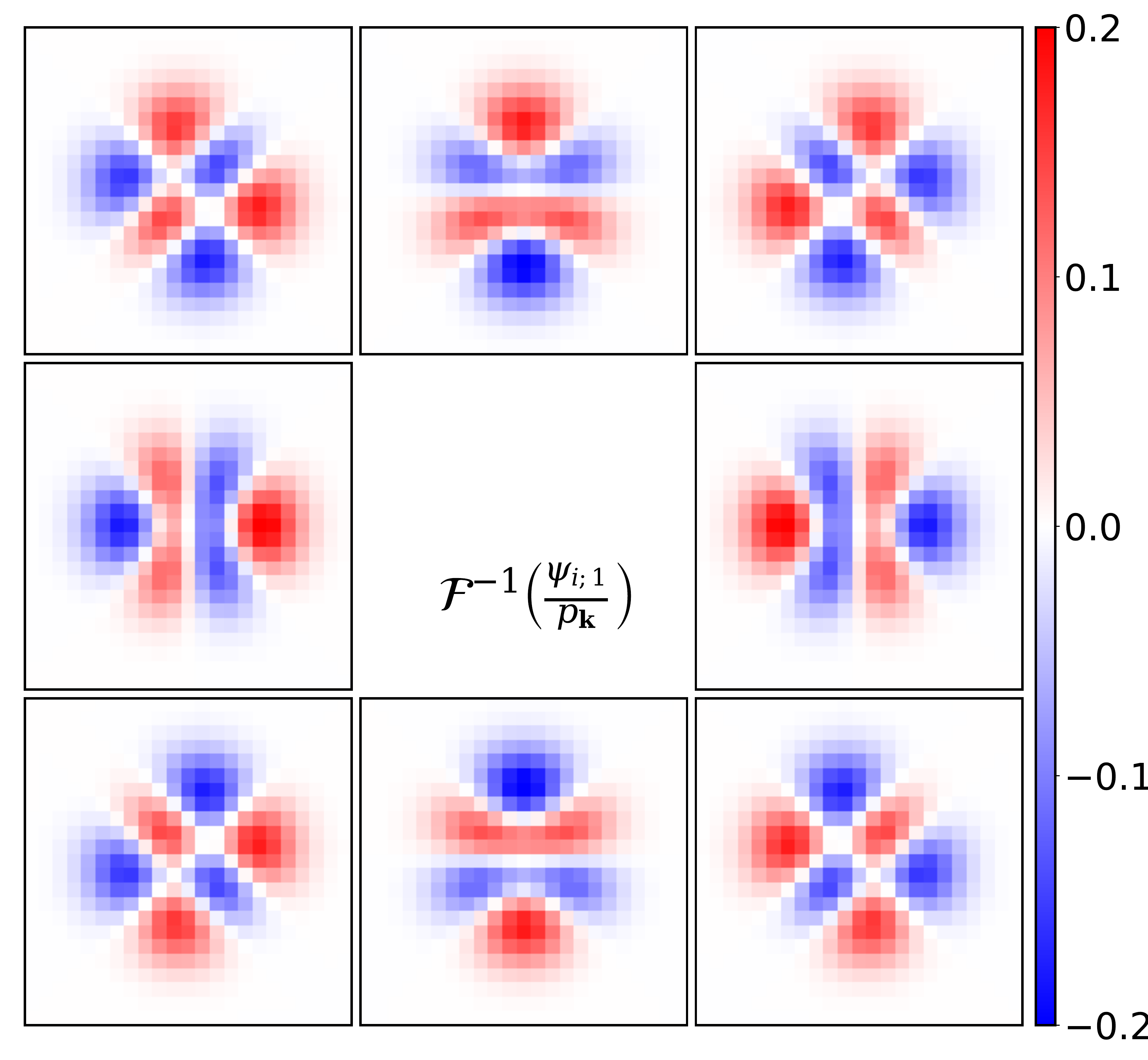}
\end{center}
\caption{
    The first component shear response of peak basis functions (shown in
    Figure~\ref{fig:psi}) defined in configuration space with respect to
    postPSF images, which is the inverse Fourier transform of $\psi_{i;1}$
    deconvolved by the PSF, $p_\vk$\,.
    }
    \label{fig:psidg1}
\end{minipage}
\end{figure*}

In this paper, we analytically derive the correction for shear-dependent
detection bias at the pixel level based on the framework of the \FPFS{} shear
estimator, and we show, with blended galaxy image simulations, that the
analytical method reduces the detection-induced shear bias by an order of
magnitude, to the sub-percent level. To be more specific, we propose to
interpret image pixel values after smoothing as projections of the image field
onto a set of basis functions. Based on this new interpretation, we derive the
shear responses of the pixel values using the shear responses of the pixel
basis functions. Galaxy detection from images is interpreted as a selection
using these pixel values, and the selection bias due to the galaxy detection is
analytically corrected using the shear responses of the pixel values. Our
analytical method is more than $100$ times faster than \metadet{}, and does not
require the generation of multiple catalogs to correct for shear-dependent
detection bias.

This paper is organized as follows: In Section~\ref{sec:Method}, we propose an
analytical method to correct for shear-dependent detection bias and noise bias at
the image pixel level. Then, in Section~\ref{sec:sims}, we introduce the galaxy
image simulations that we used to test (not calibrate) the proposed formalism
for correcting the above-mentioned shear biases. The results of the tests on
isolated and blended galaxies are shown in Section~\ref{sec:results}. Finally,
in Section~\ref{sec:Summary}, we summarize our results and the future outlook.

\section{METHODOLOGY}
\label{sec:Method}
In Section~\ref{sec:2_formalism}, we introduce the formalism of the \FPFS{}
shear estimator, which interprets the lensing shear distortion and image noise
as generating perturbations in the galaxy number density in the space of galaxy
properties. In Section~\ref{sec:2_pixel}, we interpret image pixel values
after smoothing as projections onto \FPFS{} pixel basis functions and derive
the shear response of smoothed pixel values. In Section~\ref{sec:2_gal}, we
define \FPFS{} properties of galaxies to quantify their flux, size and shape;
in addition, the shear responses of these quantities are analytically derived.
In Section~\ref{sec:2_selection}, we apply the formalism with the \FPFS{}
properties and their shear response to correct shear-dependent detection bias
and selection bias. In Section~\ref{sec:2_covariance}, we derive the covariance
matrix for those introduced \FPFS{} properties. In Section~\ref{sec:2_summary},
we summarize the steps in applying this shear estimator.

\subsection{Formalism}
\label{sec:2_formalism}

We first rewrite the works of \citet{FPFS-Li2018} and \citet{FPFS_Li2022} into
a general formalism
(Sec.~\ref{sec:2_formalism_galnum}). The basic idea is that the statistics of a
sample of galaxies can be described by its number distribution in the space of
galaxy properties. Within this space, we derive the leading-order perturbations
due to weak-lensing shear distortion (Section~\ref{sec:2_formalism_shear}) and
image noise on the galaxy number distribution
(Section~\ref{sec:2_formalism_noise}).

\subsubsection{Intrinsic noiseless galaxy number distribution}
\label{sec:2_formalism_galnum}

Galaxy ellipticity, which has two components: $e_\alpha$ ($\alpha \in
\{1,2\}$), is widely used to infer shear from large samples of galaxies
\citep{Shapes_Bernstein2002,Regaussianization}. It is a spin-$2$ property of a
galaxy, and is normally defined using the second-order moments of galaxy light
profiles. In this subsection, we derive corrections for shear-estimation bias
within a general framework that can be applied to many different ellipticity
definitions. For this reason, we have used $e_\alpha$ without providing a
mathematical definition. Once we discuss the specific implementation of this
formalism used in this paper, we will provide our ellipticity definition used
for shear estimation (Section~\ref{sec:2_gal}).

For a galaxy sample with normalized probability density function (PDF; denoted
as $\mathcal{P}$) in galaxy property space, the expected average of the
weighted galaxy ellipticities is
\begin{equation}
\label{eq:ell_average}
\langle w e_\alpha \rangle = \int \rmd{}e_\alpha \rmd{} w \, \mathcal{P}(e_\alpha,w)
\,e_\alpha w\,,
\end{equation}
where $e_\alpha$ and $w$ are galaxy ellipticity and selection weight,
respectively. Here the selection weight is a non-negative spin-$0$ function of
galaxy properties, and a galaxy is removed from the sample if the selection
weight equals zero. We propose to estimate the shape and selection weight for
each smoothed image pixel and then select a subset of them as galaxy candidates
using the selection weights. Following this line of thought, the galaxy
detection from pixels is interpreted as applying hard-thresholding weights to
the pixels. Also, the selection of a galaxy sub-sample is interpreted as
applying hard-thresholding weights to the galaxy candidates. The origin of
detection and selection bias is that the weights have a nonzero response to
weak-lensing shear. Note, this selection weight is different from the
traditional shape weight for optimal galaxy shape estimation \citep[see
e.g.,][]{HSC1-GREAT3Sim}, and we choose not to normalize the average of
weighted ellipticity by the average of the selection weight to avoid
introducing another shear response from the average selection weight into the
shear estimator. The ellipticity and selection weight are multivariable
functions of a set of basis modes, $\vm=( v_0, v_1, \dots )$, where each
element, $v_i\,$, is a linear projection of the image onto a basis function
(e.g., a Gaussian function centered at a specific pixel). We can write the
integral in equation~\eqref{eq:ell_average} as a summation over $N_\text{gal}$
galaxies indexed by $l$, since the number of galaxies is finite in reality:
\begin{equation}
\label{eq:ell_average_finite}
    \langle w e_\alpha \rangle =
    \frac{1}{N_\text{gal}}\sum_{l=1}^{N_\text{gal}} w^{(l)} e^{(l)}_\alpha\,.
\end{equation}
Again, this is an average of weighted ellipticity, not the weighted average of
ellipticity, since the denominator in equation~\eqref{eq:ell_average_finite} is
fixed to $N_\text{gal}$ rather than the summation over the weights.
Note, $N_\text{gal}$ refers to the total number of galaxies in the Universe,
which is not changed by the lensing distortion. Although $N_\text{gal}$ is
unmeasurable, as will be shown in our shear estimator
(equation~\eqref{eq:shear_estimator_noiseless}), the $N_\text{gal}$
normalization for the nominator and the denominator cancel with each other, so
that we do not need to know $N_\text{gal}$ for shear estimation.

The first assumption (\textbf{assumption~1}) in shear estimation is that
intrinsic (unlensed) galaxies are randomly oriented; as a result, the
expectation values of  non-spin-$0$ properties of intrinsic galaxy light
profiles are zero due to rotation symmetry (see Appendix~\ref{app:spin_property}
for details). Ellipticity is a  spin-$2$ property; therefore, the average
ellipticity of intrinsic galaxy light profiles equals zero. In addition, given
that the selection weight does not include any spin-$2$ component, the average
of the intrinsic (unlensed) weighted ellipticity equals zero.

However, due to weak-lensing shear distortion on galaxies, which is caused by a
foreground inhomogeneous mass distribution, the average of the (weighted)
ellipticity of observed galaxy light profiles deviates from zero. Here
\begin{equation}
    \bm{A}=\begin{pmatrix}
    1- \gamma_1   &  -\gamma_2\\
    -\gamma_2      &  1+\gamma_1
    \end{pmatrix}
\end{equation}
is the Jacobian matrix of the mapping from the lensed sky to the true sky. The
quantities $(\gamma_1, \gamma_2)$ represent the shear distortion: $\gamma_1$
stretches the image in the horizontal direction of the sky coordinate system,
and $\gamma_2$ stretches the image in the direction at an angle of $45\deg$
with the horizontal direction. Note that we set the convergence from lensing
distortion \citep{WL-rev-Bartelmann01} to zero to simplify the notation. We
denote the intrinsic ellipticity and intrinsic selection weight as
$\bar{e}_\alpha$ and $\bar{w}$, respectively. Symbols with bars are for
intrinsic (unlensed) properties. A summary of our notation, and how we indicate
intrinsic, lensed, and other types of properties, are  summarized in
Table~\ref{tab:notation}, which is the same as Table~1 of \citet{FPFS_Li2022}.

In addition to the shear distortion, image noise (which includes read noise,
and photon noise from background and sources) also changes the average
ellipticity. The second assumption (\textbf{assumption~2}) in shear estimation
is that the PDF of image noise is symmetric with respect to zero; as a result,
for a basis mode defined with a linear operation on the image, the PDF of the
measurement error of the basis mode, $\delta{v}_i$, is symmetric with respect
to zero, and the expectation values of any odd-order statistics of the
measurement error are also zero. Although a Poisson distribution is not
symmetric about its mean, in the background dominated regime, the asymmetry can
be neglected and the assumption holds true. The covariance (second-order
statistics) between two different modes with index $i$ and $j$ is denoted as
$K_{v_i}^{v_j} = \langle \delta{v}_i \delta{v}_j \rangle\,$. It is worth
mentioning that one can also define basis modes as a linear operation on the
power of the image after subtracting the expectation value of the power of the
image noise \citep{Z15,Li17Auto}.

The lensed images are convolved with the PSF, and galaxy shapes are changed by
the PSF. As a result, the PSF changes the distribution in the galaxy number
density space, $\mathcal{P}(e_\alpha,w)$. However, we measure the basis modes
$v_i$ from galaxies after deconvolving the PSF in Fourier space
\citep{Z08,FPFS-Li2018} -- so the influence of the PSF is removed by the
deconvolution, assuming that the PSF is accurately determined.

\begin{table}
\caption{
    Table for accent notations. The examples are for the ellipticity, but the
    notation also applies to other quantities.
    }
\begin{center}
\begin{tabular}{cl} \hline
    Accented ellipticity & Definition \\ \hline
    $\bar{e}_{1,2}$         & intrinsic (unlensed) galaxy ellipticity \\ \hline
    $e_{1,2}$               & ellipticity of lensed galaxies \\ \hline
    $\tilde{e}_{1,2}$       & ellipticity of noisy lensed galaxies \\ \hline
    $\widehat{e}_{1,2}$     & ellipticity after noise bias correction\\ \hline
\end{tabular}
\end{center}
\label{tab:notation}
\end{table}

\subsubsection{Shear perturbation}
\label{sec:2_formalism_shear}

The average of the weighted ellipticity transforms under the shear distortion
as
\begin{equation}
\label{eq:ell_sheared}
\begin{split}
\langle w e_\alpha\rangle= \langle \bar{w} \bar{e}_\alpha \rangle
+ \sum_{\beta=1,2} \frac{\partial \langle w e_\alpha \rangle}{\partial \gamma_\beta} \gamma_\beta
+ \mathcal{O}\left(\gamma^3\right)\,.
\end{split}
\end{equation}
Assuming that the selection weight, $w$, is a spin-$0$ scalar, the average of
the intrinsic weighted ellipticity, $\langle \bar{w} \bar{e}_\alpha \rangle$\,,
is identically zero (\textbf{assumption~1}). Since the shear distortion
satisfies $\gamma_\alpha \ll 1$, the leading order in
equation~\eqref{eq:ell_sheared} is the first order in shear, which is a vector
perturbation in the galaxy number space. The off-diagonal terms in the
$2\times2$ shear response matrix, $\frac{\partial (w e_\alpha)}{\partial
\gamma_\beta}$ ($\alpha\ne\beta$), are spin-$4$ properties of intrinsic galaxy
light profiles, the expectation values of which are identically zero
(\textbf{assumption~1}). The term with the second order in shear is also
identically zero due to the rotation symmetry. We find that, in the
weak-lensing regime with typical amplitude of shear $\sim0.03$, the neglected
third-order term is about $0.03^3$ (the corresponding bias is $<1\times10^{-3}$
relative to the shear), which is less than the requirement for the stage-IV
weak-lensing survey and hence negligible. We refer the reader to
Appendix~\ref{app:shear_perturb} for detailed discussions of the second-order
and thid-order shear perturbations. In addition, we summarize the spin number
of galaxy properties and review their rotation symmetries in
Appendix~\ref{app:spin_property}.

According to \citet{metacal-Huff2017}, the shear response of the average of
weighted ellipticity is the first-order derivative of the average of weighted
ellipticity to the shear distortion, and the shear can be estimated by
\begin{equation}
\label{eq:shear_estimator_noiseless}
\widehat{\gamma}_\alpha= \frac{\langle w e_\alpha\rangle}
{\mathcal{R}_\alpha}\,,
\end{equation}
where $\mathcal{R}_\alpha \equiv \frac{\partial\langle w e_\alpha
\rangle}{\partial \gamma_\alpha}$ are the shear responses (the diagonal
terms) of the average of the  weighted ellipticity\footnote{ \metadet{}
\citep{metaDet-Sheldon2020} retains the off-diagonal spin-$4$ terms.}. From the
rotational symmetry arguments, $\mathcal{R}_1$ and $\mathcal{R}_2$ are expected
to be identical; however, we treat them independently in the analysis as a
sanity check.

We substitute equation~\eqref{eq:ell_average_finite} into the definition of
$\mathcal{R}_\alpha$ and apply the derivative chain rule to write the shear
response in terms of our basis modes $v_i$ as
\begin{equation}
\label{eq:ell_average_response}
\begin{split}
\mathcal{R}_\alpha
&=  \langle w e_{\alpha;\alpha} + w_{;\alpha} e_{\alpha} \rangle \\
&= \frac{1}{N_\text{gal}}\sum_{l=1}^{N_\text{gal}} \sum_i
\left( w^{(l)} \frac{\partial e^{(l)}_\alpha}{\partial v_i}v^{(l)}_{i;\alpha}
+ e_\alpha^{(l)}\frac{\partial w^{(l)}}{\partial v_i} v^{(l)}_{i;\alpha} \right)\,,
\end{split}
\end{equation}
where the subscript `$;\alpha\,$' refers to the partial derivative with respect
to one component of shear, $\gamma_{\alpha}\,$. Similar to the weighted
ellipticity, we can apply the derivative chain rule to get the first-order
(leading-order) shear response for other weighted observables.

One can follow the approach of \metacal{} to obtain the shear responses of the
observables (e.g.,\ $v_{i;\alpha}$, $e_{\alpha;\alpha}$) in real observations
by creating counterfactual images with different input shears and measuring the
difference in the observables for different shears. This is a finite-difference
construction of a shear response as a derivative. In this paper, we take a
different approach, consistent with that of  \citet{FPFS-Li2018}: we use the
shear response of a set of basis functions (in other words, basis coordinates),
e.g.,\ shapelets \citep{shapeletsI-Refregier2003,Shapes_Bernstein2002}, to
derive the shear response of any observables constructed with these basis
functions. This approach of analytically calculating the impact of shear on the
coordinate system (basis functions) avoids repeatedly shearing each galaxy to
derive shear responses; therefore, it saves significant computational time.

\subsubsection{Noise perturbations}
\label{sec:2_formalism_noise}
Generally speaking, the weighted ellipticity, $w e_\alpha$, and its shear
response are nonlinear functions\footnote{ Some shear estimators (see, e.g.\,
\citealt{Z17}) avoid using any nonlinear observables.} of $v_i$. As a result of
this non-linearity, image noise biases the measurement of these nonlinear
observables \citep{noiseBiasRefregier2012}. As shown in
\citet{metacal-Sheldon2017}, such a noise bias can be statistically corrected by
adding artificially sheared noise fields with the same statistical properties
but different realizations to the observed images. We take the alternate
approach proposed by \citet{FPFS_Li2022}, Taylor expanding the noisy
observables as functions of $\vec{\delta{v}}$ and analytically deriving the
leading order noise bias correction using the covariance of the measurement error,
which can be determined from the images.

To be more specific, we take the expectation value of the noisy weighted ellipticity,
denoted as $\langle \tilde{w}\tilde{e}_{\alpha}\rangle$ (noisy observables are
denoted with tilde), as an example:
\begin{equation}
\label{eq:ell_noisy}
\begin{split}
\langle \tilde{w} \tilde{e}_\alpha\rangle= \langle w e_\alpha \rangle
+ \frac{1}{2}\sum_{i,j} \left\langle \frac{\partial^2
    (\tilde{w} \tilde{e}_\alpha)}{\partial v_i \partial v_j}
    \delta v_i \delta v_j \right\rangle
+ \mathcal{O}\left((\delta v_i)^4\right)\,,
\end{split}
\end{equation}
where $\frac{\partial^2 (\tilde{w} \tilde{e}_\alpha)}{\partial v_i \partial
v_j}$ is the Hessian matrix of the weighted ellipticity with respect to the
basis modes. The odd-order terms (e.g., the first- and third-order terms) of
the  measurement error reduce to zero after averaging over a large number of
galaxies (\textbf{assumption~2}). \citet{FPFS-Li2018} constructed a weighted
ellipticity using shapelets mode to ensure that
\begin{equation}
    \abs{\frac{\partial^{(n+1)} (w e_\alpha)}{(\partial v_i)^{(n+1)}} (\delta
    v_i)^{(n+1)}}\ll
    \abs{\frac{\partial^n (w e_\alpha)}{(\partial v_i)^n} (\delta v_i)^n} \,,
\end{equation}
so that the second-order term of the noise residual, which is a tensor
perturbation in the space of galaxy properties, is the dominant term beyond
$\langle w e_\alpha\rangle$ in equation~\eqref{eq:ell_noisy}, and higher
even-order terms can be neglected.

We introduce the debiased weighted ellipticity, $\widehat{w e_\alpha}$,
following \citet{FPFS_Li2022}.  Its expectation value is
\begin{equation}
\label{eq:ell_noisy_correct}
\langle \widehat{w e_\alpha}\rangle \equiv \langle \tilde{w} \tilde{e}_\alpha \rangle
-\frac{1}{2}\sum_{i,j} \left\langle
    \frac{\partial^2 (\tilde{w} \tilde{e}_\alpha)}{\partial v_i \partial v_j}
    K_{v_j}^{v_i}\right\rangle\,.
\end{equation}
The correction term is proportional to the expectation value of the contraction
between the Hessian matrix of the nonlinear observable: $\frac{\partial^2
(\tilde{w} \tilde{e}_\alpha)}{\partial v_i \partial v_j}$ and the noise
covariance matrix of linear variables $v_i$ and $v_j$: $K_{v_j}^{v_i} = \langle
\delta v_i \delta v_j \rangle\,$. We investigate the Hessian matrix in detail in
Appendix~\ref{app:noirev} by separating its elements into zeroth-, first- and
higher-order derivatives in the selection weight. The covariance matrix of
these linear observables is derived for homogeneous noise in
Section~\ref{sec:2_covariance} using the correlation function of noise between
pixels. It worth mentioning that in \citet{imPT_Li2023} we use
auto-differentiation in \texttt{jax}
\footnote{\url{https://github.com/google/jax}} to automatically derive the
shear response and second-order noise bias correction following
equations~\eqref{eq:ell_average_response} and \eqref{eq:ell_noisy_correct}.

Similar to the weighted ellipticity, we can derive the second-order (i.e.,
leading-order) noise bias correction for the expectation value of its shear
response (see Appendix~\ref{app:noirev} for details). We denote the debiased
shear response as $\widehat{\mathcal{R}}_\alpha\,$, and the shear estimator is
\begin{equation}
\label{eq:shear_estimator}
\widehat{\gamma}_\alpha=\frac{\langle \widehat{w e_\alpha}\rangle}
    {\widehat{\mathcal{R}}_\alpha}\,.
\end{equation}

In the following context, we will compress images into a basis vector space of
$( v_0,v_1, \dots )\,$, which meets the following requirements:
\begin{enumerate}
    \item For each element, $v_i$, its shear response, $\partial
        v_i/\partial \gamma_\alpha\,$, can be analytically derived and
        measured from images;
    \item The covariance matrix of measurement errors, $K_{v_i}^{v_j}$, can be
        analytically derived and measured from images.
\end{enumerate}
With these basis modes, we can derive the shear response of the average weighted
ellipticity and correct the noise bias.

\subsection{Shear response of re-smoothed pixels}
\label{sec:2_pixel}

\begin{figure}
\begin{center}
    \includegraphics[width=0.48\textwidth]{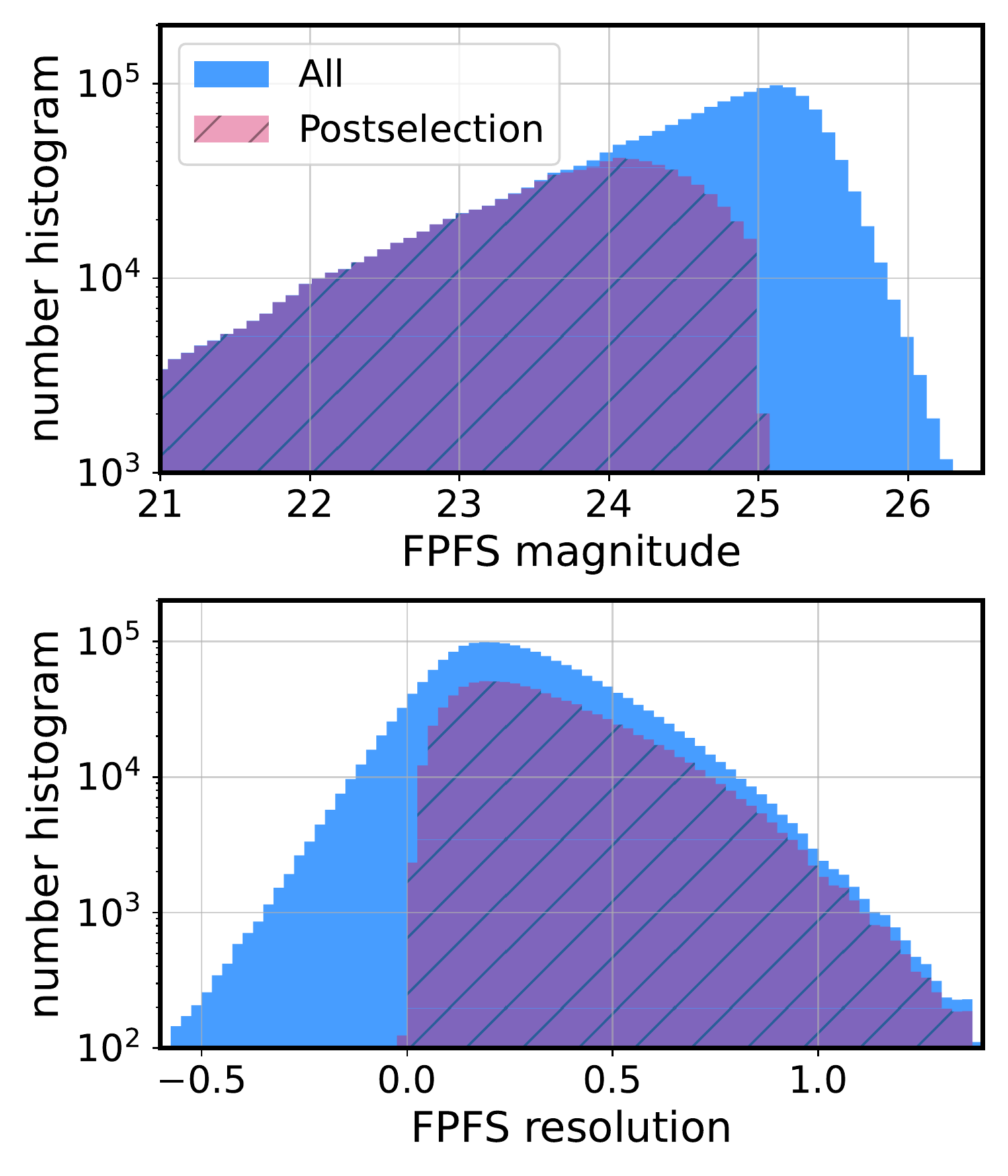}
\end{center}
\caption{
    Number histograms as functions of \FPFS{} magnitude ($m_{\text{F}}$, upper
    panel) and \FPFS{} resolution ($R_2$, lower panel) measured from isolated
    noisy galaxies with HSC-like noise, pixel size and PSF seeing. The
    unhatched histograms are for all of the galaxies in the isolated simulation
    described in Section~\ref{sec:sims}; hatched histograms are after the
    selection with soft cuts (Section~\ref{sec:2_selection}) at
    $m_{\mathrm{F}}=25$ and $R_2=0.05$ to select galaxies that are sufficient
    for weak-lensing science.
    }
    \label{fig:obsHist_fixed}
\end{figure}

In order to derive the shear response of the detection process, we reinterpret
the re-smoothed pixel values as projections of the image signal onto a set of
pixel basis functions in Section~\ref{sec:2_pixel_basesfun}. Benefiting from
this interpretation, we derive the linear shear response of the smoothed pixels
in Section~\ref{sec:2_pixel_response}. Based on the pixel basis functions and
their shear response, we define a set of peak basis functions to determine
whether there is a detected peak at the location of each pixel and derive the
shear response of these peak basis functions in
Section~\ref{sec:2_pixel_pdetect}. The peak modes and their responses will be
used to define detection and selection, and to correct for shear-dependent
biases in these processes in Section~\ref{sec:2_selection}.

\subsubsection{Pixel basis functions}
\label{sec:2_pixel_basesfun}

The observed astronomical images are smoothed by the PSF from the  atmosphere
and the telescope optics. We focus on  well-sampled images, so that one can
transform the pixelated images into a continuous image signal without loss of
information according to the Shannon sampling theorem.

The intrinsic image signal  (prelensing, prePSF) at a position $\vx$ in
configuration space is denoted as $\bar{f}_\vx$, the lensed image signal
distorted by shear $\gamma$ is denoted as $f_\vx$, and the observed image
signal smeared by a PSF, $p_\vk$, is denoted as $f^{p}_\vx$\,. Note that in
this paper, $f_\vx \equiv f(\vx)$ and $f_\vk \equiv f(\vk)$ are used to denote
the signal in configuration space and Fourier space, respectively.

Here we define basis modes for the pixels of re-smoothed images, neglecting the
shear distortion. The transformation of the pixel values under a shear
distortion will be studied in Section~\ref{sec:2_pixel_response}.

An image signal, $f_\vx$, can be transformed into Fourier space and the Fourier
transform at wave number $\vk$ is
\begin{equation}
f_\vk=\iint \dd[2]{x} f_\vx e^{-\rmi \vk\cdot\vx}\,,
\end{equation}
where $\rmi^2=-1$ is used to denote the complex number symbol to distinguish it
from the indexing symbol, $i\,$.

We re-smooth the observed image, which is originally smoothed by the PSF,
with a smoothing kernel to transform the PSF to an isotropic Gaussian. The
re-smoothing kernel is defined in Fourier space as multiplication by
    $h_\vk/p_\vk\,,$
which is an isotropic Gaussian kernel, denoted as
$h_\vk=\text{exp}(-\abs{\vk}^2\sigma_h^2/2)\,$, deconvolved by the PSF
of the observed image, $p_\vk$. $\sigma_h$ is the scale radius of the Gaussian
kernel in configuration space. Note, the typical scale radius of the target
Gaussian kernel should be greater than the typical scale radius of the original
PSF in configuration space so that the convolution does not amplify the noise on
small scales (large $\abs{k}$). This re-smoothed image is denoted as
$f^{h}_\vx$, where the superscript indicates that it is the prePSF signal
convolved with Gaussian, $h_\vx$.

We will detect peaks from the re-smoothed image as galaxy candidates and
measure shear from the detected galaxy sample. In order to correct the shear
estimation bias caused by shear-dependent detection at the pixel level, we need
to derive the shear response of each pixel on the re-smoothed image. We
reinterpret the pixel values of the re-smoothed image as a projection of the
continuous image signal onto a \textit{pixel basis function}; here we define
this interpretation, and the lensing shear distortion is taken into account in
Section~\ref{sec:2_pixel_response}. The re-convolved image signal in
configuration space is the inverse Fourier transform of $f^h_\vk$:
\begin{equation}
f^{h}_\vx= \frac{1}{(2\pi)^2} \iint \dd[2]{k}\, f^{h}_\vk e^{\rmi \vk \cdot \vx}
= \frac{1}{(2\pi)^2} \iint \dd[2]{k}\, f_\vk h_\vk e^{\rmi \vk \cdot \vx} \,.
\end{equation}

We focus on one specific pixel located at position $\vx^0$\,. The value of the
image at the location of this pixel is
\begin{equation}
    \label{eq:pixval_onepix}
    f_{\vx^0}^h= \iint \dd[2]{k}\, f_\vk
    \left\{ \frac{1}{(2\pi)^2} e^{-\abs{\vk}^2\sigma_h^2/2}
    e^{\rmi \vk \cdot \vx^0} \right\}\,.
\end{equation}
By interpreting the term in the bracket as the \textit{basis function} for the
pixel at $\vx^0$:
\begin{equation}
\label{eq:resmooth_kernel_prepsf}
\phi^*_{\vx^0}(\vk) = \frac{1}{(2\pi)^2}e^{-\abs{\vk}^2\sigma_h^2/2}
e^{ \rmi \vk \cdot \vx^0}\,,
\end{equation}
we redefine the pixel value as a projection of the image signal in Fourier
space onto the pixel basis function:
\begin{equation}
    f_{\vx^0}^h= \iint \dd[2]{k}\, \phi^*_{\vx^0}(\vk) f_\vk \,\,.
\end{equation}
With this interpretation, we can derive the shear response of a pixel value
using the shear response of its pixel basis function.

\subsubsection{Shear response of re-smoothed pixels}
\label{sec:2_pixel_response}

\begin{figure}
\begin{center}
    \includegraphics[width=0.45\textwidth]{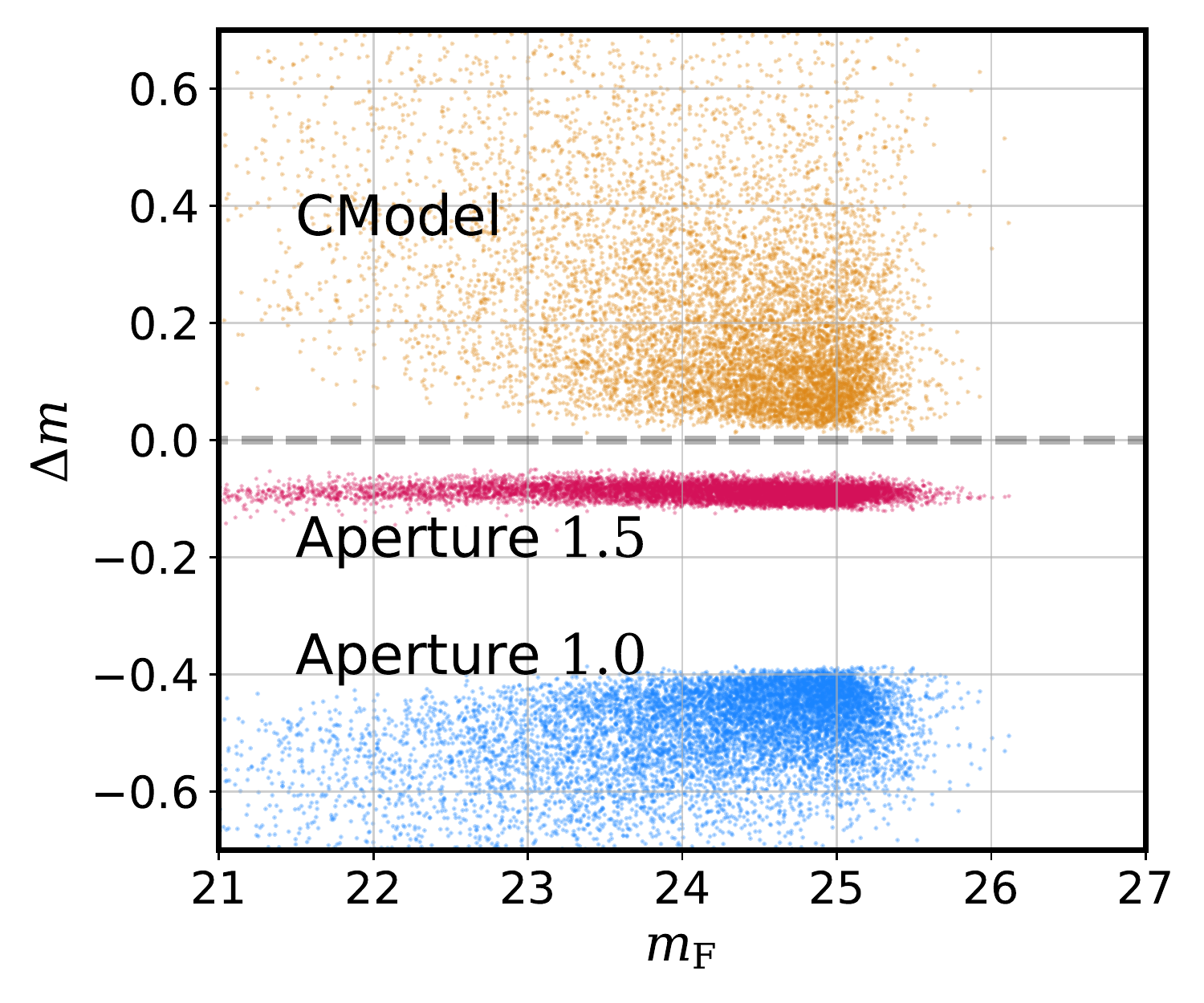}
\end{center}
\caption{
    Scatter plot of the difference between \FPFS{} magnitude and other
    magnitude measurements, $m_\text{other}$, versus \FPFS{} magnitude,
    $m_\text{F}\,$. $\Delta m= m_\text{F} -m_\text{other}$, where
    $m_\text{other}$ is measured by the HSC pipeline \citep{HSC1-pipeline}.
    The three magnitude measurements from the HSC pipeline include aperture
    fluxes with diameter $1\farcs0$ (lower) and $1\farcs5$ (middle), and CModel
    (upper) fluxes. The horizontal dashed line is for $\Delta m=0$\,. As shown
    in the figure, the $1\farcs5$ galaxy aperture flux has the least scatter
    and the smallest offset with respect to the \FPFS{} magnitude since the
    effective smoothing scale of the \FPFS{} galaxy flux is very close to the
    aperture with diameter $1\farcs5$\,.
    }
    \label{fig:compare_fluxes}
\end{figure}

We study the influence of the weak-lensing shear distortion on the re-smoothed
pixel values. In configuration space, shear distorts the local coordinate
system: $\vx=\bm{A^{-1}}\bar{\vx}\,$, and it conserves surface brightness
density: $f_\vx=\bar{f}_{\bar{\vx}}$. With the assumption that the convergence
$\kappa=0$, we have $\vk=\bm{A}\bar{\vk}\,$ and $f_\vk=\bar{f}_{\bar{\vk}}$. Here
$\bar{\vx}$ and $\bar{\vk}$ refer to the coordinates before shear distortion in
configuration space and Fourier space, respectively.

Following our new interpretation introduced in
Section~\ref{sec:2_pixel_basesfun}, the pixel value of the distorted signal is
a projection of the distorted signal onto the pixel basis function defined on
the distorted coordinates:
\begin{equation}
\label{eq:pixval_sheared}
    f^{h}_{\vx^0} =
    \iint \dd[2]{k} \,\phi^*_{\vx^0}(\vk)\,f_\vk=
    \iint \dd[2]{\bar{k}} \,\phi^*_{\vx^0}(\bm{A}\bar{\vk})\,\bar{f}_{\bar{\vk}} \,.
\end{equation}
This equation suggests that we can determine how the smoothed pixel value,
$f^{h}_{\vx^0}\,,$ transforms by studying how the pixel basis function,
$\phi_{\vx^0}(\bm{A}\vk)$, transforms under a shear distortion, keeping the cut value fixed.

Since the lensing shear distortion is small in the weak-lensing regime, we
Taylor expand the basis as a function of $\gamma_{1,2}$ and only keep the first
order of $\gamma_{1,2}$:
\begin{equation}
\label{eq:transformEq_phi}
    \phi_{\vx^0}(\bm{A}\vk)=
    \left( 1 + \gamma_1 \bm{S}_1 + \gamma_2 \bm{S}_2 \right)
    \phi_{\vx^0}(\vk) \,,
\end{equation}
where $\bm{S}_{1,2}$ are the two components of the linear shear distortion
operator in Fourier space. Based on the derivative chain rule, we have:
\begin{equation}
\label{eq:linear_shear_operator}
\bm{S}_1 = -\left(k_1\frac{\partial}{\partial k_1} - k_2\frac{\partial}{\partial k_2}\right)\,, \qquad
\bm{S}_2 = -\left(k_1\frac{\partial}{\partial k_2} + k_2\frac{\partial}{\partial k_1}\right)\,.
\end{equation}
By substituting equation~\eqref{eq:linear_shear_operator} into
equation~\eqref{eq:transformEq_phi}, we find that the responses of the pixel
basis to the two components of shear distortion, ($\gamma_1$,$\gamma_2$), are
\begin{equation}
\label{eq:pixbasis_shearRes}
\begin{aligned}
    \phi_{\vx^0;1}&\equiv \frac{\partial \phi_{\vx^0}}{\partial \gamma_1}=
    \left( (k_1^2-k_2^2)\sigma_h^2
    +\rmi x^0_1\,k_1 - \rmi x^0_2\,k_2 \right) \phi_{\vx^0} \,,\\
    \phi_{\vx^0;2}&\equiv \frac{\partial \phi_{\vx^0}}{\partial \gamma_2}=
    \left( \left(2\,k_1\,k_2\right)\sigma_h^2
    +\rmi x^0_2\,k_1 + \rmi x^0_1\,k_2 \right) \phi_{\vx^0}\,.
\end{aligned}
\end{equation}
Using equations~\eqref{eq:pixval_sheared} and \eqref{eq:pixbasis_shearRes}, we
derive the responses of the pixel value to the two components ($\alpha=1,2$) of
shear distortion
\begin{equation}
\label{eq:pixvalue_shearRes}
    f^{h}_{\vx^0;\alpha} \equiv \frac{\partial f_{\vx^0}^{h}}{\partial \gamma_\alpha}
    =\iint \dd[2]{k} \,\phi^*_{\vx^0;\alpha}\, f_\vk\,.
\end{equation}

\subsubsection{Peak detection from images}
\label{sec:2_pixel_pdetect}

We propose to use eight basis modes for each pixel in order to characterize the
pixel value with respect to other nearby pixels, and thereby identify peaks
which could serve as galaxy candidates. We refer to these as `peak modes'.
Taking a pixel centered at $\vx^0$ as an example, we define the peak modes of
this pixel as
\begin{equation}
    q_{i}= f^{h}_{\vx^0}-f^{h}_{\vx^0
    +\left(\cos{(i \pi/4)}\,,\,\sin{(i \pi/4)}\right)}\,,
\end{equation}
where $i\in \{0,1, \dots, 7\}$, and $\left(\cos{(i \pi/4)}\,,\,\sin{(i
\pi/4)}\right)$ are shifting vectors in the image plane with length equals one.

The shifting vectors have eight different directions separated by $\pi/4$ to
ensure that the peak detection based on these peak modes does not induce any
non-zero spin-$2$ or spin-$4$ anisotropy. The modes with shifting vectors along
the horizontal and vertical directions are the values of the four adjacent
pixels. If we only use these four modes to do galaxy detection, there is a
spin-$4$ leakage in the shear estimation. This leakage would have three
consequences: (1) the average amplitudes of the two components of intrinsic
ellipticity would differ: $\langle \bar{e}_1^2\rangle \neq
\langle\bar{e}_2^2\rangle$\,; (2) the expectation values of the off-diagonal
terms of the shear response matrix, $\langle e_{1;2}\rangle$ and $\langle
e_{2;1}\rangle$, would be nonzero; and (3) the expectation values of the
diagonal terms of the response matrix would not be identical. If this spin-$4$
leakage is not treated correctly (e.g., using the average between
$\mathcal{R}_1$ and $\mathcal{R}_2$ for shear estimation), it would cause a
shear-estimation bias close to $1\%$. Therefore, we use eight basis modes
instead of four.

The corresponding peak basis functions for the peak modes are
\begin{equation}
\label{eq:peakfuncs_define}
\psi_{i}= \phi_{\vx^0}-\phi_{\vx^0+\left(\cos{(i \pi/4)}\,,\,\sin{(i \pi/4)}\right)}\,,
\end{equation}
and the shear responses of these peak basis functions are combinations of the pixel
response functions, $\phi_{\vx;\alpha}$ defined in
equation~\eqref{eq:pixbasis_shearRes}:
\begin{equation}
\label{eq:peakfuncs_response}
\psi_{i;\alpha}=
    \phi_{\vx^0;\alpha}
    -\phi_{\vx^0+\left(\cos{(i \pi/4)}\,,\,\sin{(i \pi/4)}\right);\alpha}\,,
\end{equation}
where $\alpha \in \{1,2\}\,.$ The shear responses of the peak modes,
$q_{i;\alpha}$, can be measured by projecting the prePSF galaxy image signal
onto the shear response functions:
\begin{equation}
\label{eq:peakmode_response}
    q_{i;\alpha}
    =\iint \dd[2]{k} \, \psi_{i;\alpha} \, f_\vk\,.
\end{equation}
We will discuss in detail how to correct the bias from shear-dependent
detection using the shear response of the peak modes in
Section~\ref{sec:2_selection}.

The definitions of peak basis functions and their shear responses are given in
Fourier space with respect to images deconvolved from the PSF (prePSF images).
Since the operations of the PSF deconvolution and the projections onto the
basis functions are commutative and associative, we can combine the PSF
deconvolution with the projection operator, and the combined peak basis
functions and their shear responses are $\psi_i/p_\vk$ and
$\psi_{i;1}/p_\vk$\,. The peak modes and their shear responses can also be
estimated by directly projecting PSF-convolved (postPSF) images onto the
combined bases that are defined with respect to the postPSF images. In
Figures~\ref{fig:psi} and~\ref{fig:psidg1}, we show the combined basis
functions in configuration space, which are the inverse Fourier transforms of
$\psi_i/p_\vk$ and $\psi_{i;1}/p_\vk$\,.

We identify local peaks as potential galaxies by adding a selection weight on
the peak modes, and use the corresponding shear responses of the peak modes to
correct the bias from shear-dependent detection, which will be covered in
Section~\ref{sec:2_selection}.

In real observations, the PSF is spatially varying on an image. To mitigate the
influence of PSF variations, we first convolve the image with the re-smoothing
kernel, and conduct a preselection to find candidates of local peaks from the
re-smoothed images with \textit{loose} selections on peak modes and re-smoothed
pixel vales. After the preselection, we conduct a postselection by measuring
peak modes and the galaxy properties that will be introduced in
Section~\ref{sec:2_gal} for each peak candidate and applying \textit{strict}
cuts on them to select galaxies. For the convolution in the preselection, we
use the average PSF of the image to define the re-smoothing kernel; whereas in
the postselection, we use the PSF model for each peak candidates. If the strict
cuts from the postselection select a conservative subset of the initial peak
candidates, the PSF variation in real observations does not lead to a bias in
the peak identification.

We note that for our detection algorithm, a galaxy that is close to a pixel
edge or corner can be ``detected'' 2 or 4 times, respectively, but with weights
approximately $0.5$ or $0.25$ each time. In addition, the measurement center
for every detection is at the center of the corresponding peak pixel, and
refining the estimated galaxy centroid is not possible since we cannot
analytically compute the shear response of centroid refinement. In future
works, we will further optimize these aspects of our detection algorithm.

\subsection{Shear response of galaxy properties}
\label{sec:2_gal}

\begin{figure}
\begin{center}
    \includegraphics[width=0.45\textwidth]{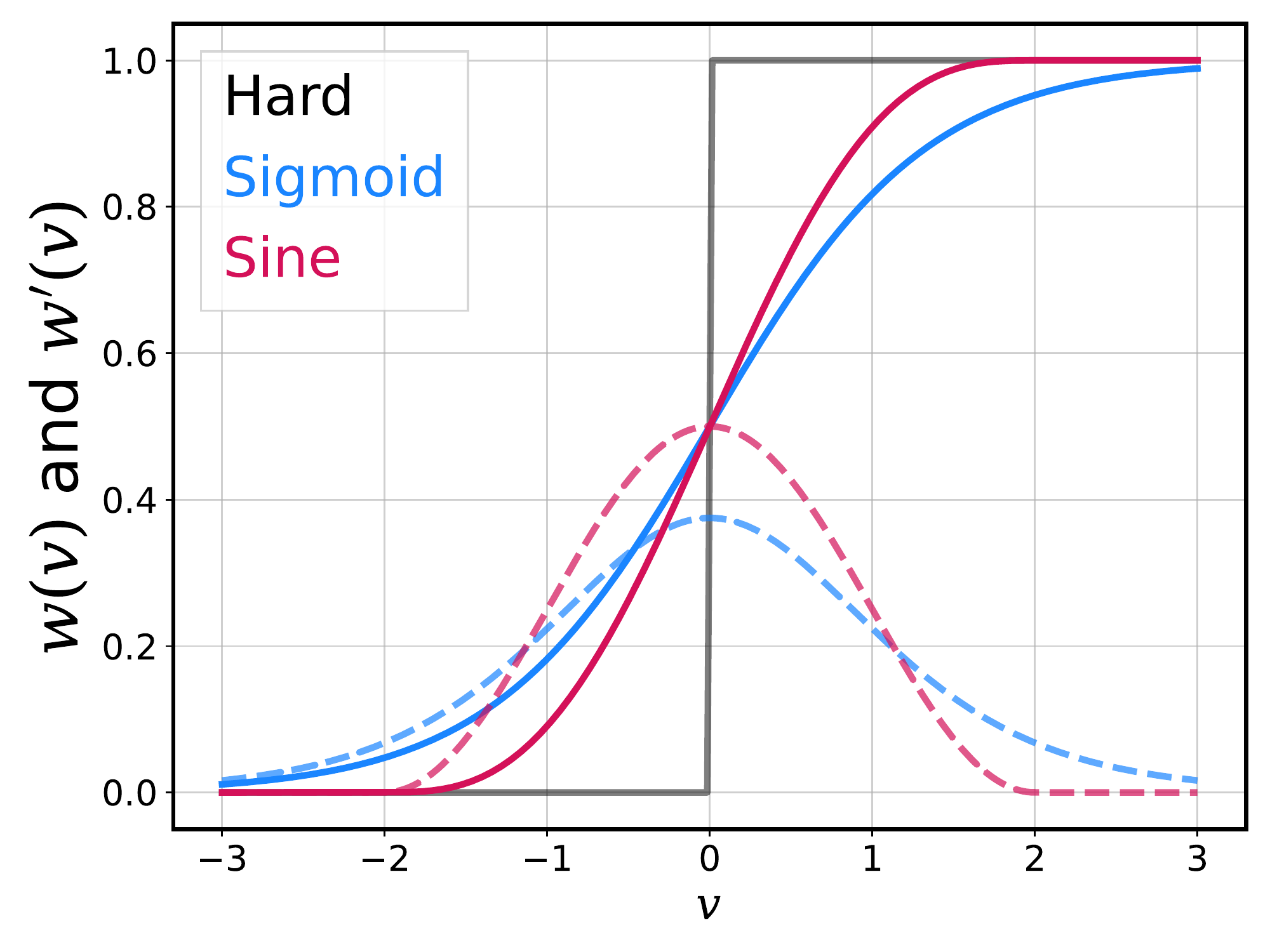}
\end{center}
\caption{
    Selection functions (solid lines) and their first-order derivative (dashed
    lines) of hard, sigmoid (equation~\eqref{eq:sel_func_cinf}) and truncated
    sine (equation~\eqref{eq:sel_func_c2}) cuts. The smoothness parameter is
    set to $\omega=2$\,. We do not show the first-order derivative of the hard
    cut since that goes to infinity at $v=0\,$.
    }
    \label{fig:selfun}
\end{figure}

In addition to the peak modes defined in Section~\ref{sec:2_pixel}, we use a
few observables to quantify the properties of the detected galaxies using polar
shapelet modes (Section~\ref{sec:2_gal_shapelets}). The galaxy properties
studied here include flux (Section~\ref{sec:2_gal_flux}), size
(Section~\ref{sec:2_gal_size}) and shape (Section~\ref{sec:2_gal_ell}).

\subsubsection{Polar shapelets}
\label{sec:2_gal_shapelets}

We construct galaxy properties corresponding to the flux, size, and shape out
of combinations of shapelet modes. Shapelet modes are projections of the galaxy
image signal onto a set of Gaussian weighted orthogonal functions
\citep{shapeletsI-Refregier2003,Shapelets_Massey}. Furthermore, we derive the
first-order shear responses of these galaxy properties using the shear
responses of shapelet modes given by \citet{Shapelets_Massey}. This is
essential to deriving and correcting shear-estimation biases induced by galaxy
selection process using the detected galaxy properties.

The polar shapelet basis functions
\citep{Shapelets_Massey,Shapes_Bernstein2002} are defined as
\begin{equation}\label{eq:shapeletmodes}
\begin{split}
\chi_{nm}(\vx \,|\, \sigma_h)&=(-1)^{(n-|m|)/2}\left\lbrace
    \frac{[(n-|m|)/2]!}{[(n+|m|)/2]!}\right\rbrace^\frac{1}{2}\\
    &\times
    \left(\frac{\rho}{\sigma_h}\right)^{|m|}
    L^{|m|}_{\frac{n-|m|}{2}}\left(\frac{\rho^2}{\sigma_h^2}\right)e^{-\rho^2/2\sigma_h^2}
    e^{-im\theta},
\end{split}
\end{equation}
where $L^{|m|}_{\frac{n-|m|}{2}}$ are the Laguerre polynomials, $n$ is the
radial number and $m$ is the spin number. $n$ can be any non-negative integer,
and $m$ is an integer between $-n$ and $n$ in steps of two. $\sigma_h$
determines the scale of shapelet functions in configuration space, which is set
to the same scale as the Gaussian kernel used to smooth images in the detection
process of Section~\ref{sec:2_pixel_basesfun}. ($\rho$, $\theta$) is used to
denote locations in $2$D polar coordinates, and the center of the coordinate
system is set to the center of the peak of detected source, which does not have
sub-pixel offsets. We scale the polar shapelet basis functions in
\citet{Shapelets_Massey} by $\sigma_h\sqrt{\pi}$ so that $\chi_{00}(\vx)$
(which is an isotropic Gaussian function) is normalized to integrate to one:
\begin{equation}
\iint \dd[2]{x} \chi_{00}(\vx)=1\,,
\end{equation}
and the re-smoothing kernel also integrates to one as the PSF integrates to one.

One important property of shapelet basis functions is that under a Fourier
transform, the shapelet basis functions change as
\begin{equation}
\chi_{nm}(\vx \,|\, \sigma_h) \longrightarrow \tilde{\chi}_{nm}(\vk)
= \rmi^n\chi_{nm}\left(\vk\,|\,1/\sigma_h\right) \,.
\end{equation}
The shapelet function in Fourier space, $\tilde{\chi}_{nm}\,$, has the same
functional form as $\chi_{nm}$ but the scale radius is the inverse of that of
the function in configuration space \citep{shapeletsI-Refregier2003}.

We measure shapelet modes from galaxy light profiles in Fourier space after PSF
deconvolution:
\begin{equation}
\label{eq:shapelets_modes}
    M_{nm} \equiv \iint \dd[2]{k} \, \left(\tilde{\chi}_{nm}(\vk)\right)^{*}
    \frac{f^p_\vk}{p_\vk}\,.
\end{equation}
Again, we set the scale radius of the Gaussian weight in shapelets, $\sigma_h$,
to be greater than the scale radius of the PSF in configuration space. Note
that we use an isotropic Gaussian kernel, and its size and shape are {\it not}
adapted to those of the galaxies or PSFs. We do not adapt the basis functions
to the size of the galaxy light profile, in order to simplify the derivation of
shear responses of basis modes without assumptions regarding galaxy morphology.
In real observations, we can separate the survey into small patches and choose
a fixed smoothing scale, $\sigma_h$, for each patch. $M_{00}$ is the same as
the smoothed pixel value of the central peak; we refer to it as the \FPFS{}
peak value.

Note that the shapelet modes in this paper are different from the \FPFS{}
shapelet modes defined in \citet{FPFS-Li2018,FPFS_Li2022}.  Here we measure
shapelet modes from the deconvolved galaxy profiles in Fourier space, rather
than from the power of the galaxy's Fourier transform. The reason for this
difference is as follows: The old \FPFS{} method follows \citet{Z08} to avoid
shear-dependent off-centering bias by measuring observables on the power in
Fourier space after subtracting the average noise power \citep{Z15}. In this
paper, we set the peaks detected from images as the center of the coordinate
system for the measurement. Since we derive the shear response of the peak
detection, the anisotropy in the centering definition is corrected with the
shear response. This non-power based estimator significantly simplifies the
derivation of the covariance between different \FPFS{} shapelet modes, not only
from read noise and background photon noise, but also from source photon noise.
We will expand and elaborate on this in Section~\ref{sec:2_covariance}.

Shapelet modes change when a galaxy image is distorted by a shear,
$\gamma_{1,2}\,$. We use the transform formula for shapelets --- equation~(41)
of \citet{Shapelets_Massey} --- to derive the shear responses of the flux, size
and shape defined in terms of shapelet modes. In summary, to the first order in
shear, a \textit{finite} number of shapelet modes (separated by $|\Delta n|=2$
and $|\Delta m|=2$) are coupled under shear distortion; therefore, one can
write the shear response of any shapelet mode as a linear combination of a {\it
finite} number of other shapelet modes \citep{FPFS-Li2018}. As we will define
galaxy size and shape using finite combinations of shapelet modes, their shear
responses are also finite combinations of shapelet modes. It is worth
mentioning that the \FPFS{} shapelet modes are measured from deconvolved
galaxies; therefore, PSFs do not bias the derivation, assuming that they are
accurately determined at the positions of the galaxies.

\subsubsection{Galaxy flux}
\label{sec:2_gal_flux}

\begin{figure*}
\begin{center}
    \includegraphics[width=0.90\textwidth]{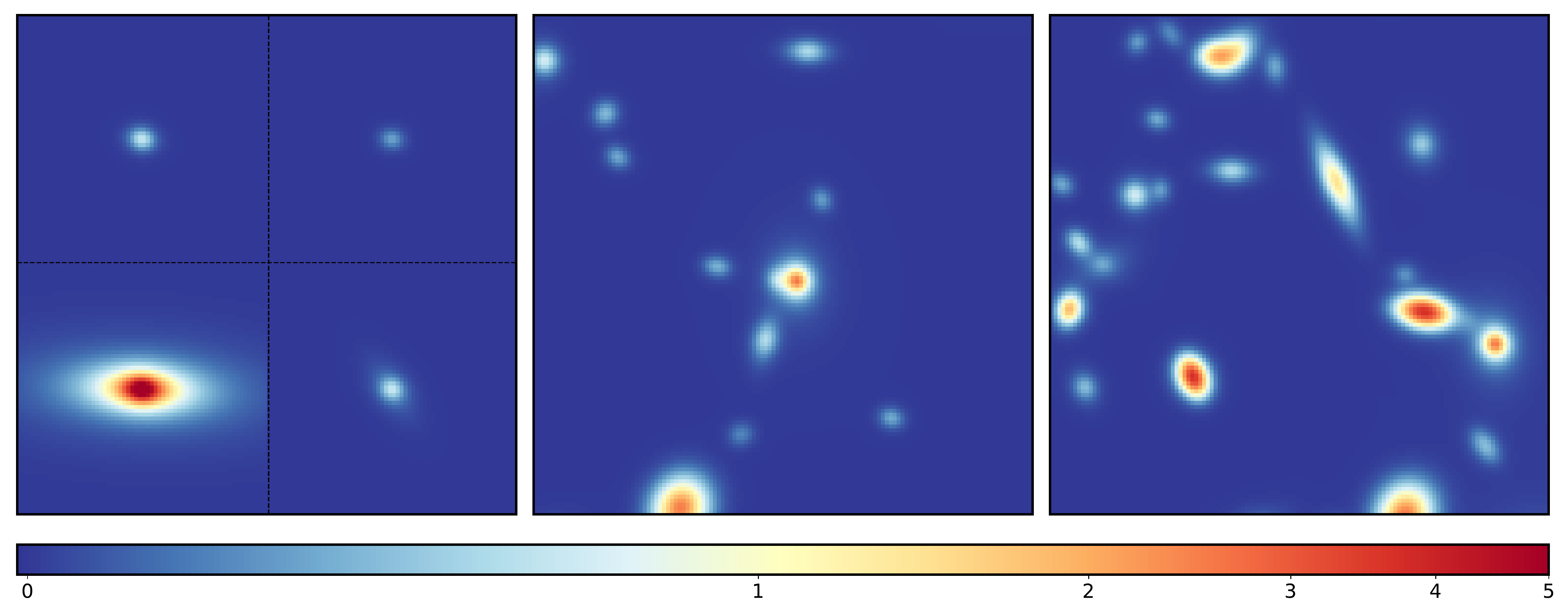}
\end{center}
\caption{
    The left panel shows a $128~\mathrm{pix}\times128~\mathrm{pix}$ (or,
    equivalently, $0.36\times 0.36~\mathrm{arcmin}^2$) region of the isolated
    stamp-based galaxy image simulation, where the black lines indicate the
    boundaries of the $64~\mathrm{pix}\times64~\mathrm{pix}$ stamps. The middle
    and the right panels are for blended galaxy image simulations with input
    densities of $85~ \text{arcmin}^{-2}$ and $170~\text{arcmin}^{-2}$,
    respectively. The images shown here are noiseless for demonstration, but
    different noise realizations are added to the galaxy images during the
    tests of shear estimation.
    }
    \label{fig:sim_isoblend}
\end{figure*}

The zeroth order \FPFS{} shapelet mode, $M_{00}$\,, is the value of the peak
pixel, which can be used to quantity the brightness of galaxies. Since the
re-smoothing kernel integrates to one, we define the \FPFS{} flux as a
rescaling of the peak value:
\begin{equation}
\label{eq:flux_define}
   F = \frac{M_{00}}
   {\iint \dd[2]{k} \abs{\tilde{\chi}_{00}(\vk)}^2/\abs{p_\vk}^2}\,,
\end{equation}
where the denominator is the square of the $L^2$ norm of the re-smoothing
kernel. $F$ is the flux of the best fit re-smoothing kernel to the observed
galaxy profile assuming the noise variance is a constant over the galaxy scale.
This is mathematically similar to the prePSF Gaussian
flux\footnote{\url{https://github.com/esheldon/ngmix}} proposed by
\citet{prePSF_moments2022}.

On raw astronomical images before photometric calibration, $M_{00}$ is in units
of [photon counts per $\mathrm{arcsec}^{2}]$; in contrast, in this paper, we
focus on coadded images of HSC / LSST after photometric calibration, and
$M_{00}$ is reported in units of $[\mathrm{nano\,Jy~arcsec}^{-2}]$. Since the
denominator in equation~\eqref{eq:flux_define} is in unit of
$[\mathrm{arcsec}^{-2}]$, $F$ is in unit of $[\mathrm{nano\,Jy}]$\,. The
\FPFS{} magnitude is defined as
\begin{equation}
    \label{eq:def_mF}
    m_\text{F} \equiv m_\text{zero} - 2.5 \log(F)\,,
\end{equation}
where $m_{\text{zero}}$ is the zero point of the survey (e.g., for HSC coadded
images\footnote{Coadded images are the weighted sum of images (after smoothing
with a warping kernel) from multiple visits at the same position in the same
optical band \citep{HSC1-pipeline}.}, the zero point is $27$). The galaxy
number histogram as a function of \FPFS{} magnitude for isolated galaxies in
the simulation that will be introduced in Section~\ref{sec:3_sim_iso} is shown
in the top panel of Figure~\ref{fig:obsHist_fixed}. When producing this figure,
we do not run the detection or selection processes during the image processing,
but rather tell the pipeline the centroids of the galaxies. In this paper, we
refer to this setup as the forced-center setup.

Under a shear distortion, $M_{00}$ (\FPFS{} peak value) changes from its
intrinsic value, $\bar{M}_{00}$\,, and the linear shear responses of the
\FPFS{} peak value for two shear components are
\begin{equation}
\label{eq:shear_response_m00}
    M_{00;1}=-\sqrt{2}M_{22c}\,, \qquad
    M_{00;2}=-\sqrt{2}M_{22s}\,,
\end{equation}
where $M_{22c}$ and $M_{22s}$ refer to the real (`$\cos{}$') and imaginary
(`$\sin{}$') components of the complex shapelet mode $M_{22}$, respectively.
These shear responses for the \FPFS{} peak value will be used to derive the
shear response of the galaxy detection/selection using the \FPFS{} magnitude
and to correct the detection/selection bias induced by the shear-dependence of
$M_{00}$ in Section~\ref{sec:2_selection_multi}.

In Figure~\ref{fig:compare_fluxes}, we show the relationship between the
\FPFS{} magnitude and the other magnitudes measured by the HSC pipeline
\citep{HSC1-pipeline} on noiseless, isolated galaxies with HSC-like pixel scale
($0\farcs168$) and seeing size ($0\farcs59$). The Gaussian shapelet kernel used
to measure $M_{00}$ has a FWHM of $1\farcs4$\,. The \FPFS{} galaxy magnitudes
are fainter than the CModel magnitudes by about $0.3$ magnitude on average
since the effective aperture scale of the shapelets Gaussian kernel is less
than the typical galaxy scale; as a result, not all of the light from galaxies
is included in the effective PSF-deconvolved Gaussian window. The aperture
magnitudes with diameter $1\farcs5$ have the smallest average offset from and
least scatter with respect to the \FPFS{} magnitudes. This is because the
aperture size is close to the effective smoothing scale of re-smoothing kernel
adopted by the \FPFS{} magnitudes.

\subsubsection{Galaxy size}
\label{sec:2_gal_size}

Second-order Gaussian weighted moments are widely used to quantify the size of
galaxies (see, e.g., \citealt{Regaussianization}). We decompose the spin-0
second-order radial moment as a combination of shapelet modes:
\begin{equation}
    \iint \dd[2]{x}\, f_\vx \, \left(\frac{\rho}{\sigma_h}\right)^2
    e^{-\rho^2/2\sigma_h^2}= M_{00}+M_{20}\,,
\end{equation}
and define the \FPFS{} resolution using shapelet modes:
\begin{equation}
    \label{eq:def_r2}
    R_2\equiv \frac{M_{00}+M_{20}}{M_{00}}\,.
\end{equation}
Note that this shapelet resolution is conceptually similar to the \reGauss{}
resolution defined in \citet{Regaussianization}. The galaxy number histogram as
a function of \FPFS{} resolution for the isolated galaxy image simulations with
the forced-center setup,  which assumes the galaxies' centroids are known
during the image processing, is shown in the lower panel of
Figure~\ref{fig:obsHist_fixed}.

Since the shear responses of $M_{20}$ for the two shear components are
\begin{equation}
\label{eq:shear_response_m20}
M_{20;1}=-\sqrt{6}M_{42c}\,, \qquad
M_{20;2}=-\sqrt{6}M_{42s}\,,
\end{equation}
according to \citet{Shapelets_Massey}, the linear shear responses of the
\FPFS{} resolution can be derived with equation~\eqref{eq:shear_response_m00}:
\begin{equation}
\begin{split}
R_{2;1} &=
    \sqrt{2} \left(
    \frac{M_{22c}M_{20}}{\left(M_{00}\right)^2}\right)
    -\sqrt{6} \frac{M_{42c}}{M_{00}}\,,\\
R_{2;2} &=
    \sqrt{2} \left(
    \frac{M_{22s} M_{20}}{\left(M_{00}\right)^2}\right)
    -\sqrt{6} \frac{M_{42s}}{M_{00}}\,.
\end{split}
\end{equation}

\subsubsection{Galaxy ellipticity}
\label{sec:2_gal_ell}

We define galaxy ellipticity as
\begin{align}
\label{eq:ellipticity_define_v}
e_1\equiv\frac{M_{22c}}{M_{00}+C}\,,\qquad
e_2\equiv\frac{M_{22s}}{M_{00}+C}\,.
\end{align}
The weighting parameter $C$, introduced by \citet{FPFS-Li2018}, adjusts the
relative weight between galaxies with different brightness; moreover, it
ensures that the second-order noise bias in equation~\eqref{eq:ell_noisy} is
the leading-order term and we can neglect the higher-order terms. While this
ellipticity definition follows the basic form suggested by \citet{FPFS-Li2018},
it differs in that this ellipticity is measured from the galaxy's Fourier
transform, rather than from the Fourier power.

To the first order in shear, $e_{\alpha}$ transforms under the shear
distortion, where $\alpha,\beta \in \{1,2\,\}$, as follows:
\begin{equation}\label{eq:ellipticity_transform}
    \bar{e}_\alpha\rightarrow{e}_\alpha=
    \bar{e}_\alpha+ \sum_{\beta=1,2} e_{\alpha;\beta} \gamma_\beta \,,
\end{equation}
and the \textit{shear response} matrix is $e_{\alpha;\beta} \equiv \partial
e_\alpha/\partial\gamma_\beta\,$ \citep{metacal-Huff2017}.
In \citet{FPFS-Li2018}, we use the linear shear responses of shapelets
given by \citet{Shapelets_Massey} to derive the expectation values of the
diagonal elements of the matrix:
\begin{align}\label{eq:response_define}
\left\langle e_{\alpha;\alpha} \right\rangle=
\frac{1}{\sqrt{2}}\left\langle s_0-s_4+2e_{\alpha}^2\right\rangle\,,
\end{align}
where $s_{0,4}$ are spin-$0$ properties:
\begin{align}
    s_{0} \equiv \frac{M_{00}}{M_{00}+C}\,,\qquad
    s_{4} \equiv \frac{M_{40}}{M_{00}+C}\,.
\end{align}
The off-diagonal terms of the matrix are composed of spin-$4$ shapelet modes,
which go to zero when averaging over a large number of galaxies if the
selection of the galaxy sample does not cause any spin-$4$ leakage.

The definitions of ellipticity, in  equation~\eqref{eq:ellipticity_define_v},
and its shear response, equation~\eqref{eq:response_define}, indicate that
their expectation values go to zero for a star sample if the PSF is precisely
and accurately modelled.
Moreover, as shown in Figure~(8) of \citet{FPFS_Li2022}, our noise bias
correction can remove the non-zero second-order term of measurement error,
$(\delta e_\alpha)^2$ in the shear response. Therefore, stellar contamination
in the galaxy sample should not bias shear estimation.

This shear response of the ellipticity, $\left\langle e_{\alpha;\alpha}
\right\rangle$, can only be used to estimate the ensemble weak-lensing shear if
there is no shear-dependent detection/selection, shear-dependent weighting, or
image noise. The shear estimator for this over-simplified case is
\begin{equation}
    \widehat{\gamma}_\alpha=\frac{\langle e_\alpha \rangle}
    {\left\langle e_{\alpha;\alpha} \right\rangle}\,,
\end{equation}
and we will revisit this in detail when deriving corrections for detection and
selection biases in Section~\ref{sec:2_selection}.

\subsection{Shear response of detection/selection}
\label{sec:2_selection}

We define the selection weight functions, $w$ in
equation~\eqref{eq:ell_average}, corresponding to cuts on \FPFS{} peak modes
(Section~\ref{sec:2_pixel_basesfun}), \FPFS{} magnitude
(Section~\ref{sec:2_gal_flux}) and \FPFS{} resolution
(Section~\ref{sec:2_gal_size}), respectively. The cuts on the peak modes are
used to identify peaks in images as galaxy candidates; the cuts on \FPFS{} peak
flux and resolution are used to select galaxies that are sufficient for
weak-lensing science. Normally, cuts on peak modes are regarded as part of the
galaxy detection process, and cuts on magnitude and resolution are regarded as
part of the galaxy sample selection process. For our purpose we describe these
using a common formalism and treat them both as a form of selection.

We begin from selection using one observable following \citet{FPFS_Li2022} in
Section~\ref{sec:2_selection_one}, and then generalize the formalism to cuts on
multiple observables in Section~\ref{sec:2_selection_multi}.

\subsubsection{One-observable selection}
\label{sec:2_selection_one}

In \citet{FPFS_Li2022}, we analytically derived and corrected for the selection
bias caused by a hard cut on one observable, $v$:
\begin{equation}
    v > \mu \,,
\end{equation}
by studying how the edge of the cut changes under shear distortion. A different
approach is adopted here: we follow the formalism introduced in
Section~\ref{sec:2_formalism_shear} to derive the detection/selection bias
correction by considering how the histogram of galaxy properties is shifted
under a shear distortion.

Following \citet{KaiserFlow2000}, we interpret a hard cut on one observable as
a Heaviside step selection weight applied to the galaxy sample:
\begin{equation}
w_\text{H}(v)=
\begin{cases}
    1 & \text{if } v > \mu\\
    0 & \text{if } v \leq \mu
\end{cases}\,.
\end{equation}
From equations~\eqref{eq:ell_sheared} and \eqref{eq:ell_average_response}, the
shear dependency of the selection weight leads to an additional anisotropy that
is proportional to the shear:
\begin{equation}
\begin{split}
\Delta\langle  w e_\alpha \rangle_{\text{sel}}
    =\gamma_\alpha
    \iint \rmd e\rmd v \,\mathcal{P}(e,v) \frac{\partial w_\text{H}(v)}{\partial v}
    \frac{\partial v}{\partial \gamma_\alpha}e_\alpha \,,
\end{split}
\end{equation}
where $\mathcal{P}(e,v)$ is the $2$D PDF of the galaxy sample.
Since $\partial w(v)/\partial v=\delta_\mathrm{D}(v-\mu)$, where
$\delta_\mathrm{D}$ is the Dirac delta function, we have
\begin{equation}
\label{eq:selbias_onesel_hard}
\Delta \langle w e_\alpha \rangle_{\text{sel}}
    =\gamma_\alpha
    \left. \mathcal{P}\left( v \right) \right|_{v=\mu}
    \left.\left\langle e_\alpha
    \frac{\partial v}{\partial \gamma_\alpha}
    \right\rangle \right|_{v=\mu}\,,
\end{equation}
where $\left. \mathcal{P}(v)\right|_{v=\mu}$ is the marginal PDF at $v=\mu\,$.
Equation~\eqref{eq:selbias_onesel_hard} is identical to equation~(31) of
\citet{FPFS_Li2022}, although we derive them from different perspectives.

\subsubsection{Multi-observable selection}
\label{sec:2_selection_multi}

In real observations, we need to apply cuts on multiple galaxy properties.
However, we find that the correction for noise bias in
equation~\eqref{eq:selbias_onesel_hard} is very noisy and unstable especially
when applying a hard selection weight function to multiple galaxy cuts.  This
instability arises because the hard selection weight, $w_\text{H}$, is
discontinuous at the selection boundary, and the estimate of the marginal
number density at multiple cuts is unstable. To address this issue, we explore
whether the application of a selection weight that is differentiable,
corresponding to a soft cut, might make the process of correcting for the
selection bias due to multiple cuts more stable.

The sigmoid function is a smooth function with continuous derivatives up to
infinite order (as indicated by the subscript in $w_\infty$):
\begin{equation}
\label{eq:sel_func_cinf}
w_\infty (v\,|\, \omega)=
    \frac{1}{1+\exp\left(-\frac{v}{\omega}\right)} \,,
\end{equation}
which is used to avoid the discontinuity in the hard selection function. The
parameter $\omega>0$, which we refer to as the smoothness parameter, changes
the average slop of the sigmoid function --- the sigmoid function approaches a
step function as $\omega$ approaches 0. However, the sigmoid function is not
compact --- that is, it does not go precisely to zero even for $v\ll 0$.

We propose to use a truncated sine function for the selection weight:
\begin{equation}
\label{eq:sel_func_c2}
w_2 (v\,|\, \omega)=
\begin{cases}
    0   & \text{if }  v \in (-\infty, -\omega)\\
        \frac{1}{2}+ \frac{v}{2\,\omega}
        + \frac{1}{2\pi}\sin\left( \frac{v \pi}{\omega} \right)
        & \text{if } v \in [-\omega, \omega]\\
    1   & \text{if }  v \in (\omega, +\infty)
\end{cases}\,.
\end{equation}
This truncated sine function has continuous derivatives up to second-order (as
indicated by the subscript in $w_2$), and it goes precisely to zero below
$-\omega\,$. The hard, sigmoid and truncated sine selection functions and their
first-order derivatives are shown in Figure~\ref{fig:selfun}. The compactness
of the truncated sine function enables us to neglect the galaxies with
observable $v<-\omega$ during the detection and selection process.

We define the cut on peak modes, $q_i$, where $i \in \{0,..,7\}$, with a
smoothness parameter $\omega_q$, as
\begin{equation}
\label{eq:sel_cut_peak}
    w_{\mu_q \omega_q}=\prod_{i=0}^{7}\,
    w_2(q_i - \mu_q M_{00} - \omega_q \,|\, \omega_q)\,.
\end{equation}
For this selection, the mean cutoff is $\mu_q M_{00} + \omega_q$, and the width
is $\omega_q$, so that this soft cut removes all galaxies with $q_i < \mu_q
M_{00}$ and select peaks from the image. In addition, this cut downweights
galaxies between $\mu_q M_{00}$ and $\mu_q M_{00} + 2\omega_q$. Note that here
we introduce $\mu_q M_{00}$ to avoid false detections near bright sources. We
fix $\mu_q$ to a small value ($5\times 10^{-3}$) in this paper.

For a cut on \FPFS{} magnitude, $m_\text{F}$, at $\mu_m$, with a smoothness
parameter $\omega_m$, the selection weight function is given by
\begin{equation}
\label{eq:sel_cut_mag}
    w_{\mu_m \omega_m}=
    w_2\left( F - 10^{\frac{m_\text{zero}-\mu_m}{2.5}} \,|\, \omega_m \right)\,.
\end{equation}
Note that to reduce one level of non-linearity, we transform the cut on
magnitude into a cut on $F$, which is a linear function of the image signal.
The mean cutoff on $F$ is $10^{\frac{m_\text{zero}-\mu_m}{2.5}}$, and the
cutoff width is $\omega_m$\,.

For a cut on \FPFS{} resolution, $R_2$, at $\mu_R$, with a smoothness parameter
$\omega_R$, the selection weight function is given by
\begin{equation}
\label{eq:sel_cut_res}
    w_{\mu_R \omega_R}=
    w_2\left( M_{20}+(1-\mu_R)M_{00}\,|\, \omega_R \right)\,.
\end{equation}
Here, too, we reduce one level of non-linearity by transforming the cut on
resolution, $R_2$, into a cut on $M_{20}+(1-\mu_R)M_{00}$, which is a linear
function of the image signal. Following its definition, a cut on \FPFS{}
resolution at $\mu_R$:
\begin{equation}
    \frac{M_{20}+M_{00}}{M_{00}}>\mu_R\,
\end{equation}
is equivalent to a cut:
\begin{equation}
    M_{20}+(1-\mu_R)M_{00}>0\,.
\end{equation}
Then we approximate this hard cut on the linear observable,
$M_{20}+(1-\mu_R)M_{00}$\,, with the soft truncated sine function.

The final selection weight used to select the galaxy sample is
\begin{equation}
\label{eq:sel_cutw_final}
w=w_{\mu_q \omega_q} w_{\mu_m \omega_m} w_{\mu_R \omega_R}\,.
\end{equation}
Since we have calculated the first order shear responses of peak modes and
shapelet modes in Sections~\ref{sec:2_pixel} and \ref{sec:2_gal}, we can derive
the derivative of the selection weight function to shear. Therefore, the shear
response of the average of weighted ellipticity with the correction for
shear-dependent detection/selection bias {\em in the absence of noise} can be derived following
equation~\eqref{eq:ell_average_response}, and the shear estimator is
\begin{equation}
\label{eq:shear_estimate_noiseless}
    \widehat{\gamma}_\alpha=
    \frac{\left\langle w e_{\alpha}\right\rangle}
    {\left\langle w e_{\alpha;\alpha} + w_{;\alpha} e_{\alpha} \right\rangle}\,,
\end{equation}
where $e_\alpha$ is defined in equation~\eqref{eq:ellipticity_define_v} and
$e_{\alpha;\alpha}$ is in equation~\eqref{eq:response_define}.

Our method of correction for detection and selection bias differs from that of
\metadet{} \citep{metaDet-Sheldon2020}. They correct for biases from
shear-dependent detection and selection by shearing each galaxy image and
rerunning the detection and selection processes. In contrast, we adopt a
coordinate-based approach by analytically deriving the basis functions' shear
responses, which is functionally equivalent to shearing the basis functions
(but analytically rather than via a numerical re-simulation process).  However,
the common aspect to these approaches is that they seek an empirical correction
for selection and detection biases, rather than relying on calibration from
simulations.

\subsection{Covariance of measurement error}
\label{sec:2_covariance}

The shear estimator in equation~\eqref{eq:shear_estimate_noiseless} neglects
the bias from image noise. Due to the non-linearity in the galaxy ellipticity
and selection weight, image noise biases both the measured ellipticity and the
selection weight. In order to estimate the second-order noise bias corrections
for the observables in equation~\eqref{eq:shear_estimate_noiseless} (e.g., the
noise bias correction for the average of weighted ellipticity in
equation~\eqref{eq:ell_noisy_correct}), we analytically derive the covariance
of the measurement errors on the basis modes defined in
Section~\ref{sec:2_pixel} (peak modes) and \ref{sec:2_gal} (shapelet modes).

We denote the image noise in configuration space and Fourier space as $n_\vx$
and $n_\vk\,$, respectively. Here we focus on noise fields that are homogeneous
on the length scales of galaxies, but that have  pixel-to-pixel correlations.
Image noise on coadded images of ground-based observations (e.g., HSC and LSST)
is approximately homogeneous on the length scales of typical galaxies, since
the image noise of ground-based observations is dominated by sky background
noise, which varies on larger scales. The correlation of noise between pixels
is caused by the warping kernel when resampling the images to a common pixel
grid in the coaddition process (see, e.g., \citealt{HSC1-pipeline}). Noise
correlation on single exposures can also be caused by the brighter-fatter
effect and inter-pixel capacitance of CMOS detector
\citep{sysRoman_Givans2022}. Additionally, undetected faint galaxies can
effectively also lead to correlation across pixels
\citep{undetSource_Eckert2020}.

The noise homogeneity on galaxy length scales means that there is no
correlation between the noise on different wave numbers in Fourier space. The
Fourier power spectrum of noise is denoted as $N_\vk$, and the covariance of
noise in Fourier space is
\begin{equation}
    \langle {n}_{\vk} n_{\vk'}^{*} \rangle= N_\vk \delta_\mathrm{D}(\vk-\vk')\,,
\end{equation}
where $\delta_\mathrm{D}$ is the Dirac delta function.

Since the \FPFS{} basis modes are linear functions of images, the corresponding
measurement errors are the noise projected onto the shapelet bases after
deconvolution:
\begin{equation}
\begin{split}
    \delta{M}_{nm}&= \iint \dd[2]{k} \,
    \left(\tilde{\chi}_{nm}(\vk)\right)^{*} \frac{n_\vk}{p_\vk}\\
    \delta{q}_{\vx;\alpha}& = \iint \dd[2]{k}
    \left(\psi_{\vx;\alpha}(\vk)\right)^* \frac{n_\vk}{p_\vk}\,.
\end{split}
\end{equation}
Taking the covariances between shapelet modes $M_{nm}$ and $M_{n'm'}$ as an
example, the covariances are
\begin{equation}
\begin{split}
K_{M_{nm}}^{M_{n'm'}}\equiv&
    \left \langle \delta{M}_{nm} \delta{M}^*_{n'm'} \right\rangle=
    \iint \dd[2]{k} \tilde{\chi}_{nm}^{*}\,
    \tilde{\chi}_{n'm'} \frac{N_\vk}{P_\vk}\,.\\
\end{split}
\end{equation}
In the real observations, these covariances can be estimated if we can measure
the noise power function, $N_\vk$, from blank pixels, and the Fourier power
function of PSF, $P_\vk\,$, from the PSF model:
\begin{equation}
P_\vk=p_\vk p^*_\vk\,.
\end{equation}
The covariance between the other basis modes has the same form. It is worth
noting that the covariances of measurement errors do not depend on galaxy
properties, unlike those derived from the galaxy Fourier power in
\citet{FPFS_Li2022}. Moreover, the above equations do not assume that the noise
is Gaussian.

It is worth noting that, for images of space-based observations, image noise is
dominated by galaxy photon noise that is not homogeneous. As we show in
Appendix~\ref{app:inhomogeneous_noise}, the covariances of basis modes can be
analytically derived for inhomogeneous noise if it does not have pixel-to-pixel
correlation. This is approximately true for single exposures of space-based
observations, as noise on single exposures has very small pixel-to-pixel
correlations. This paper focuses on homogeneous noise; we leave the discussion
of inhomogeneous noise to our future work.

\subsection{Shear estimation recipe}
\label{sec:2_summary}

Finally, we summarize the process of applying our shear estimator defined in
equation~\eqref{eq:shear_estimator} with our specific implementation of
ellipticity and selection weights. This code for the pipeline is
public\footnote{\url{https://github.com/mr-superonion/FPFS}}, and it can carry
out the detection, selection, and shear measurement for $\sim$$1000$ galaxies
per CPU~second.

We briefly summarize the pipeline for ensemble shear estimation as follows:
\begin{enumerate}
    \item
        Carry out a loose preselection of peaks as galaxy candidates after PSF
        deconvolution and re-smoothing with a target Gaussian kernel;
    \item Compute the selection weight defined in
        equation~\eqref{eq:sel_cutw_final} for each preselected peak. This
        selection weight is constructed with shapelet modes
        (Section~\ref{sec:2_gal_shapelets}) and peak modes
        (Section~\ref{sec:2_pixel_pdetect}) and used for the postselection of
        galaxies sufficient for weak-lensing science;
    \item Measure the spin-$2$ ellipticity defined in
        equation~\eqref{eq:ellipticity_define_v} for each postselected galaxy.
        The ellipticity is constructed with shapelet modes
        (Section~\ref{sec:2_gal_shapelets});
    \item Estimate the shear response of the ellipticity according to
        equation~\eqref{eq:response_define} and the shear response of the
        selection weight following Section~\ref{sec:2_selection_multi}. These
        estimates use the shear responses of shapelet modes
        (equations~\eqref{eq:shear_response_m00} and
        \eqref{eq:shear_response_m20}) and those of peak modes
        equation~\eqref{eq:peakmode_response};
    \item Estimate the shear response of the average weighted ellipticity using
        equation~\eqref{eq:ell_average_response};
    \item Correct for the noise biases in the expectation values of the
        weighted ellipticity and its shear response following
        equations~\eqref{eq:app_noicor_n}--\eqref{eq:app_noicor_d2}. The noise
        bias correction uses the covariance matrix of measurement errors on
        basis modes introduced in Section~\ref{sec:2_covariance} and the
        Hessian matrix of the ellipticity and its shear response as functions
        of basis modes summarized in Appendix~\ref{app:noirev};
    \item Shear is estimated with equation~\eqref{eq:shear_estimator}, which
        incorporates corrections for detection and selection bias, and for
        noise bias.
\end{enumerate}

\section{IMAGE SIMULATION}
\label{sec:sims}
\begin{figure*}
\centering
\begin{minipage}[t]{0.45\textwidth}
\begin{center}
    \includegraphics[width=1.\textwidth]{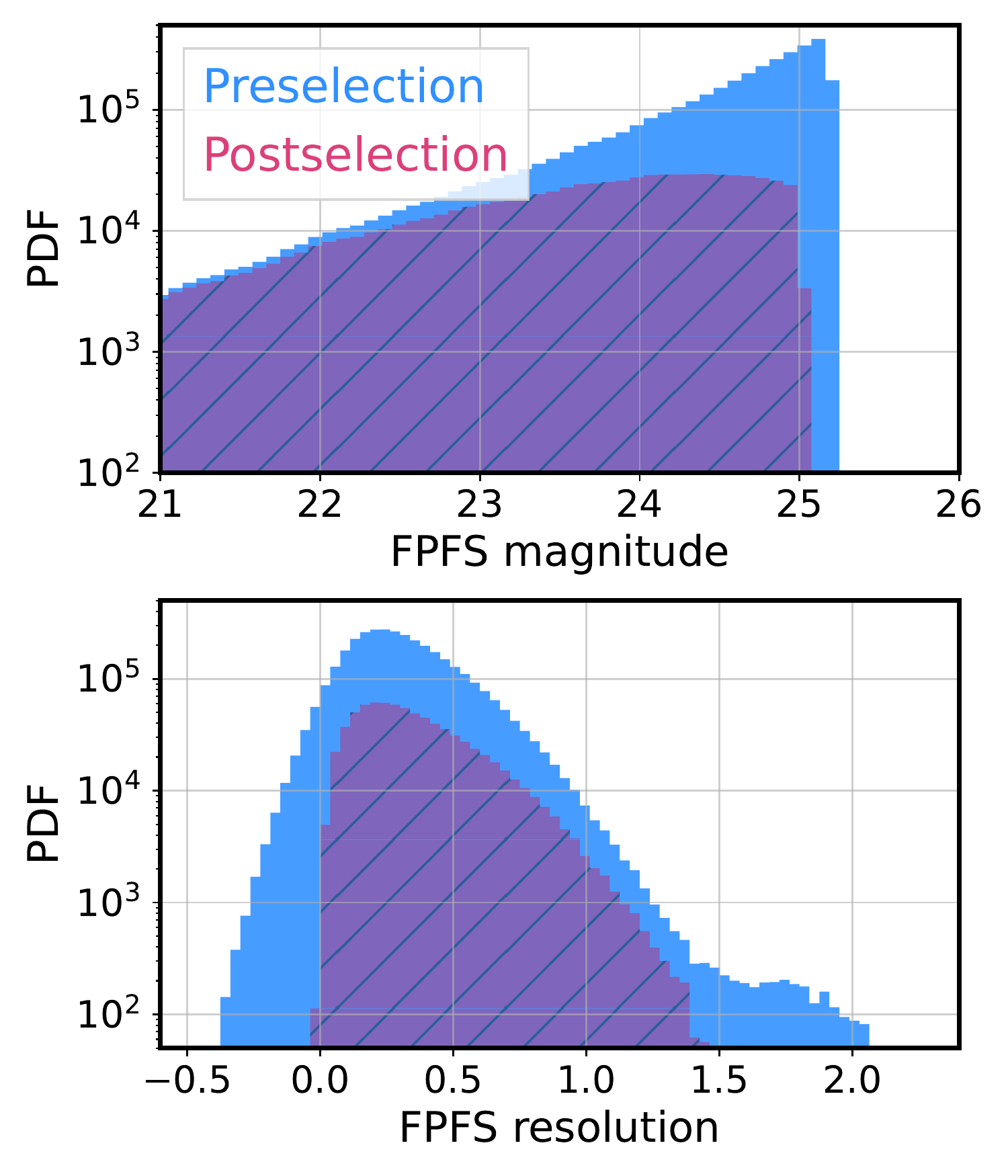}
\end{center}
\caption{
    The \FPFS{} magnitude ($m_\text{F}$; defined in equation~\eqref{eq:def_mF})
    and \FPFS{} resolution ($R_2$; defined in equation~\eqref{eq:def_r2})
    distributions for the preselection (unhatched) and postselection (hatched)
    galaxy samples from the stamp-based image simulations, where galaxies have
    a random sub-pixel offset from stamp's center.
    }
    \label{fig:obsHist_shifted}
\end{minipage}
\hfill
\begin{minipage}[t]{0.45\textwidth}
\begin{center}
    \includegraphics[width=1.\textwidth]{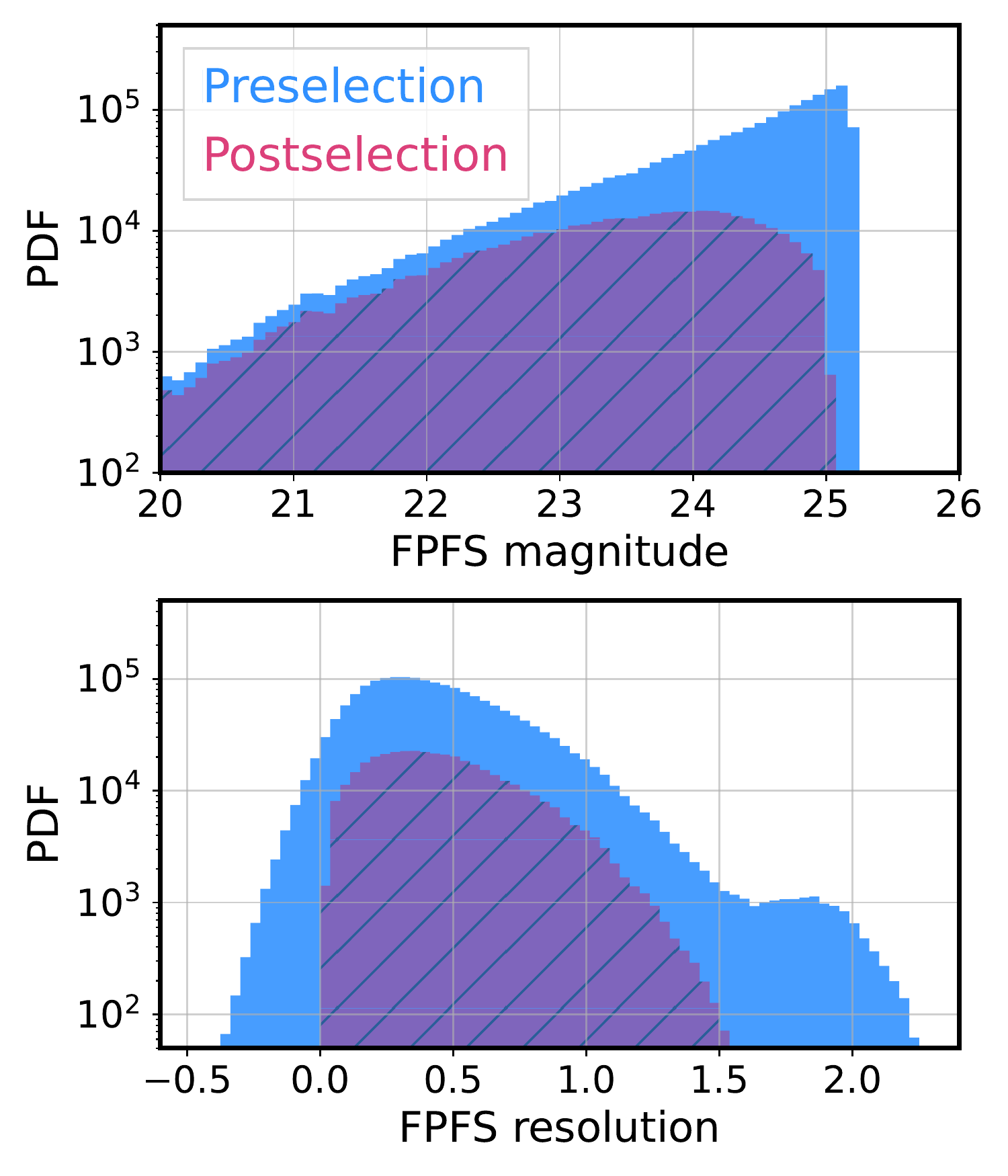}
\end{center}
\caption{
    Same as Figure~\ref{fig:obsHist_shifted}, but for galaxies in the blended
    image simulation with input number density $85~\mathrm{arcmin}^{-1}$\,.
    }
    \label{fig:obsHist_blended}
\end{minipage}
\end{figure*}

We test the performance of the shear estimator after correcting for noise bias,
selection bias and detection bias by analyzing mock astronomical images that
have been distorted by known input shears. The estimated shear,
$\widehat{\gamma}_{1,2}$, is related to the input shear, $\gamma_{1,2}$, as
\begin{equation} \label{eq_shear_biases}
    \widehat{\gamma}_{1,2}=(1+m_{1,2})\,\gamma_{1,2}+c_{1,2}\,,
\end{equation}
where $m_{1,2}$ (multiplicative bias) and $c_{1,2}$ (additive bias) are used to
quantify the accuracy of the shear estimator
\citep{shearSys_Huterer2006,galsim-STEP1}.

\subsection{Galaxies, PSF and noise}
Galaxy images are generated using the open-source package \galsim{}
\citep{GalSim}. We use the COSMOS HST Survey
catalogue\footnote{\url{https://zenodo.org/record/3242143\#.YPBGdfaRUQV}}
\citep{cosmoGalModel252} with limiting magnitude $F814W=25.2$\, as our input
galaxy catalog. The galaxies' light profiles are approximated with the
best-fitting single S\'ersic model \citep{Sersic1963} or two-component
bulge-disk model \citep[with the bulge component
following][]{deVauProfile1948}. This parametric galaxy catalogue can be used
directly for image rendering by \galsim{} (for details, see
\citealt{galsim-GREAT3}). We truncate each input parametric galaxy model at
five times its half-light radius. Each parametric model is expanded by a random
number uniformly distributed between $0.95$ and $1.05$ to change its size,
while not adjusting its flux. Then galaxies are rotated by a random angle
before they are distorted by the input shear.

The pixel scale is set to $0\farcs168\,$, which is the pixel scale of HSC
coadded images. For these simulations, the PSF image is modelled
with a
\cite{Moffat1969} profile,
\begin{equation}\label{Moffat PSF}
    p_{m}(\mathbf{x})=\left[1+c\left(\frac{|\mathbf{x}|}
    {r_\mathrm{P}}\right)^2\right]^{-3.5},
\end{equation}
where $c=2^{0.4}-1$ and $r_\mathrm{P}$ is set such that the full width half
maximum (FWHM) of the PSF is $0\farcs60$, matching the mean seeing of the HSC
survey in the $i$-band \citep{HSC3-catalog}. However, we note that the Moffat
PSF profile is different from the actual PSFs in HSC. We use the HSC
pipeline to measure the FWHM of the input Moffat PSF with a Gaussian weighted
moment, and the resulting FWHM is $0\farcs65\,$. The difference between the
input FWHM and measured FWHM is due to the fact that the measurement algorithm
assumes a Gaussian profile. The PSF profile is truncated at a radius four times
of its FWHM. In order to test whether our algorithm can remove shear estimation
bias from PSF anisotropy, we shear the PSF so that it has ellipticity
$(e_1=0.02, e_2=-0.02)\,$.

We add image noise composed of photon noise from a constant sky background and
read noise, neglecting the contribution of photon noise from galaxy. This is
because photon noise from the sky background dominates over that from galaxies
for ground-based observations. We leave detailed treatment of source Poisson
noise to future work. Our noise model includes anisotropic correlation between
pixels matching the autocorrelation function of a third-order Lanczos kernel,
i.e.,\ $a=3$ in
\begin{equation}
    L(x,y)=\begin{cases}
        \sinc{\frac{x}{a}}\,\sinc{x}\,\sinc{\frac{y}{a}}\,\sinc{y}
        & \mathrm{if}~\abs{x},\abs{y}<a\\
        0
        & \mathrm{otherwise,}
    \end{cases}
\end{equation}
where $\sinc{x}=\sin{(\pi x)}/\pi x$. This kernel was used to warp and coadd
exposures for the HSC survey dataset \citep{HSC1-pipeline}. Ignoring
pixel-to-pixel correlations, our resulting noise variance is $7\times 10^{-3}
~\mathrm{nano\,Jy}$, which is approximately two times the average noise
variance on HSC coadds in \citet{HSC3-catalog}.  We chose this higher noise
level to ensure that our algorithm can be applied to the HSC survey even for
its noisiest images.

\subsection{Shape noise and image noise cancellation}

To reduce the intrinsic shape noise in our tests, enabling us to tightly
constrain shear biases with fewer simulations, we generate two images for each
galaxy distorted by different shears --- ($\gamma_1=0.02$, $\gamma_2=0$) and
($\gamma_1=-0.02$, $\gamma_2=0$). However, the images share exactly the same
realisation of image noise
\citep[following][]{preciseSim-Pujol2019,metaDet-Sheldon2020}. In addition, we
force our galaxy sample to contain orthogonal galaxies with the same morphology
and brightness but the major axes rotated by $90 \deg$ following
\citet{galsim-STEP2}.

To measure the shear measurement bias defined in
equation~\eqref{eq_shear_biases}, we measure multiplicative bias and additive
bias as
\begin{equation}
\label{eq:cbias_estimator}
    {c_1} = \frac{\langle \widehat{w e_1}^+ + \widehat{w e_1}^- \rangle}
    {( \widehat{\mathcal{R}}_{1}^{+} + \widehat{\mathcal{R}}_{1}^{-} )}
\end{equation}
and
\begin{equation}
\label{eq:mbias_estimator}
    {m_1} = \frac{\langle \widehat{w e_1}^+ - \widehat{w e_1}^- \rangle}
    {0.02(\widehat{\mathcal{R}}_{1}^{+} + \widehat{\mathcal{R}}_{1}^{-})}-1\,,
\end{equation}
where $\widehat{w e_1}^+$ and $\widehat{\mathcal{R}}^{+}_{1}$ are the first
component of the weighted ellipticity and the shear response for its
expectation value, respectively. They are estimated from the images distorted
by the positive shear, $(\gamma_1=0.02, \gamma_2=0)\,$. $\widehat{w e}_1^-$ and
$\widehat{\mathcal{R}}^{-}_1$ are from images with the negative applied shear,
$(\gamma_1=-0.02, \gamma_2=0)$\,. The bias estimators in
equations~\eqref{eq:cbias_estimator} and \eqref{eq:mbias_estimator} assume that
$\widehat{\mathcal{R}}_{1}^{+} = \widehat{\mathcal{R}}_{1}^{-}$\, which is true
for our simulation since the input galaxy sample is the same for the images
with positively and negatively distorted galaxies.
For these very well-sampled images, anisotropy in the horizontal/vertical
versus diagonal directions due to the pixel response function is considered to
be part of the effective PSF. Given that our correction for the impact of PSF
dilution on shear inference is accurate, the multiplicative and additive biases
we find for $\widehat{\gamma}_2$ should be comparable to those for
$\widehat{\gamma}_1\,$. Therefore, we only confirm that the results for
$\widehat{\gamma}_2$ are consistent with those for  $\widehat{\gamma}_1$ for
noiseless images. To save computational time, we do not test the noise bias
correction for estimation of $\widehat{\gamma}_2$, which requires a large
number of additional image simulations. It is worth noting that this testing
scheme reduces statistical error in multiplicative bias due to image noise. The
errors on the means of multiplicative bias and additive bias are estimated with
jackknife resampling of the galaxies.

The galaxies in each orthogonal galaxy pair and galaxies with different applied
shears are selected (weighted) independently, and we apply the shear estimator
to the selected sample to test our corrections for detection bias and selection
bias.

\subsection{Isolated and blended setups}

We prepare both isolated and blended image simulations so that we can
separately quantify the shear estimation biases for isolated galaxies and those
related to blending.

\subsubsection{Isolated galaxies}
\label{sec:3_sim_iso}
For the isolated image simulations, we randomly select galaxies from the HST
parametric galaxy catalogue. After the shear distortion and PSF convolution,
galaxies are rendered in $64\times 64$\,pixel postage stamp images.
The convolved galaxies are further truncated by the boundaries of their postage
stamp. Each simulated image contains $100\times100$ postage stamps, and each
postage stamp contains an isolated galaxy randomly selected from the input
COSMOS galaxy sample with replacement (each galaxy is selected repeatedly).
Each image has $5\times10^3$ orthogonal galaxy pairs with identical
morphologies and fluxes, but with major axis directions separated by
$90\deg$\,. We show a small region of one simulated image in the left panel of
Figure~\ref{fig:sim_isoblend}.
We prepare simulations with two setups: one puts the galaxy centroid at the
center of each postage stamp; the other shifts galaxies with random sub-pixel
offsets. For each setup, we generate $3000$ images with different realizations
for the image noise, galaxy sample, random galaxy rotation and size expansion
factor. Galaxies in each of the images are distorted by $\gamma_1=\pm 0.02$,
meaning that there are two versions of each image.

We run the \FPFS{} detection and selection process on $200$ of the $3000$
simulated stamp-based images, and show the preselection and postselection
galaxy number histograms as functions of \FPFS{} magnitude and resolution in
Figure~\ref{fig:obsHist_shifted}. As mentioned in
Section~\ref{sec:2_pixel_pdetect}, the preselection process applies hard cuts
on the re-smoothed pixel values and \FPFS{} magnitude using loose threshold
values: $q_i > -0.12~[\mathrm{nano\,Jy}]$ and $m_\text{F}<25.2$\,. However, the
postselection process is a soft selection with stricter thresholds on
magnitude, peak modes and resolution to select galaxies that are sufficient for
weak-lensing science, as defined in Section~\ref{sec:2_selection_multi}. The
smoothness parameters for the soft postselection are set to
$\omega_q=0.2~\mathrm{[nano~Jy]}$, $\omega_R=0.2~\mathrm{[nano~Jy]}$ and
$\omega_m=0.2~\mathrm{[nano~Jy]}$\,, and the cutoff centers for magnitude and
resolution are $m_\text{F}=25$ and $R_2=0.05$\,. Note that we do not add a very
conservative resolution cut since we are not particularly worried about star
contamination in our shear estimation since, as shown in Figure (8) of
\citet{FPFS_Li2022}, the ellipticity and shear response of noisy stars all
average to zero. The magnitude cut corresponds to SNR$\sim$$12.5$\,. In the
following tests, we apply these soft cuts for galaxy detection and selection by
default.

\subsubsection{Blended galaxies}
\label{sec:3_sim_blended}

For the simulations to test the impact of blending, the HST parametric galaxies
are rendered into random positions on the images instead of dividing images
into postage stamps. We generate $10^4$ images with different realizations for
the galaxy sample, galaxy positions, random rotation angle, size expansion
factor and image noise. The galaxies are evenly distributed within a circle
centered at the image's center. The size of the image is designed to be $10\%$
larger than the diameter of the circle. The number density of the input
galaxies is set to $85$~arcmin$^{-2}$, which is similar to the default setup of
\citet{DESY3-BlendshearCalib-MacCrann2021}. In \citet{HSC3-catalog}, we checked
that the number density of galaxies with CModel magnitude brighter than $24.5$
and \reGauss{} resolution greater than $0.3$ detected from simulations with
this input galaxy number density matches the number density of the
magnitude-limited galaxy sample in the HSC survey. In addition, we produce
images with a higher number density, $170$~arcmin$^{-2}$, for an approximate
stress test of the algorithm under extreme conditions (e.g., near the center of
a galaxy cluster -- albeit without attempting to simulate in detail the
different galaxy population that would be present in a cluster). The center of
the shear distortion is fixed to the image center, so that it changes both the
shape and position of each galaxy. The images of blended galaxy simulations
with different number densities are shown in the middle and right panel of
Figure~\ref{fig:sim_isoblend}.

It is worth noting that \citet{DESY3-BlendshearCalib-MacCrann2021} tested
\metacal{} using simulations with redshift-dependent shear by dividing galaxies
into four redshift bins and applying different shears to galaxies in each bin.
They found that blended galaxies at different redshifts change the galaxies'
effective number density distribution as a function of redshift since the shape
measured from one galaxy contains information from other galaxies blended with
the measured one but located at different redshifts. However, in this paper, we
do not investigate the changes in effective redshift distribution, and only
focus on the case that all galaxies in one image are distorted by the same
shear. In addition, we do not include clustering in the galaxy's spatial
distribution. With clustering, it is more likely for blended galaxies to be
located in the same redshift bin; therefore, clustering can change the
effective redshift distribution. However, we leave the tests related to
redshift-dependent shear to the future work.

We run the \FPFS{} detection and selection process on $200$ images of the
blended image simulation with input galaxy number density
$85~\mathrm{arcmin}^{-1}$\,, and we show the preelection and postselection
galaxy number histograms as functions of \FPFS{} magnitude and resolution in
Figure~\ref{fig:obsHist_blended}. Comparing the bottom panels of
Figures~\ref{fig:obsHist_shifted} and \ref{fig:obsHist_blended}, we find that
for the blended image simulation, there are many preselection galaxies with
extremely large resolution and there is a secondary peak in the resolution
histogram at $R_2=2$\,. Many of these extremely large galaxies are false
detections near bright sources. However, those galaxies in the secondary peak
are removed after the postselection since we have a conservative cut on peak
modes as shown in equation~\eqref{eq:sel_cut_peak}.

\section{RESULTS}
\label{sec:results}
\begin{figure*}
\centering
\begin{minipage}[t]{0.45\textwidth}
\begin{center}
    \includegraphics[width=1.\textwidth]{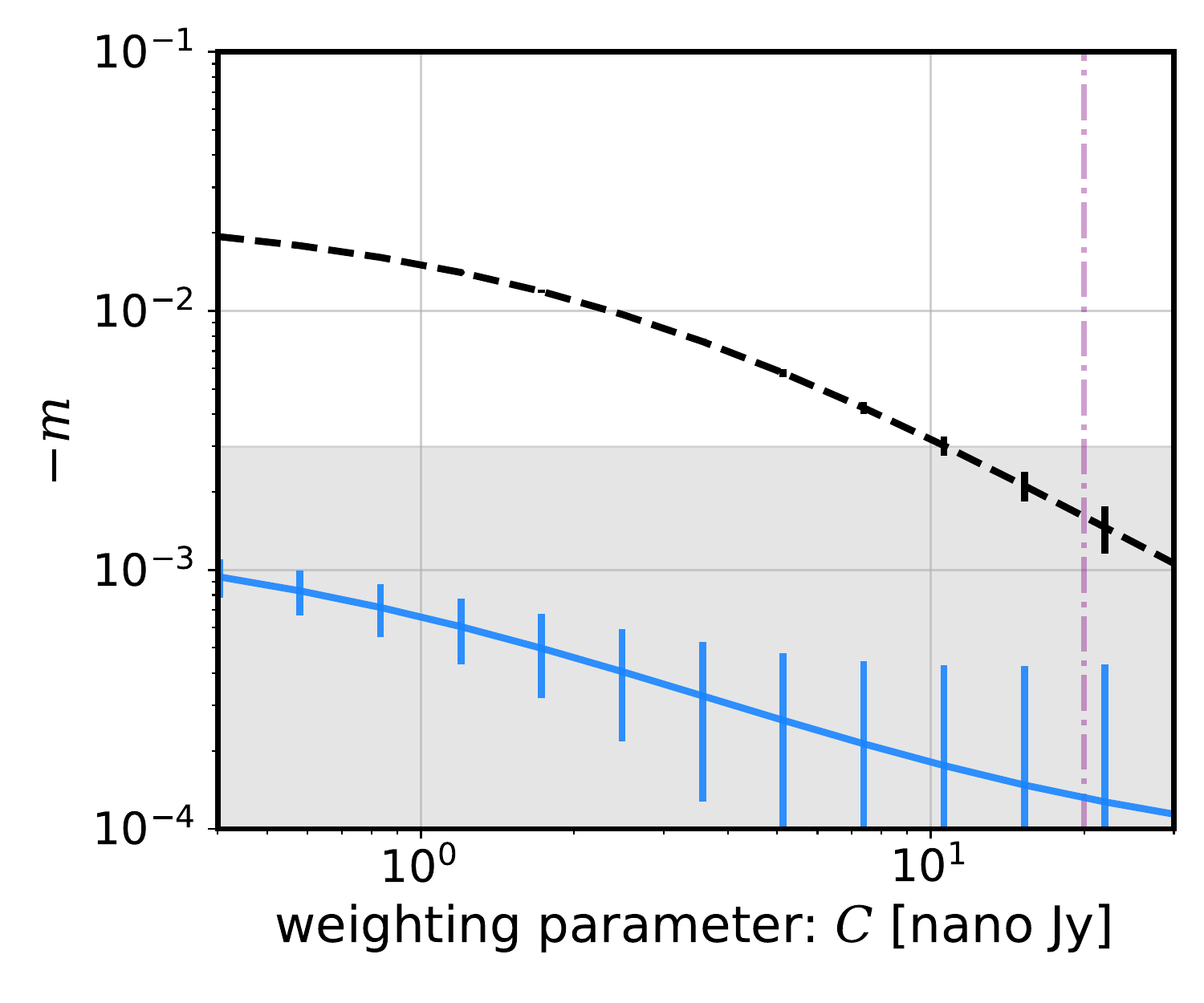}
\end{center}
\caption{
    The multiplicative bias caused by image noise (i.e., noise bias) as a
    function of weighting parameter, $C\,$. The solid and dashed lines show the
    results with and without the second-order noise bias correction,
    respectively. All measured multiplicative biases are negative. The gray
    region denotes the LSST ten-year requirement on the control of
    multiplicative bias, which is defined for redshift-dependent multiplicative
    bias \citep{LSSTRequirement2018}. For this figure, galaxies are isolated,
    and no detection or selection processes were carried out, which has enabled
    us to isolate the impact of noise bias more specifically. The vertical
    dash-dotted line is the default value for $C$\,.
    }
    \label{fig:res_noiBias}
\end{minipage}
\hfill
\begin{minipage}[t]{0.45\textwidth}
\begin{center}
    \includegraphics[width=1.\textwidth]{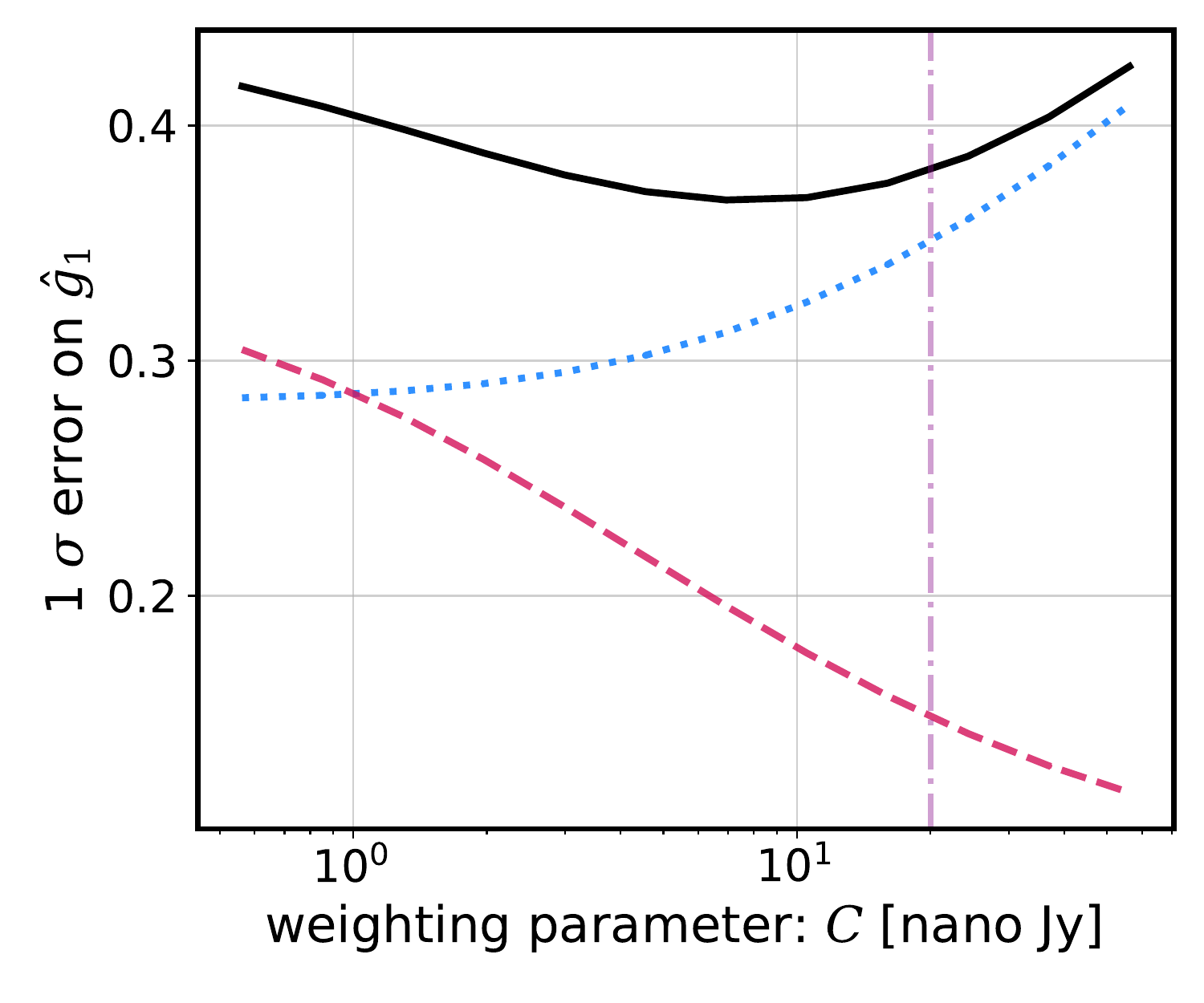}
\end{center}
\caption{
    The $1\sigma$ statistical uncertainty on shear measurements
    $\widehat{\gamma}_1$ at the individual galaxy level (solid line), as a
    function of weighting parameter, $C\,$. The total uncertainty has
    contributions due to measurement error (dashed line) and intrinsic shape
    noise (dotted line). We separate shape noise from measurement error using
    noisy galaxy images and the corresponding noiseless galaxy images. The
    vertical dash-dotted line is the default setup for $C$\,.
    }
    \label{fig:res_optimalC}
\end{minipage}
\end{figure*}

\subsection{Isolated galaxies}
\label{sec:4_isogals}

In this subsection, we focus on isolated galaxies simulated within postage
stamps (Section~\ref{sec:3_sim_iso}). We test the \FPFS{} shear estimator under
two different conditions. For the first setup, we do not run the \FPFS{}
detection process, but rather place each galaxy at the center of the postage
stamp and force a measurement for each galaxy based on that known center. We
test the correction for noise bias on galaxy shape estimation without applying
any cut on the galaxy sample, and then we apply flux- and resolution-based
selection criteria to the galaxy sample to test our corrections for selection
biases. For the second setup, each galaxy has a random sub-pixel offset from
the center of the postage stamp, and we run the \FPFS{} peak detection
algorithm and measure galaxy properties using the detected peak as the
centroid. We apply flux- and resolution-based selection criteria to the
detected galaxy sample and compare the results with those for the forced-center
condition, to isolate the biases related to galaxy detection.

\subsubsection{Forced center}
\label{sec:4_iso_force}
We first use the forced-center simulations to test the noise bias correction
without galaxy selection and detection. We will use our results to set the
parameter $C$, which enters into the denominator of the ellipticity definition
in equation~\eqref{eq:ellipticity_define_v}.
To be more specific, we use all of the galaxies in the simulation by setting
the selection weight function, $w$, to 1 for all galaxies. As shown in
Figure~\ref{fig:res_noiBias}, after the second-order noise bias correction with
equation~\eqref{eq:ell_noisy_correct}, the amplitudes of multiplicative biases
are reduced by at least one order of magnitude. In addition, the noise biases,
both before and after the second-order correction, decrease as functions of the
weighting parameter, $C$. This is consistent with the mathematical derivation
in \citet{FPFS_Li2022}: the noise bias residuals are proportional to the second
and fourth powers of $\delta M_{nm}/(M_{00}+C)$ before and after the
second-order correction, respectively. We find that the additive bias is below
$10^{-4}$\, for both corrected and uncorrected shear estimators when
$C>1~[\mathrm{nano~Jy}]$\,; therefore, we do not plot the additive bias here.

According to \citet{FPFS-Li2018}, not only the accuracy, but also the precision
of the estimated shear, depends on the weighting parameter, $C$\,. As shown in
Figure~\ref{fig:res_optimalC}, the shape noise increases as a function of
$C$\,, whereas the measurement error due to image noise decreases as a function
of $C$\,. The trends are consistent with what we found in Figure~(3) of
\citet{FPFS_Li2022}. Note, since the non-power-based shear estimator
adopted here has lower measurement error compared to the power-based shear
estimator adopted in \citet{FPFS_Li2022}, the total statistical uncertainty
(including both shape noise and measurement error) reduces by about $6\%$\,. To
focus on the shear estimation bias from detection and sample selection, we
conservatively set the default weighting parameter for the following tests to
$C=20~[\text{nano~Jy}]$, with the intention of controlling the noise bias in
the shear estimates to the level of $\sim$$1\times10^{-4}$. As shown in
Figure~\ref{fig:res_optimalC}, this choice only increases the statistical
uncertainty by $5\%$ from the optimal value.

In Figures~\ref{fig:res_center_magcut} and \ref{fig:res_center_r2cut}, we apply
different selections on \FPFS{} magnitude and \FPFS{} resolution to select
samples from the simulated galaxies, and run the shear estimator on the
selected samples with and without correction for selection bias (including
noise bias in the selection). For both cases, noise biases in the ellipticity,
$\tilde{e}_{\alpha}$\,, and in its shear response,
$\tilde{e}_{\alpha;\alpha}$\,, are corrected. As discussed in
Section~\ref{sec:2_selection_multi}, we are not using hard cuts for galaxy
sample selection but rather a soft selection weight defined in
equation~\eqref{eq:sel_func_c2} to approximate the normal hard selection cut.
Since our adopted selection weight is differentiable up to second order, the
second-order selection bias from the soft selection cut can be analytically
corrected very effectively. The characteristics of the detected sample given
the default setup of the soft selection used here are shown in
Figure~\ref{fig:obsHist_shifted}.

We find the amplitude of multiplicative selection bias before the analytical
correction is $\sim$$2\%$\,. After the analytical bias correction, the
amplitudes of multiplicative biases shown in the top panels of
Figures~\ref{fig:res_center_magcut} and \ref{fig:res_center_r2cut} are below
$0.1\%$, which is within the LSST DESC science requirement ($\abs{m}<0.3\%$,
see \citealt{LSSTRequirement2018} for details). After applying our corrections,
the amplitudes of  additive biases in the bottom panel of
Figure~\ref{fig:res_center_magcut} and \ref{fig:res_center_r2cut} are below
$1\times 10^{-4}$\,. Therefore, we conclude that our analytical correction
reduces the selection bias to the sub-percent level for isolated galaxies. This
result is consistent with the selection bias correction with a hard selection
cut on a single property, shown in Figure~5 of \citet{FPFS_Li2022}.

Note that the tests above only include galaxy sample selection using one
observable. We will test for shear biases due to selections combining \FPFS{}
magnitude, resolution and peak modes together in the following sections.

\subsubsection{Detected center}
\label{sec:4_iso_detect}
\begin{figure*}
\centering
\begin{minipage}[t]{0.45\textwidth}
\begin{center}
    \includegraphics[width=1.\textwidth]{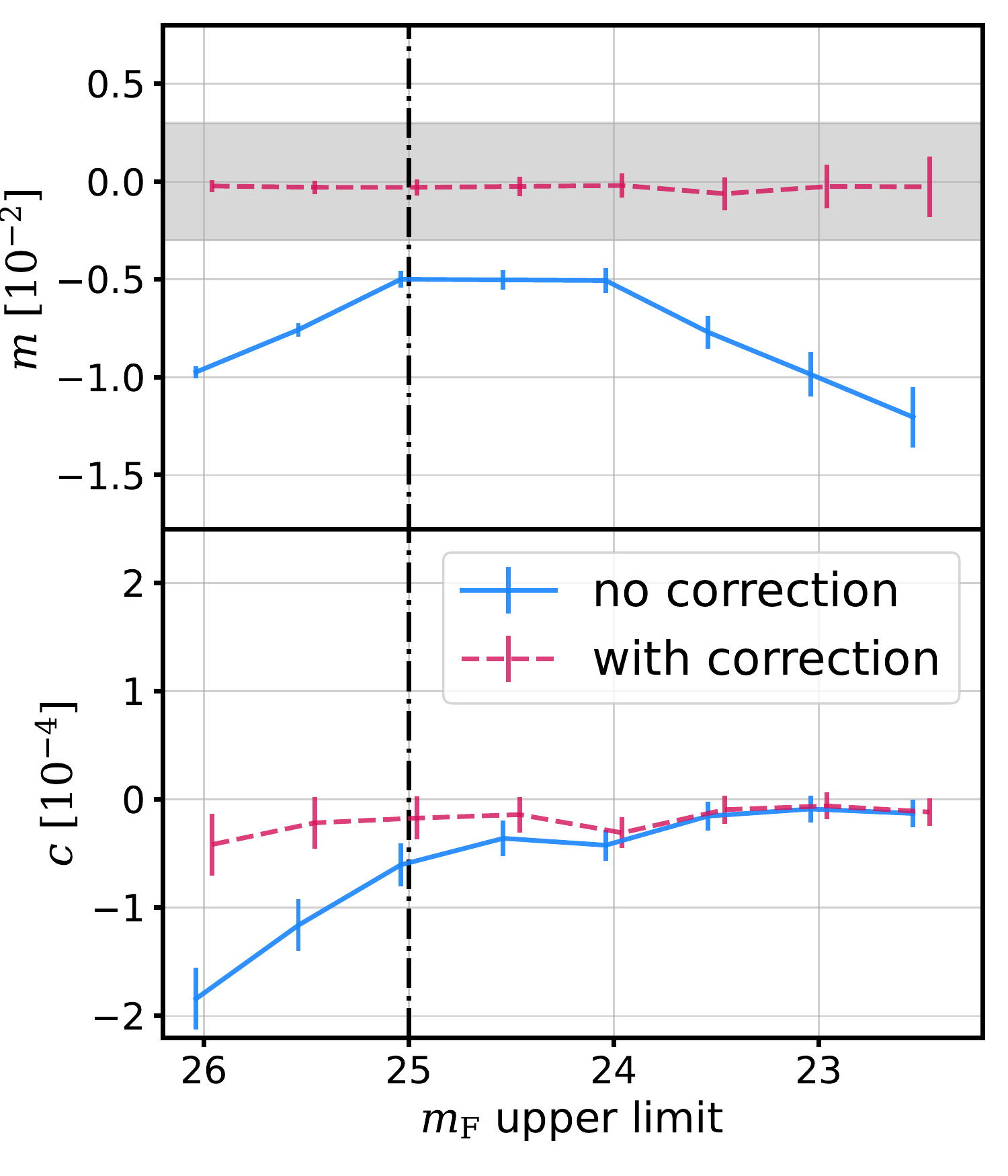}
\end{center}
\caption{
    The multiplicative bias (top panel) and additive bias (bottom panel) in
    shear as functions of the upper limit of \FPFS{} magnitude ($m_\text{F}$)
    measured from the isolated image simulation. The coordinate center for the
    measurement is set to the truth, so no detection process was carried out.
    We do not apply any cut on other observables. The solid (dashed) lines are
    results before (after) the correction for shear-dependent selection bias.
    The dark shaded region is the LSST DESC requirement on the control of
    multiplicative shear bias. The vertical dash-dotted line is the default cut
    on magnitude of the postselection.
    }
    \label{fig:res_center_magcut}
\end{minipage}
\hfill
\begin{minipage}[t]{0.45\textwidth}
\begin{center}
    \includegraphics[width=1.\textwidth]{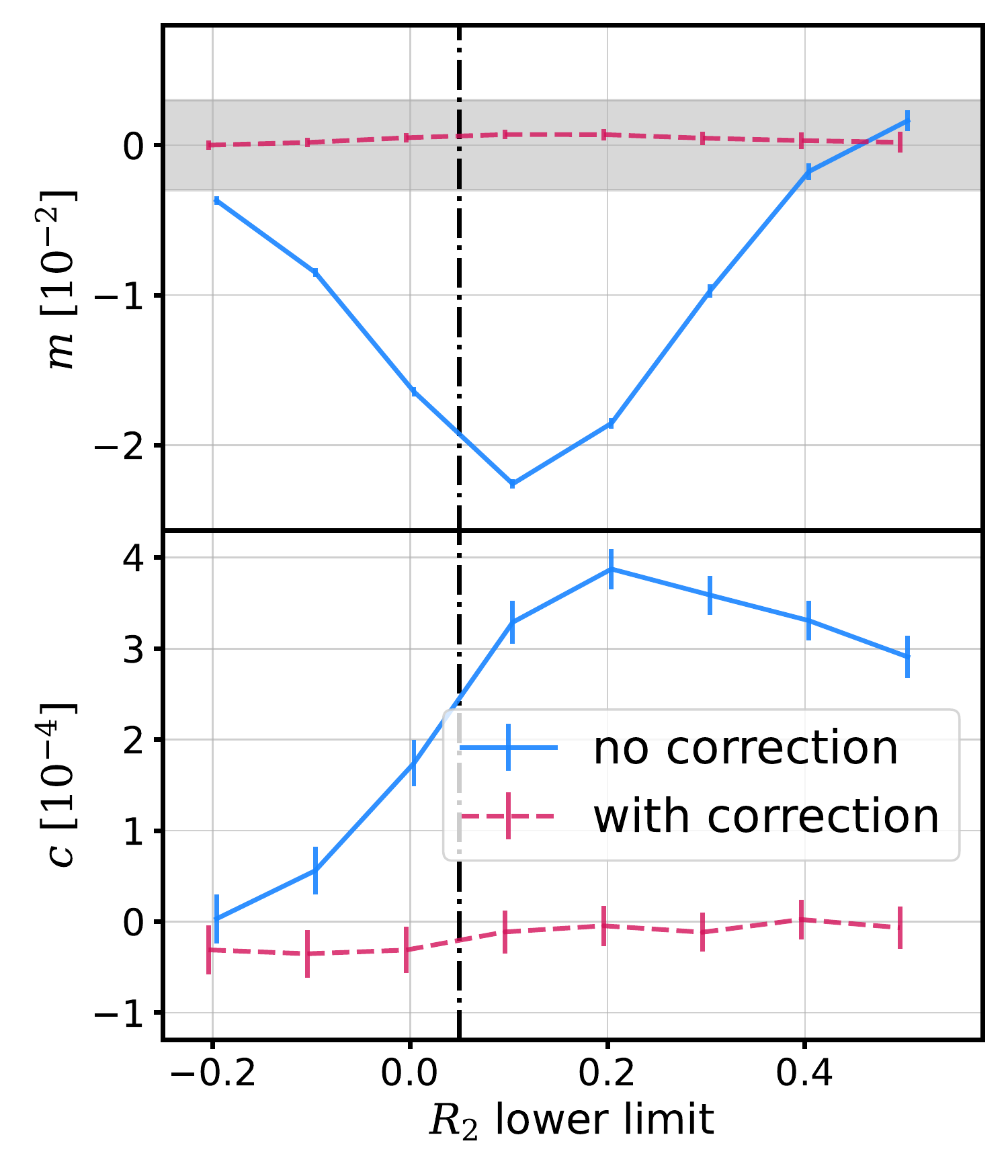}
\end{center}
\caption{
    Similar to Figure~(\ref{fig:res_center_magcut}), but here the shear
    estimation biases measured from the isolated image simulation are shown as
    functions of the lower limit of \FPFS{} resolution ($R_2$). The vertical
    dash-dotted line is the default cut on resolution of the postselection.
    }
    \label{fig:res_center_r2cut}
\end{minipage}
\end{figure*}

\begin{figure*}
\centering
\begin{minipage}[t]{0.45\textwidth}
\begin{center}
    \includegraphics[width=1.\textwidth]{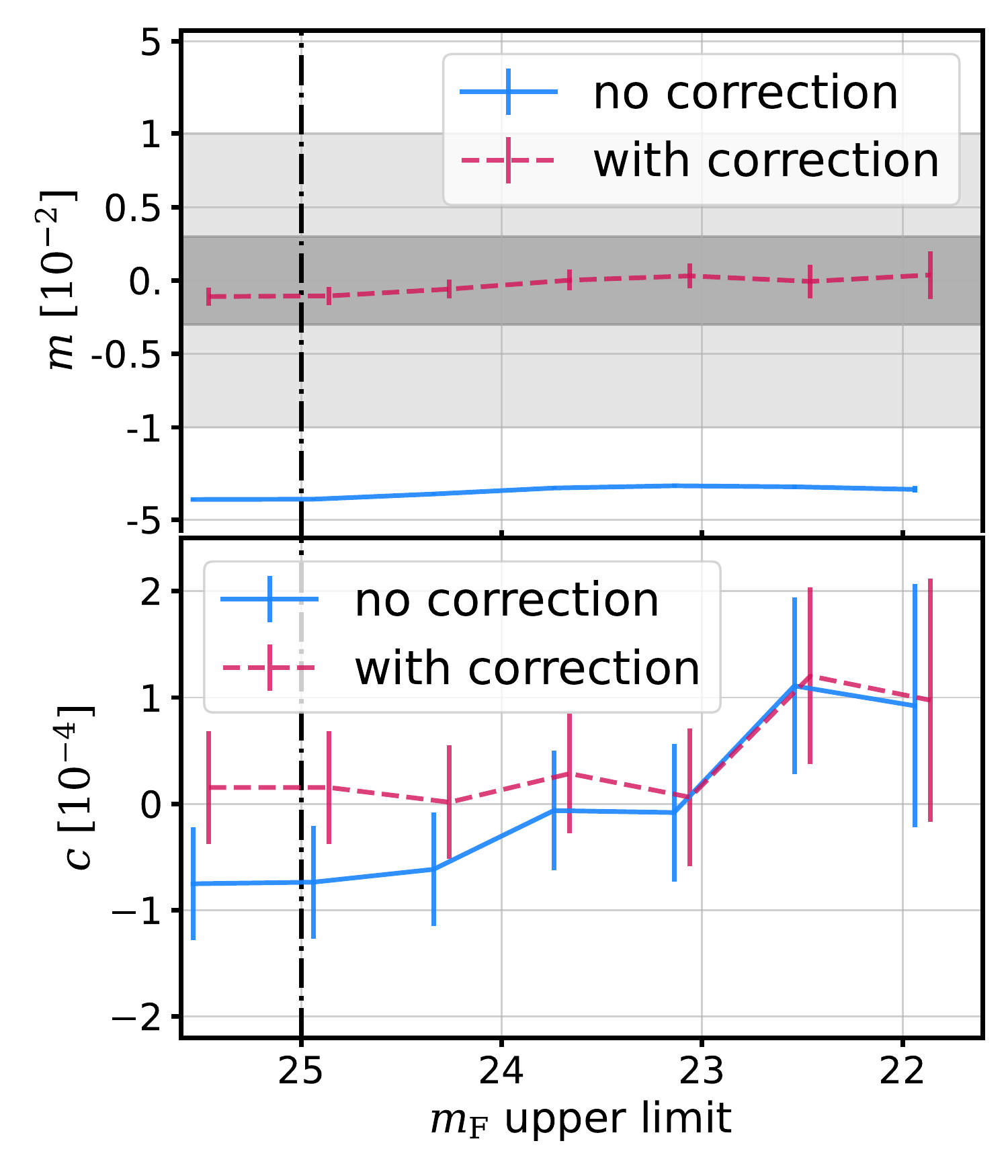}
\end{center}
\caption{
    Similar to Figure~(\ref{fig:res_center_magcut}) on the isolated image
    simulation, but the coordinate center for the measurement is set to the
    peaks identified with \FPFS{} detector. The \FPFS{} resolution cut is set
    to $R_2>0.05$\,. The $y$-axis for the light light shaded region of the top
    panel is in linear scale; while the other regions are in log scale. The
    errorbars are much smaller than the log scale so that they cannot be
    visualized in the log region. In addition, the smoothing scale of the
    Gaussian kernel, $\sigma_h$, is set to $0\farcs59$\,. The vertical
    dash-dotted line is the default cut on magnitude of the postselection.
    }
    \label{fig:res_detect_magcut}
\end{minipage}
\hfill
\begin{minipage}[t]{0.45\textwidth}
\begin{center}
    \includegraphics[width=1.\textwidth]{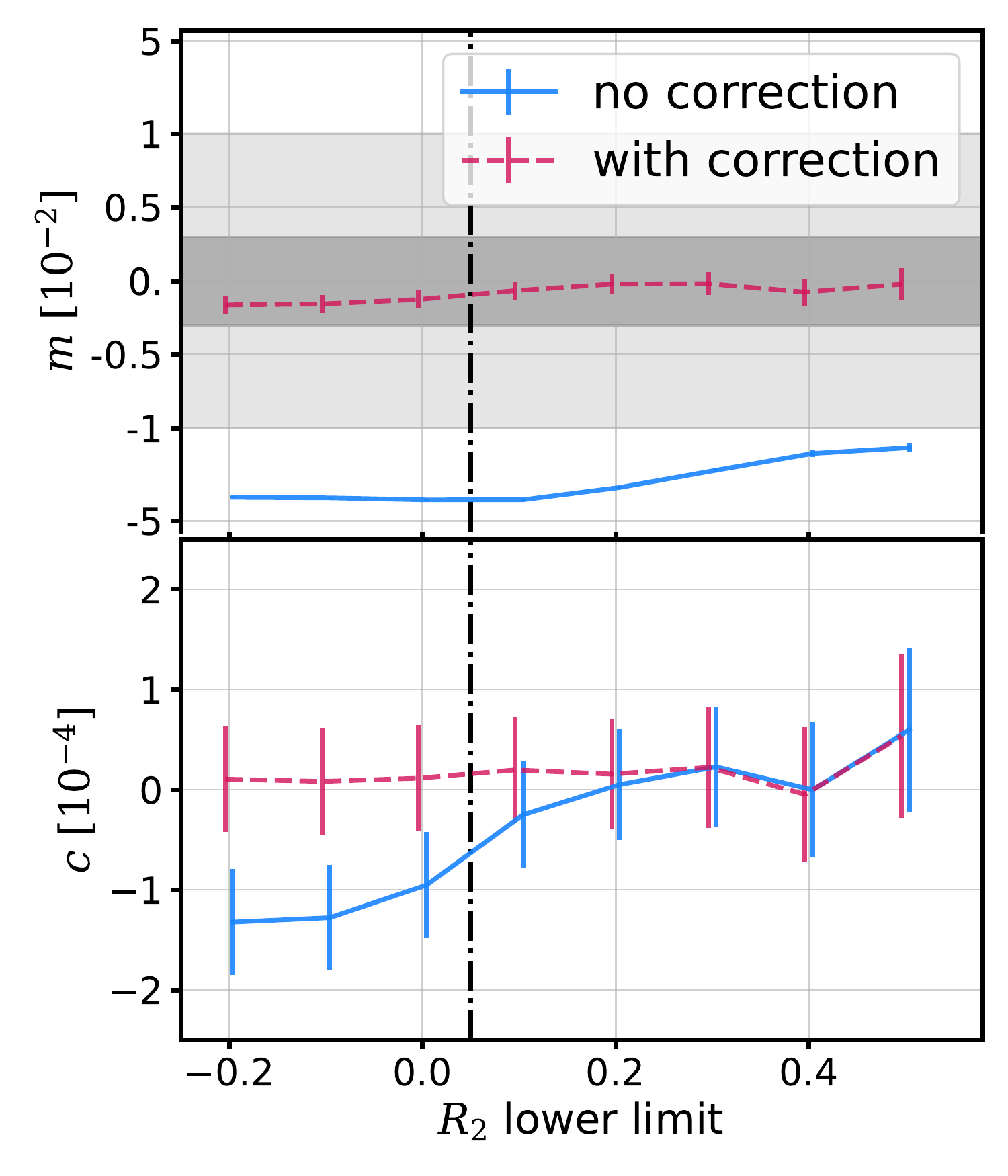}
\end{center}
\caption{
    Similar to Figure~(\ref{fig:res_center_r2cut}), but here the shear
    estimation biases measured from the isolated image simulation are shown as
    functions of the lower limit of \FPFS{} resolution ($R_2$). The \FPFS{}
    magnitude cut is fixed at $m_\text{F}<25$\,. In addition, the smoothing
    scale of the Gaussian kernel, $\sigma_h$, is set to $0\farcs59$\,. The
    vertical dash-dotted line is the default cut on magnitude of the
    postselection.
    }
    \label{fig:res_detect_r2cut}
\end{minipage}
\end{figure*}

We next test the performance of our method for the galaxies detected from the
isolated galaxy-image simulations where galaxies have sub-pixel offsets from
the center of the postage stamps. The default setup for the soft selection is
the same as in Figure~\ref{fig:obsHist_shifted}. The results are shown in
Figures~\ref{fig:res_detect_magcut} and \ref{fig:res_detect_r2cut}. Note, these
plots show the results as a function of cuts on one parameter; however, cuts
are being made simultaneously on the other parameters. In addition, in our
formalism, one galaxy can be detected several times with different centers with
detection weight $<1$\,.

Before the analytical correction for the detection bias and selection bias, the
multiplicative biases are at a level of $-5\%$\,. Comparing the multiplicative
biases before analytical correction to those with fixed centers in
Figures~\ref{fig:res_center_magcut} and \ref{fig:res_center_r2cut}, we conclude
that the detection process itself causes approximately $-4\%$ multiplicative
biases for the isolated galaxy image simulations. This multiplicative bias is
caused by the selection of noisy sheared pixels above a threshold, peak
identification from noisy sheared pixels and setting the peaks as galaxy
centroids.

After the analytical correction for noise bias, selection bias and detection
bias, the  multiplicative biases are less than $0.2\%$ (with
statistical error $\sim$$0.1\%$). Note, the multiplicative shear biases are
within the LSST requirements on the control of multiplicative bias
($\abs{m}<0.3\%$). The measured additive biases after our corrections are below
$1\times 10^{-4}$\,. Therefore, we conclude that our analytical correction is
able to reduce the shear-estimation biases, including the bias from detection,
to $\abs{m}<0.3\%$ for isolated galaxies.

\subsection{Blended galaxies}
\label{sec:4_blend_0}

\subsubsection{Number density}
\label{sec:4_blend_number}

Before showing the accuracy of the algorithm on blended galaxy image
simulations, we characterize the detected/selected galaxy sample in those
simulations. The detected number density after the postselection step is
strongly dependent on the smoothing scale of the Gaussian kernel, $\sigma_h$\,.
Therefore, we test two smoothing scales, namely $\sigma_h=0\farcs59$
(FWHM=$1\farcs4$) and $\sigma_h=0\farcs45$ (FWHM=$1\farcs1$) for the image
simulation with input galaxy number density of $85~\mathrm{arcmin}^{-2}$\,.

For smoothing scale of $\sigma_h=0\farcs59$, the number density of galaxies
with non-zero selection weight after the postselection step is
$31~\mathrm{arcmin}^{-2}$. Since many of the galaxies are downweighted by the
selection weight, we define the effective number of galaxies as
\begin{equation}
N_{\text{eff}}= \sum_{l=1}^{N_{\text{gal}}} w_2^{(l)}\,,
\end{equation}
assuming that the unweighted ellipticity, $e_\alpha$\,, of each galaxy
contributes equally to the shear estimation. One can divide the galaxy number
by the area to get the effective number density. The galaxy number density
after the postselection step is $9~\mathrm{arcmin}^{-2}$. For smoothing scale
of $\sigma_h=0\farcs45$, the number density of galaxies with non-zero selection
weight is $34~\mathrm{arcmin}^{-2}$, and the effective galaxy number density is
$16~\mathrm{arcmin}^{-2}$\,. Note, in this paper, we do not discuss the optimal
choice for the smoothing scale, $\sigma_h$\,.

For comparison, the effective number density of the HSC shape catalog is
$\sim$$20~\mathrm{arcmin}^{-2}$ \citep{HSC1-catalog,HSC3-catalog}. The number
density decreases as the noise variance or the PSF size increase. We note that
the image noise variance in the simulations here is twice the average of the
HSC survey in $i$-band; our PSF FWHM estimated with Gaussian weighted moments
is $10\%$ larger than the HSC's FWHM. In addition, the PSF model is not exactly
the same as HSC PSFs.  For this reason, we generally expect a lower number
detected number density in these simulations than in the actual HSC shape
catalog.

We find that, for isolated galaxies, the number of detections is not strongly
dependent on the smoothing scale, and the differences between their shear
measurement biases are consistent at the $2\sigma$ level. Therefore, we only
showed the results for $\sigma_h=0\farcs59$ for the isolated galaxy
simulations. We defer a more detailed exploration of the dependence of biases
and detected number density as a function of $\sigma_h$ and other free
parameters to future work.

\subsubsection{Shear estimation bias}
\label{sec:4_blend_bias}

\begin{figure*}
\centering
\begin{minipage}[t]{0.45\textwidth}
\begin{center}
    \includegraphics[width=1.\textwidth]{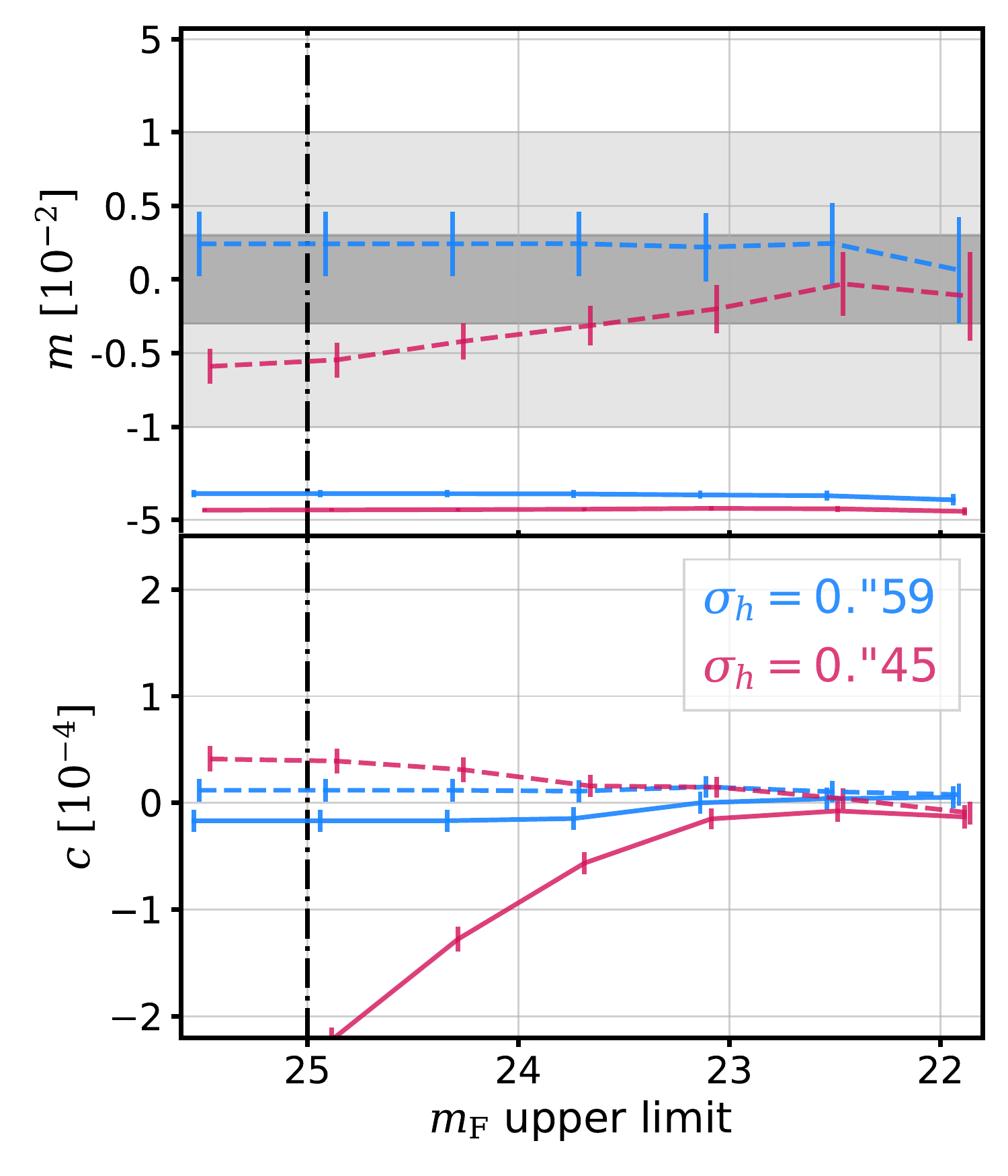}
\end{center}
\caption{
    Similar to Figure~(\ref{fig:res_detect_magcut}), but measured from the
    blended galaxy image simulation with number density
    $85~\mathrm{arcmin}^{-2}$ with two smoothing scales: $0\farcs59$ (blue) and
    $0\farcs45$ (red). The solid (dashed) lines are results before
    (after) the correction for shear-dependent detection bias and selection
    bias. The vertical dash-dotted line is the default cut on magnitude of the
    postselection.
    }
    \label{fig:res_blend1_magcut}
\end{minipage}
\hfill
\begin{minipage}[t]{0.45\textwidth}
\begin{center}
    \includegraphics[width=1.\textwidth]{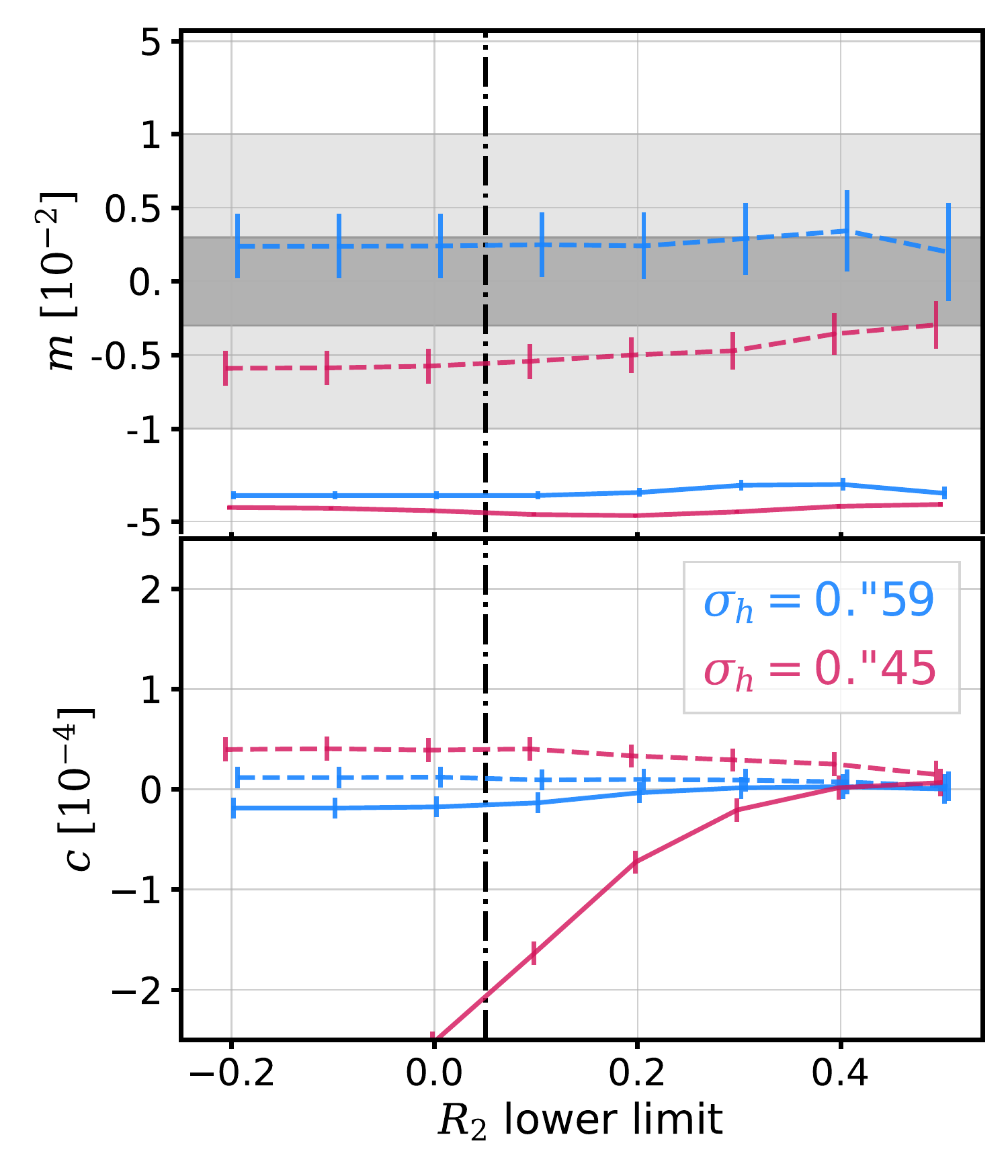}
\end{center}
\caption{
    Similar to Figure~(\ref{fig:res_detect_r2cut}), but measured from the
    blended galaxy image simulation with number density
    $85~\mathrm{arcmin}^{-2}$ with two smoothing scales: $0\farcs59$ (blue) and
    $0\farcs45$ (red). The solid (dashed) lines are results before (after) the
    correction for shear-dependent detection bias and selection bias. The
    vertical dash-dotted line is the default cut on magnitude of the
    postselection.
    }
    \label{fig:res_blend1_r2cut}
\end{minipage}
\end{figure*}

We show the multiplicative  and additive bias in the shear estimator with these
two smoothing scales for the galaxies detected from the blended image
simulations. The input galaxies number densities of the simulations are
$85~\mathrm{arcsec}^{-2}$\,. The default setup for the soft selection is the
same as in Figure~\ref{fig:obsHist_blended}. The results are shown in
Figures~\ref{fig:res_blend1_magcut}--\ref{fig:res_blend1_r2cut}.

Before the correction for the detection and selection bias, the multiplicative
shear biases are at a level of $-5\%$\,. Comparing the multiplicative biases
before analytical correction to those for isolated galaxies with fixed center
in Figures~\ref{fig:res_center_magcut} and \ref{fig:res_center_r2cut}, we
conclude that the detection process causes about $-4\%$ multiplicative biases
for the blended image simulations. This amplitude of this detection bias in
blended image simulation is consistent with what has been reported in
\citet{metaDet-Sheldon2020}. We note that the uncorrected bias is similar to
the uncorrected bias we found without blending. This is because \FPFS{} does
not make any assumptions regarding galaxy morphology, and blended galaxies at
the same redshift can be considered as a single galaxy with a complicated
morphology. However, such differences in morphology do not noticeably degrade
the accuracy of our algorithm.

After the analytical corrections for noise bias, selection bias and detection
bias, the multiplicative biases are below $0.6\%$\,, with statistical error of
$\sim$$0.15\%$\, for the two different smoothing scales that we have tested.
The measured additive biases are below $1\times 10^{-4}$ for both smoothing
scales. However, we note that there is some variation in multiplicative bias
within the range $\abs{m}<6\times10^{-3}$\, for the blended image simulations.
It is possible that this is caused by false detections at empty locations due
to fluctuations of image noise. We will study the origin of these biases
(including confirmation and mitigation of the false detection effect or other
relevant effects) in detail in our future work.

We show the results for the blended image simulation with input galaxy number
density of $170$~arcmin$^{-2}$ in Appendix~\ref{app:test_blend2}. The results
are consistent with what we found here for the simulation with input galaxy
number density of $85$~arcmin$^{-2}$\,.

The simple analytical correction reduces the multiplicative bias by an order of
magnitude. Although the bias reduction is not complete, it is a very important
start that establishes the promise of this analytical method to eventually
reach the stringent requirements of Stage IV surveys.

\subsection{HSC Year 3}
\label{sec:4_HSCY3}

In the previous subsections, we showed the performance of our algorithm with
different smoothing scales on simulations that have double the variance of HSC
survey image noise. Here, we test our algorithm, setting $\sigma_h=0.45''$, on
images with noise variance set to the mean of the noise variance in three-year
HSC (HSC-Y3) data \citep{HSC3-catalog}, and predict the performance of the
algorithm on the HSC-Y3 data.

\begin{figure*}
\centering
\begin{minipage}[t]{0.45\textwidth}
\begin{center}
    \includegraphics[width=1.\textwidth]{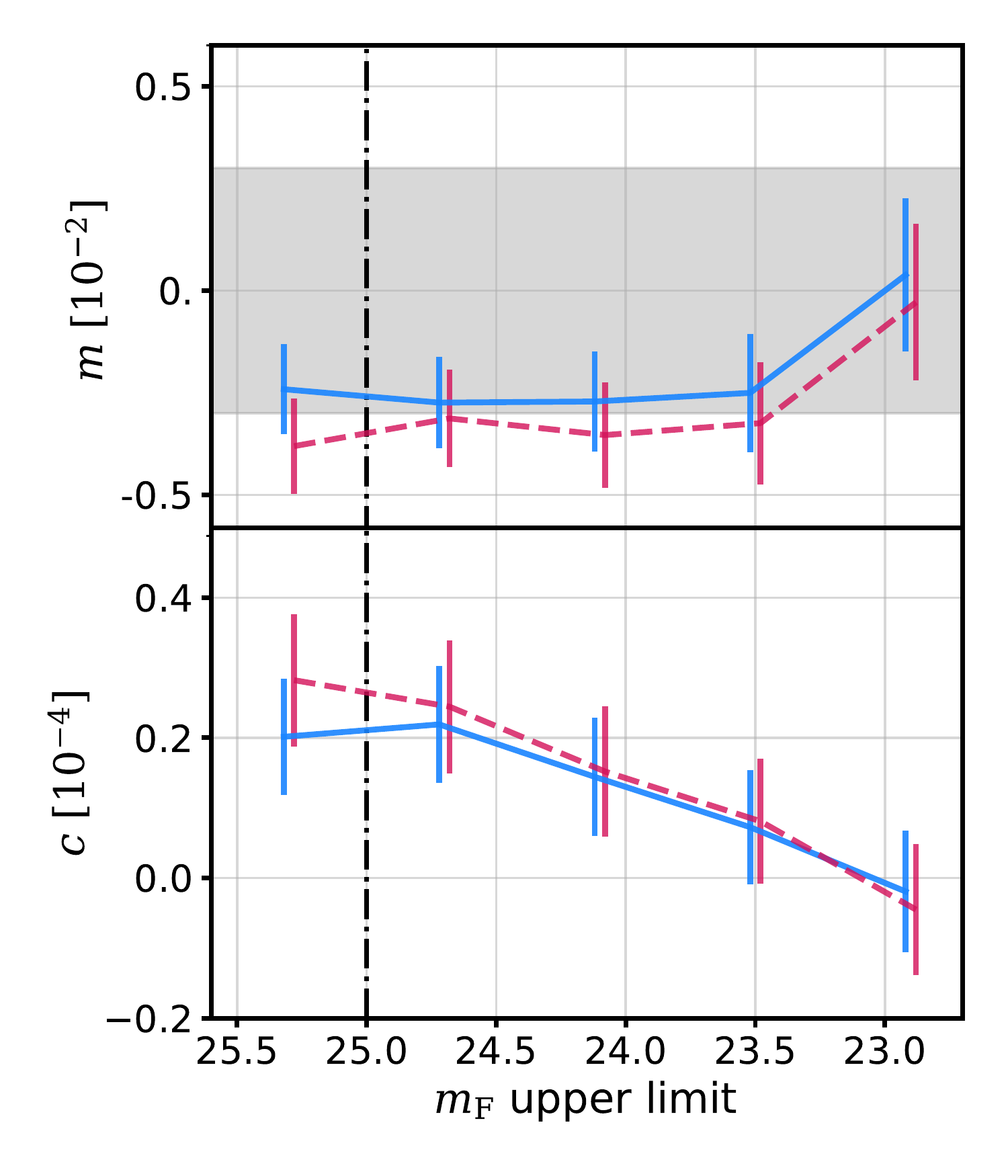}
\end{center}
\caption{
    Similar to Figure~(\ref{fig:res_blend1_magcut}), but measured from the
    blended galaxy image simulation with the average noise level of the HSC
    three-year data \citep{HSC3-catalog}. The solid lines include all the nosie
    bias correction term using auto-diff; whereas the dashed lines only include
    the noise bias correction terms given out in this paper. Shaded region is
    for LSST ten-year requirement on the control of multiplicative bias.
    }
    \label{fig:res_hsc_magcut}
\end{minipage}
\hfill
\begin{minipage}[t]{0.45\textwidth}
\begin{center}
    \includegraphics[width=1.\textwidth]{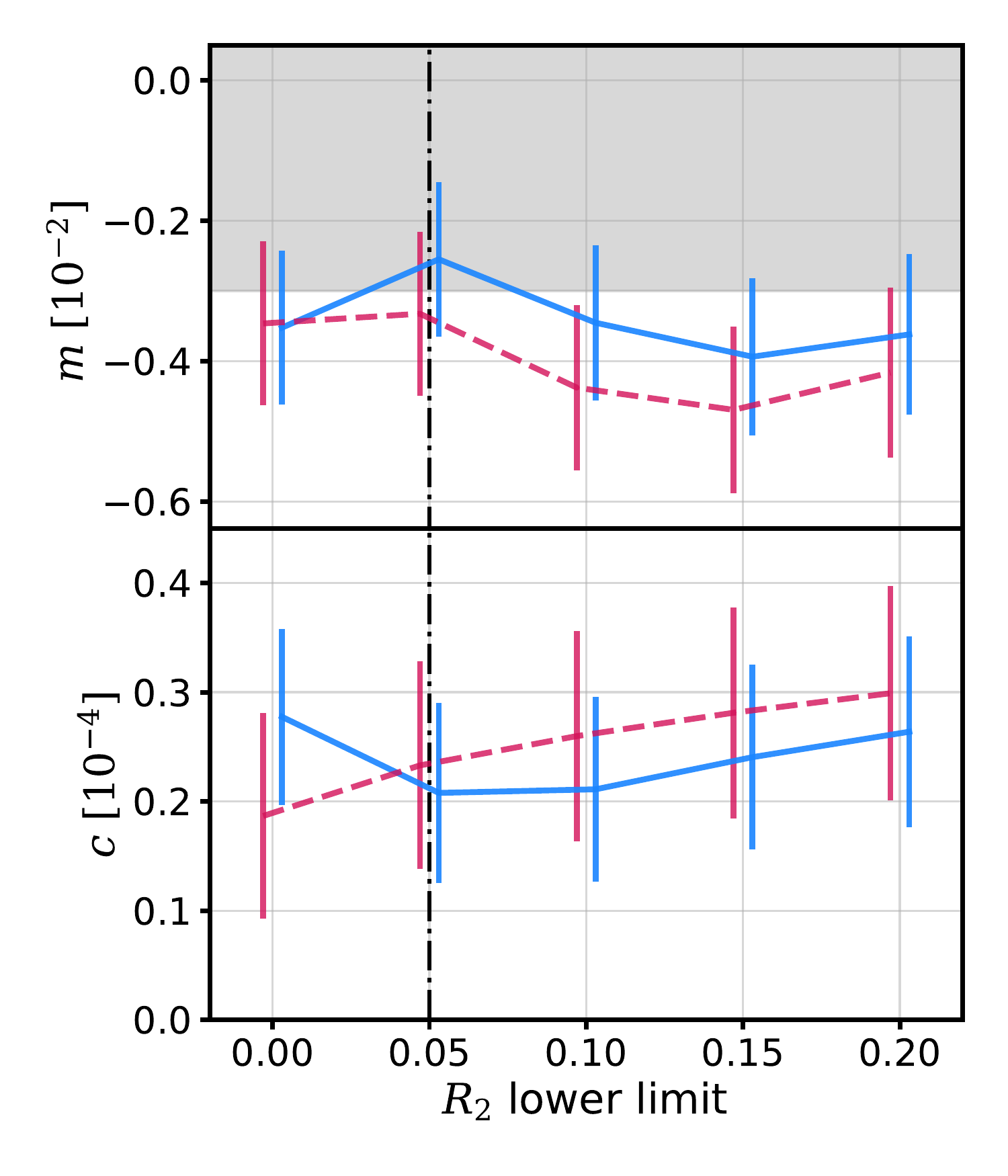}
\end{center}
\caption{
    Similar to Figure~(\ref{fig:res_blend1_r2cut}), but measured from the
    blended galaxy image simulation with the average noise level of the HSC
    three-year data \citep{HSC3-catalog}. The solid lines include all the nosie
    bias correction term using auto-diff; whereas the dashed lines only include
    the noise bias correction terms given out in this paper. Shaded region is
    for LSST ten-year requirement on the control of multiplicative bias.
    }
    \label{fig:res_hsc_r2cut}
\end{minipage}
\end{figure*}

The results for different magnitude and resolution cuts are shown as dashed
lines in Figure~\ref{fig:res_hsc_magcut} and \ref{fig:res_hsc_r2cut},
respectively. In addition, we use our auto-differentiation code
\citep{imPT_Li2023} to fully correct the second-order noise bias, and the
results for different magnitude and resolution cuts are shown as solid lines in
Figure~\ref{fig:res_hsc_magcut} and \ref{fig:res_hsc_r2cut}, respectively. The
multiplicative bias is smaller than that in the tests with double the typical
HSC noise variance, indicating that at least some of the non-zero
multiplicative bias in results earlier in this paper were caused by the image
noise. We also find that the multiplicative bias slightly reduces by
$\sim$$0.2\%$ after we include all of the second-order noise bias correction
terms \citep{imPT_Li2023}, indicating the bias from the neglected terms was not
very significant for the precision of tests in this paper. We still find a
residual multiplicative bias of about $-0.3\%$ even after including all the
second-order noise bias correction terms. We suspect this is due to (i) false
detections; or (ii) higher-order noise bias, and we will investigate the cause
of this residual bias in our future work.

\section{SUMMARY AND OUTLOOK}
\label{sec:Summary}

In this paper, we analytically correct for detection and selection bias in
shear estimation for the \FPFS{} shear estimator applied at the image pixel
level by deriving the shear response of image pixels. We also analytically
correct for the second-order noise bias in the detection process. Crucially,
the analytic correction does not depend upon calibration via external galaxy
image simulations. Moreover, unlike Bayesian Fourier Domain
\citep[\BFD{}][]{BFD-Bernstein2014, BFD-Bernstein2016}, our shear estimator
does not require a deep reference sample; and unlike \metacal{}/\metadet{}
\citep{metacal-Sheldon2017, metacal-Huff2017, metaDet-Sheldon2020}, it does not
require artificial shearing of each observed galaxy image. Our
publicly-available code (\url{https://github.com/mr-superonion/FPFS}) can
process more than a thousand galaxy images per CPU second.

Using mock images of galaxies distorted with an input known shear, we
demonstrate that our shear estimator reaches sub-percent level accuracy not
only for isolated galaxies but also for blended galaxies with an extremely
high galaxy number density under HSC-like observing conditions.

Future works should test the performance of this novel detector/shear estimator
using more complicated simulations. Here we list the assumptions in our image
simulations used to test the method in this paper, which should be the targets
of the future tests:
\begin{enumerate}
    \item The image noise is homogeneous across the sky; however, in reality,
        galaxy photon noise and the variation of background photon noise are
        inhomogeneous at small scales and large scales, respectively.
    \item The PSF is fixed rather than varying across the sky; however, in
        reality, the PSF is a function of the position on the focal plane of an
        exposure.
    \item Atmospheric chromatic effects on PSF modelling, e.g., wavelength
        dependence of seeing and atmospheric differential chromatic refraction
        \citep{chromaticPSF_Meyers2015}.
    \item The input shear is the same for every galaxy in the image; however,
        in reality, the shear is a function of galaxy redshift.
    \item The image is well sampled; however, for spaced-based surveys, images
        of single exposures are not well sampled.
    \item Sky background, stars and image artifacts are not included in our
        simulations.
\end{enumerate}
For the first two assumptions, we have proposed solutions in this paper: (i) we
derive the formalism for inhomogeneous noise in
Appendix~\ref{app:inhomogeneous_noise}; and (ii) we propose a solution to PSF
variation by conducting the postselection and measurement using the PSF
modelled for each galaxy after the preselection using the average PSF over a
field in Section~\ref{sec:2_pixel_pdetect}. We will explicitly test the
performance of the shear estimator with simulations that violate these two
assumptions in our future work. For the last three assumptions, we will
quantify their importance with realistic image simulations, and improve our
algorithm once the importance of these issues is better understood.

Another thing to note is that the formalism presented here is for estimation of
the average shear within a patch of sky. \citet{DEScatY3_Gatti2021} provided a
consistent correction for selection bias in two-point correlation functions for
\metacal{} in their Appendix~A. In our future work, we will study in detail the
correction for detection and selection bias when estimating two-point
correlation functions using our shear estimator.

Furthermore, we will attempt to understand the residual half-percent bias in
the blended simulations and test the detailed performance of the method as a
function of the choices of hyper-parameters (e.g., the smoothing scales, for
which we only considered two options in this work).

The future work outlined here is all in the spirit of getting this very
promising approach ready for direct application to Stage-IV surveys. In
addition, we are going to apply the shear estimator to the ongoing surveys
(e.g., HSC) to test its performance.

\section*{ACKNOWLEDGEMENTS}
\addcontentsline{toc}{section}{ACKNOWLEDGEMENTS}

This work was supported by a grant from the Simons Foundation (Simons
Investigator in Astrophysics, Award ID 620789).

We thank Mike Jarvis, Scott Dodelson, Arun Kannawadi, Matthew Becker, Erin
Sheldon and Gary Bernstein for their useful comments on the paper.

This paper uses the parameters of the HSC SSP observational conditions. The Hyper
Suprime-Cam (HSC) collaboration includes the astronomical communities of Japan
and Taiwan, and Princeton University. The HSC instrumentation and software were
developed by the National Astronomical Observatory of Japan (NAOJ), the Kavli
Institute for the Physics and Mathematics of the Universe (Kavli IPMU), the
University of Tokyo, the High Energy Accelerator Research Organization (KEK),
the Academia Sinica Institute for Astronomy and Astrophysics in Taiwan (ASIAA),
and Princeton University. Funding was contributed by the FIRST program from
Japanese Cabinet Office, the Ministry of Education, Culture, Sports, Science
and Technology (MEXT), the Japan Society for the Promotion of Science (JSPS),
Japan Science and Technology Agency (JST), the Toray Science Foundation, NAOJ,
Kavli IPMU, KEK, ASIAA, and Princeton University.

This paper makes use of software developed for the Vera C.\ Rubin Observatory.
We thank the Vera C.\ Rubin Observatory for making their code available as free
software at http://dm.lsst.org.

We thank the maintainers of numpy \citep{numpy_Harris2020}, scipy
\citep{scipy_Virtanen2020}, numba \citep{numba_Lam2015}, Matplotlib
\citep{matplotlib_Hunter2007}, \galsim{} \citep{GalSim} for their excellent
open-source software.

\section*{DATA AVAILABILITY}
\addcontentsline{toc}{section}{DATA AVAILABILITY}
The code  used for image processing and galaxy image simulation in this paper
is available from \url{https://github.com/mr-superonion/FPFS/tree/v3.0.2}.

\bibliographystyle{mnras}
\bibliography{citation}
\appendix
\section{Shear perturbation}
\label{app:shear_perturb}

\begin{figure}
\begin{center}
    \includegraphics[width=0.45\textwidth]{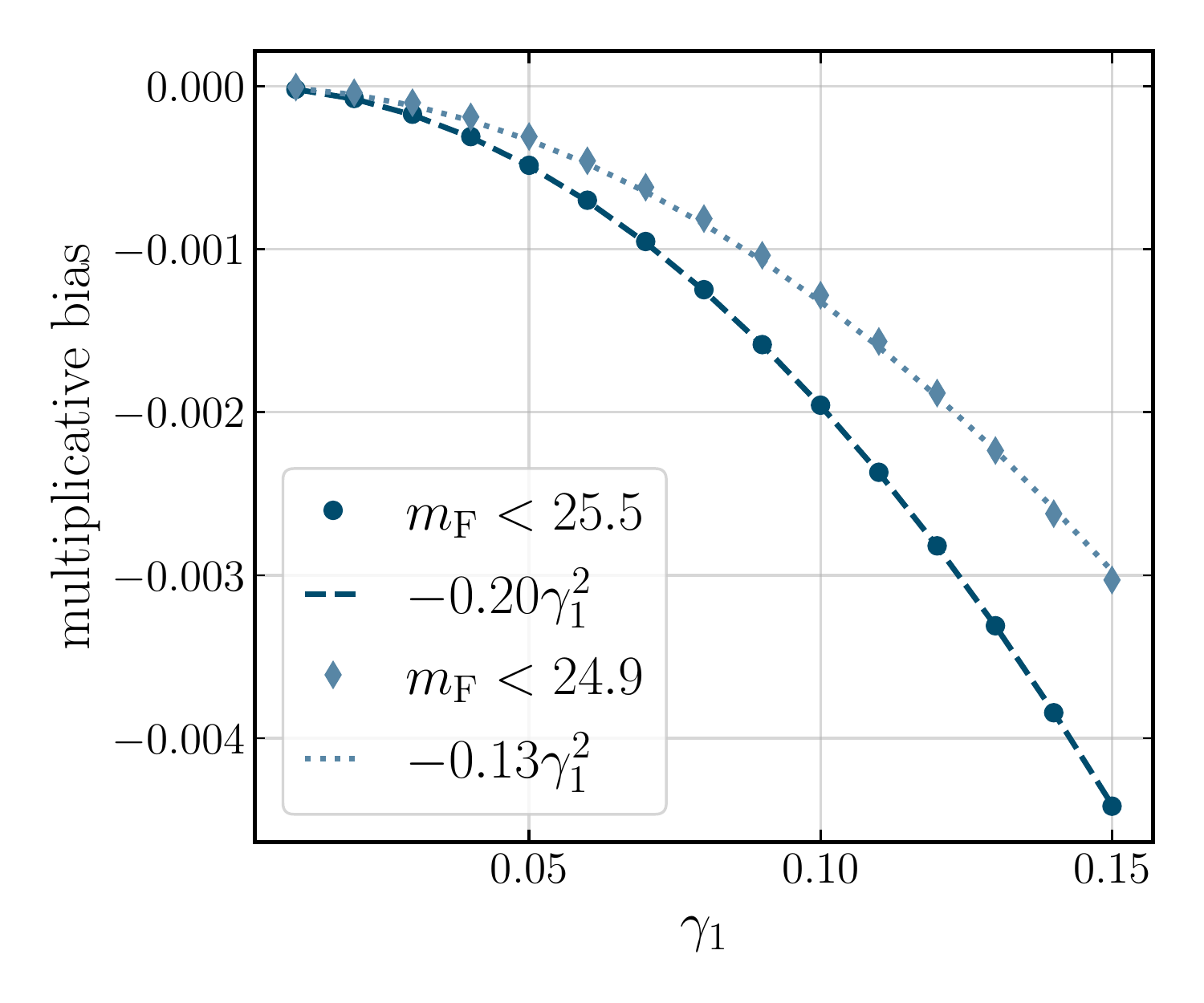}
\end{center}
\caption{
    Multiplicative bias as a function of input shear with two different \FPFS{}
    magnitude cuts. The circles (diamonds) denote the multiplicative bias in
    galaxy sample with magnitude brighter than $25.5$ ($24.9$). The biases
    scale as $-0.20 \gamma_1^2$ and $-0.13 \gamma_1^2$, respectively.
    }
    \label{fig:app_m_g2}
\end{figure}

We use noiseless isolated galaxy image simulations (as demonstrated in the left
panel of Figure~\ref{fig:sim_isoblend}) to confirm the second-order shear bias
residual is consistent with zero (as shown in equation~\eqref{eq:ell_sheared})
when galaxies are isotropically oriented. Moreover, we determine the
coefficient before the neglected third-order term in shear in
equation~\eqref{eq:ell_sheared}, which corresponds to a contribution to the
multiplicative bias that is second-order in shear. The multiplicative biases
for different galaxy samples are shown in Figure~\ref{fig:app_m_g2}. The
coefficient, which depends on galaxy properties, is approximately $-0.20$ for
galaxies with \FPFS{} magnitude brighter than $25.5$, and $-0.13$ for galaxies
with \FPFS{} magnitude brighter than $24.9$\,. We note that this third-order
shear estimation bias needs to be calibrated with external image simulation for
shear estimation in the intermediate-shear regime.

\section{Spin number and rotation symmetry}
\label{app:spin_property}

In this appendix, we discuss the connection between the spin number of galaxy
properties and the rotation symmetry. Furthermore, we derive the spin number of
the product of two properties. These concepts will be used to explain
equation~\eqref{eq:ell_sheared}, but are also relevant to some of the other
equations in this work.

The spin number of an observable that is measured from the image coordinate
system $\vx$ describes how the observable transforms when the coordinate
system, $\vx$, rotates. The reference coordinate system is set to the
two-dimensional sky coordinates adopting the flat-sky approximation with $+x$
being the horizontal axis to the west; $+y$ being the vertical axis to the
north.

Here we focus on properties that only have one spin component. In a specific
two-dimensional image coordinate system, which is a rotated, dilated and/or
weighted transform of the reference coordinates, the representation of an
observable can be written as a complex number:
\begin{equation}
    \vec{v}_{\vx} = v_{\vx1} +  \rmi v_{\vx2},
\end{equation}
where $v_{\vx1}$ and $v_{\vx2}$ are two components of the observable. These
components are
\begin{equation}
\label{eq:app_property_components}
\begin{split}
    v_{\vx1}&= \frac{\vec{v}_{\vx}+\vec{v}^{*}_{\vx}}{2}\,,\\
    v_{\vx2}&= \frac{\vec{v}_{\vx}-\vec{v}^{*}_{\vx}}{2\rmi}\,,
\end{split}
\end{equation}
where $\vec{x}^{*}$ refers to the complex conjugate of $\vec{x}$\,. Note,
$\vec{v}_{\vx}$ is not a field at the position $\vx$; rather, it is the
projection of an observable onto the image coordinate system, $\vx$\,. If the
image coordinate system $\vx$ changes (e.g., through a rotation, flip or
dilation), the representation of the observable in the image coordinates also
changes: $\vec{v}_{\vx} \longrightarrow \vec{v}_{\vx'}$\,, since the
measurement is conducted in the transformed coordinates. Note, the observable
itself does not change when the image coordinate system transforms, but its
representation with respect to the image coordinate system does change.

In this appendix, we focus on the rotation transform. The representation of a
spin-$m$ observable with respect to an image coordinate system $\vx$ can be
written as
\begin{equation}
    \label{eq:app_property_polar}
    \vec{v}_{\vx} = v \exp\left(\rmi m(\theta_0-\theta)\right)\,.
\end{equation}
$v$ is the amplitude of the observable, $\theta_0$ is the direction of the
observable with respect to the main axis of the reference coordinate system,
and $\theta$ is the direction of the main axis of the image coordinate system
with respect to the reference coordinate system.
Now we have
\begin{equation}
\begin{split}
    v_{\vx1}&= v\cos{\left(m(\theta_0-\theta)\right)}\,,\\
    v_{\vx2}&= v\sin{\left(m(\theta_0-\theta)\right)}\,.
\end{split}
\end{equation}

By definition, the representation of a spin-$m$ ($m\neq0$) observable is
negated when the coordinate system rotates by $\pi/\abs{m}$: $\theta
\rightarrow \theta + \pi/\abs{m}$\,. As a result, when the coordinate system
rotates by an odd multiple of $\frac{\pi}{\abs{m}}$, the representation
negates; when the coordinate system rotates by an even multiple of
$\frac{\pi}{\abs{m}}$, the representation transforms back to its original
value. For the case $m = 0$, the representation of the observable does not
change when the coordinate system rotates.

Since equation~\eqref{eq:ell_sheared} involves products of quantities with
different spins, in order to study equation~\eqref{eq:ell_sheared} in detail,
we determine the spin number of the product of two properties measured on the
same coordinate system. We use $\vec{v}_\vx$ for the first observable, and the
second observable is denoted as
\begin{equation}
\vec{v}'_\vx=v' \exp\left(\rmi m'(\theta'_0 - \theta)\right)\,.
\end{equation}
Since these two properties are measured in the same image coordinate, $\theta$
is the same as the one in equation~\eqref{eq:app_property_polar}. The products
of these two properties and their complex conjugates are
\begin{equation}
\begin{split}
\vec{v}_\vx^{*} \vec{v}'_\vx &= v'v \exp\left(-\rmi (m'-m)\theta+m'\theta'_0-m\theta_0)\right)\,,\\
\vec{v}_\vx \vec{v}'_\vx &= v'v \exp\left(-\rmi (m'+m)\theta+m'\theta'_0+m\theta_0)\right)\,.
\end{split}
\end{equation}
These equations tell how the representations of products transform as the
coordinate system rotates as $\theta \rightarrow \theta + \Delta \theta$\,. We
conclude that the products are combinations of spin-$(m+m')$ and spin-$(m-m')$
quantities. The terms $m'\theta'_0 \pm m\theta_0$ determine the angular phase
of the products with respect to the reference coordinates, which does not
change as the image coordinates rotate since the observable itself does not
rotate with the coordinates. According to
equation~\eqref{eq:app_property_components}, the products between the
components of these two representations can only be spin-$(m+m')$ or
spin-$(m-m')$\, since they can be written as linear combinations of
$\vec{v}_\vx^{*} \vec{v}'_\vx$ and $\vec{v}_\vx \vec{v}'_\vx$\,.

Some observables can be decomposed into many spin components; however, the weak
lensing shear, $\gamma_1+\rmi \gamma_2$, is a pure spin-$2$ quantity according
to the definition shear field \citep{WL-rev-Bartelmann01}:
\begin{equation}
    \vec{\gamma}(x,y)\propto
    \left(\frac{\partial^2}{\partial x^2} - \frac{\partial^2}{\partial y^2}\right) \Phi(x,y)
    + 2\,\rmi   \frac{\partial}{\partial x} \frac{\partial}{\partial y} \Phi (x,y)\,,
\end{equation}
where $\Phi$ is a scalar field defined on the sky coordinates. In the vector
space centered at a point $(x,y)$, $\vec{\gamma}$ is a spin-$2$ vector since
$(\frac{\partial^2}{\partial x^2} - \frac{\partial^2}{\partial y^2},
2\frac{\partial^2}{\partial x \partial y})$ negates when the coordinates of the
vector space rotate by $\pi/2$\, (while keeping the coordinates center the
same). In order to measure shear with the \FPFS{} ellipticity, we define it as
a pure spin-$2$ quantity. In addition, we require the ellipticity to not have
an angular phase offset with respect to shear after averaging over a large
number of galaxies, namely $\theta^{\langle
e\rangle}_0=\theta^\gamma_0=\theta_0$ for both shear and ellipticity.

We take a closer look at equation~\eqref{eq:ell_sheared} at the single galaxy
level and keep the perturbation terms up to the second order in shear:
\begin{equation}
\begin{split}
w e_\alpha = \bar{w} \bar{e}_\alpha
+ \left.\frac{\partial (w e_\alpha) }{\partial \gamma_\mu}\right|_{\gamma_\mu=0} \gamma_\mu
+ \left.\frac{\partial^2 (w e_\alpha)}
    {\partial \gamma_\mu \partial \gamma_\nu}\right|_{\gamma_{\mu,\nu}=0}
    \gamma_\mu \gamma_\nu
+ \mathcal{O}\left(\gamma_\alpha^3\right)\,,
\end{split}
\end{equation}
where we adopt Einstein summation notation. Since all the derivatives are
evaluated at zero shear, we neglect the $\left.\right|_{\gamma_\mu=0}$ in
the following discussion. Assuming that the selection weight is spin-$0$,
the left-hand side is a spin-$2$ observable. Since the equation should be
valid on image coordinates with any arbitrary orientation, each term on the
right-hand-side with different orders in shear should be spin-$2\,$. In
addition, the expectation of each term and $\langle e_\alpha \rangle$
should have the same angular phase with respect to the reference coordinate
system. We will discuss the first-order and second-order terms in shear,
separately.

First, we look into the term that is first order in shear: $\frac{\partial (w
e_\alpha)}{\partial \gamma_\mu} \gamma_\mu ~ (\alpha,\mu \in \{1,2\})$\,. Since
$\gamma_\mu$ is spin-$2$, any non-zero contribution from $\frac{\partial (w
e_\alpha)}{\partial \gamma_\mu}$ must be a combination of spin-$0$ and spin-$4$
properties, as we require their product to be spin-$2$\,. The diagonal
elements of the shear response matrix are composed of spin-$0$ and spin-$4$
quantities; whereas, the off-diagonal elements of the shear response matrix,
$\frac{\partial (w e_\mu)}{\partial \gamma_\nu}$ ($\mu$$\neq$$\nu$), cannot
be spin-$0$\,, since the angular phases of $\langle e_\mu \rangle$ and
$\gamma_\nu$ differ by $\pi/4$. Therefore, the off-diagonal terms of the shear
response matrix can only be spin-$4$\,.
In summary, the expectation values of the diagonal terms are nonzero, because
they are the expectations of spin-$0$ and spin-$4$ properties of intrinsic
galaxies, and the former is nonzero even for intrinsic (unlensed) properties.
In contrast, the off-diagonal terms have an expectation value of zero and do
not contribute to equation~\eqref{eq:ell_sheared}, since they are expectation
values of spin-$4$ properties of intrinsic galaxies.

Then we focus on the term that is second order in shear: $\frac{\partial^2 (w
e_\alpha)}{\partial \gamma_\mu \partial \gamma_\nu}\gamma_\mu \gamma_\nu ~
(\alpha,\mu,\nu \in \{1,2\})$\,. According to the rule for spin number of
products, $\gamma_\mu \gamma_\nu$ is a combination of spin-$0$ and spin-$4$
properties of intrinsic galaxies. Therefore, any non-zero contributions from
$\frac{\partial^2 (w e_\alpha)}{\partial \gamma_\mu \partial \gamma_\nu}$ must
be a combination of spin-$2$ and spin-$6$ properties, as we require the product
to be spin-$2$\,. The expectation values of these spin-$2$ and spin-$6$
properties of intrinsic galaxies are identically zero (\textbf{assumption~1}),
and do not contribute to equation~\eqref{eq:ell_sheared}.

\section{Noise bias correction}
\label{app:noirev}

\begin{figure}
\begin{center}
    \includegraphics[width=0.45\textwidth]{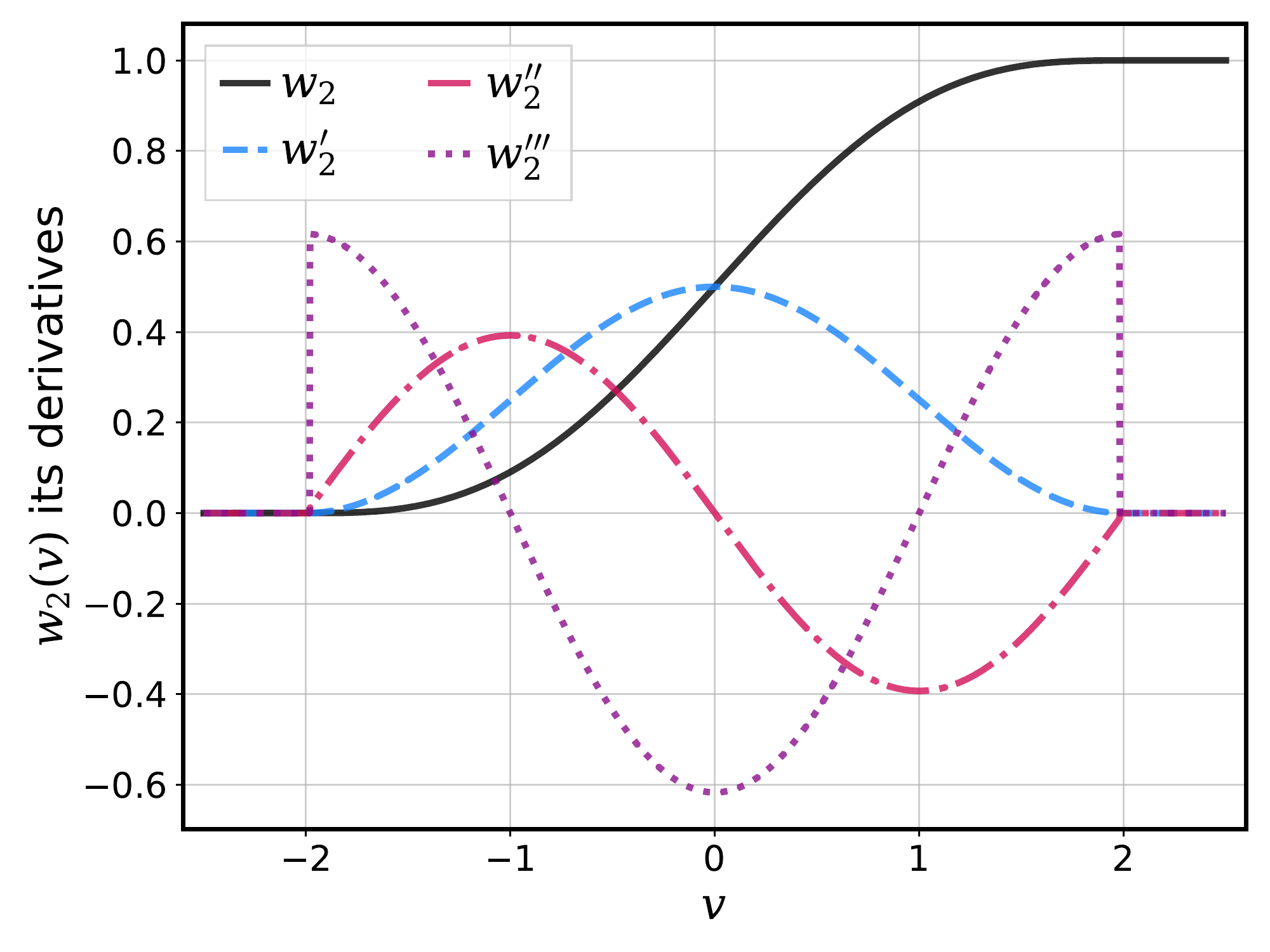}
\end{center}
\caption{
    Solid line is the truncated sine selection function, $w_2$ (defined in
    equation~\eqref{eq:sel_func_c2}). The dashed, dotted-dash and dotted lines
    are its first, second and third-order derivatives, respectively.
    }
    \label{fig:selfun_deriv}
\end{figure}

\begin{figure*}
\centering
\begin{minipage}[t]{0.45\textwidth}
\begin{center}
    \includegraphics[width=1.\textwidth]{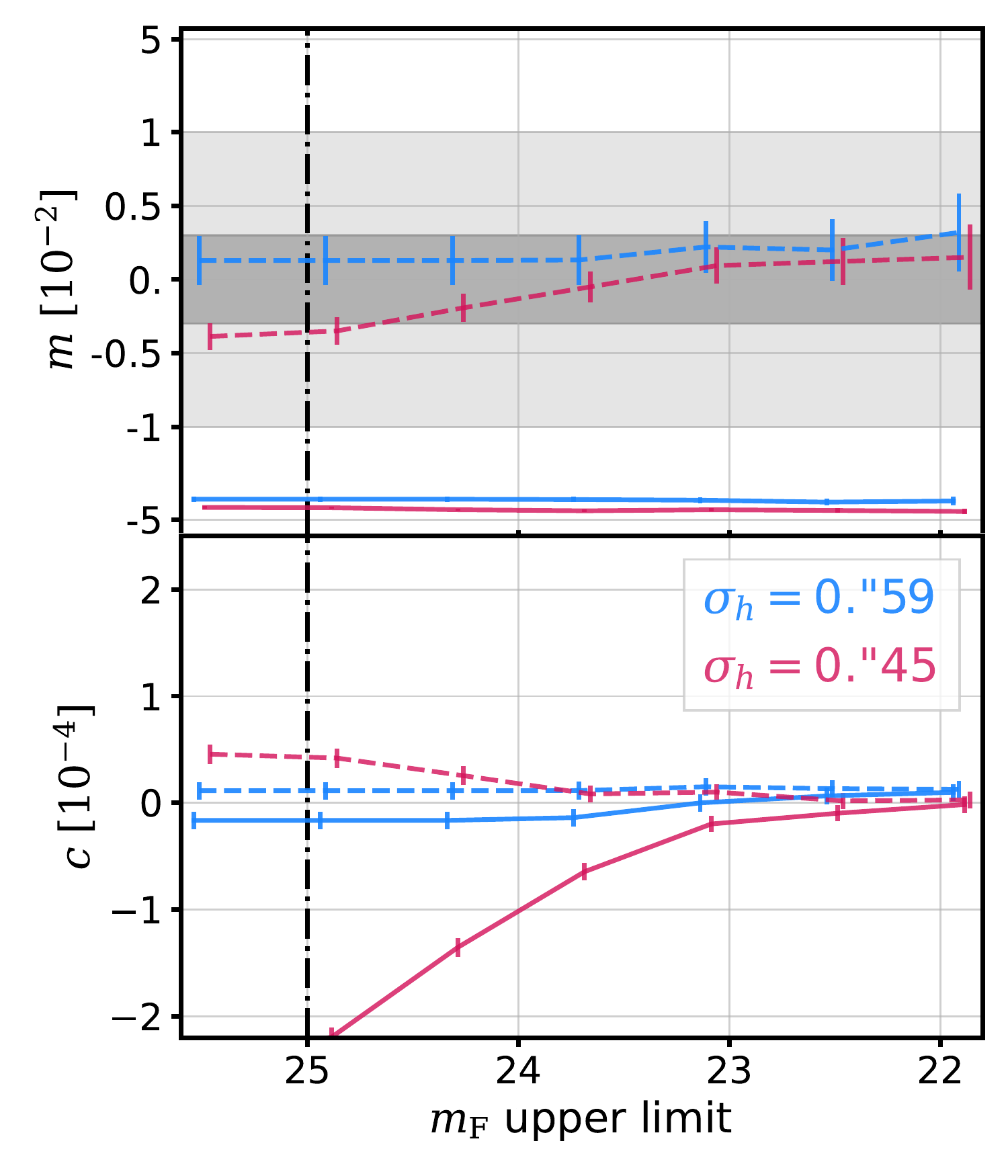}
\end{center}
\caption{
    Similar to Figure~(\ref{fig:res_blend1_magcut}), but measured from blended
    galaxy image simulation with number density $170~\mathrm{arcmin}^{-2}$\,.
    The vertical dash-dotted line is the default cut on magnitude of the
    postselection.
    }
    \label{fig:res_blend2_magcut}
\end{minipage}
\hfill
\begin{minipage}[t]{0.45\textwidth}
\begin{center}
    \includegraphics[width=1.\textwidth]{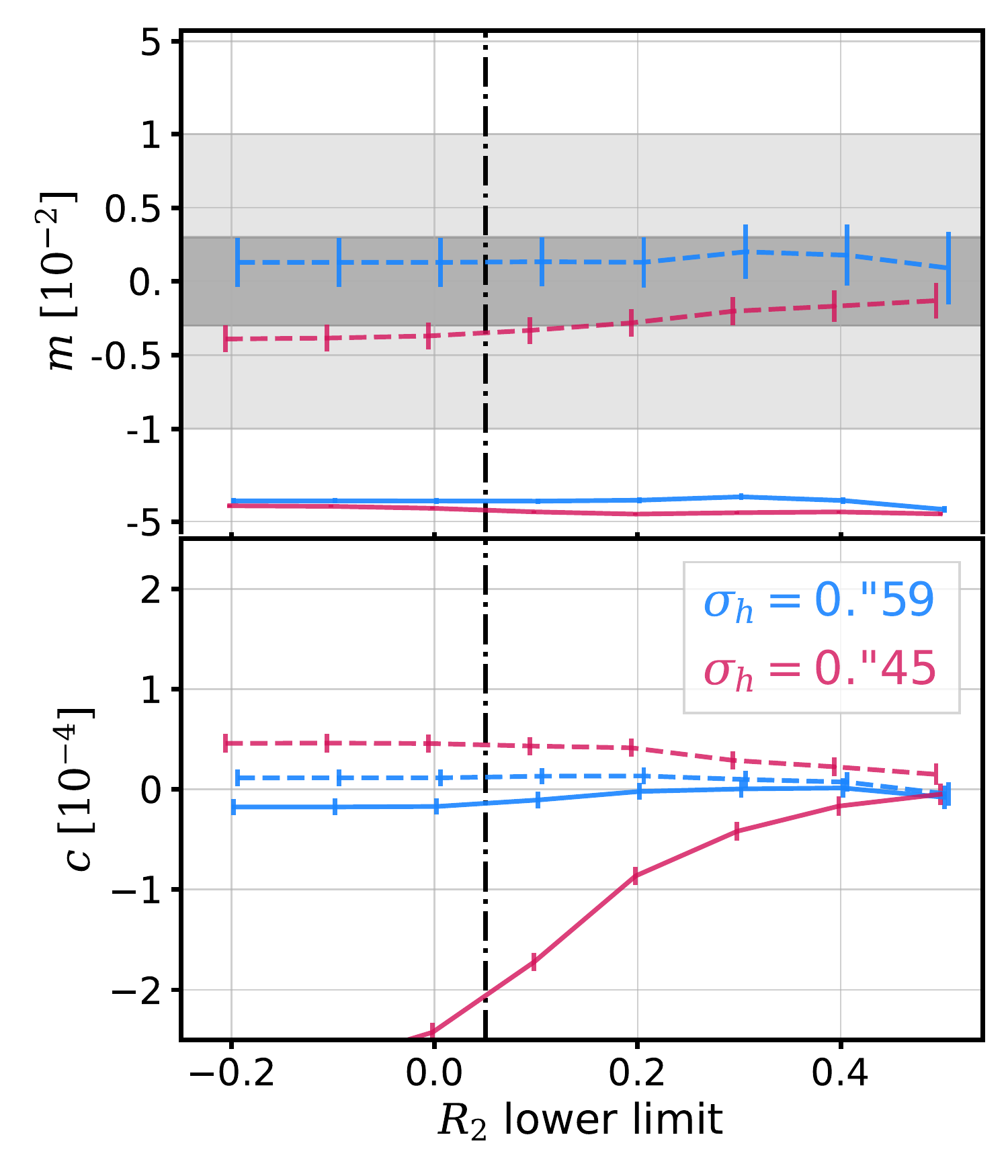}
\end{center}
\caption{
    Similar to Figure~(\ref{fig:res_blend1_r2cut}), but measured from blended
    galaxy image simulation with number density $170~\mathrm{arcmin}^{-2}$\,.
    The vertical dash-dotted line is the default cut on magnitude of the
    postselection.
    }
    \label{fig:res_blend2_r2cut}
\end{minipage}
\end{figure*}

In this appendix, we focus on the Hessian matrix in
equation~\eqref{eq:ell_noisy_correct}, $\frac{\partial^2 (\tilde{w}
\tilde{e}_\alpha)}{\partial v_i \partial v_j}$, which can be expanded to
\begin{equation}
\label{eq:app_noicor_deriv}
\frac{\partial^2 (\tilde{w} \tilde{e}_\alpha)}{\partial v_i \partial v_j}
=\tilde{w}\frac{\partial^2 \tilde{e}_\alpha}{\partial v_i \partial v_j}
+\frac{\partial{\tilde{w}}}{\partial{v_i}}\frac{\partial \tilde{e}_\alpha}{\partial{v_j}}
+\frac{\partial{\tilde{w}}}{\partial{v_j}}\frac{\partial \tilde{e}_\alpha}{\partial{v_j}}
+\tilde{e}_\alpha\frac{\partial^2 \tilde{w}_\alpha}{\partial v_i \partial v_j}\,,
\end{equation}
following the product rule for derivatives. Since the covariance matrix in
equation~\eqref{eq:ell_noisy_correct}, $K_{v_i}^{v_j}\,$, is by definition
symmetric, the noise bias correction term for the average of the weighted
ellipticity is
\begin{equation}
\label{eq:app_noicor_n}
\Delta \langle \tilde{w} \tilde{e}_\alpha \rangle_{\text{noise}}
=\frac{1}{2}\left\langle
\tilde{w}\frac{\partial^2 \tilde{e}_\alpha}{\partial v_i \partial v_j} K_{v_i}^{v_j}
+2\frac{\partial{\tilde{w}}}{\partial{v_i}}\frac{\partial \tilde{e}_\alpha}{\partial{v_j}} K_{v_i}^{v_j}
+\tilde{e}_\alpha\frac{\partial^2 \tilde{w}}{\partial v_i \partial v_j} K_{v_i}^{v_j}
\right\rangle\,,
\end{equation}
where we adopt Einstein notation on indices $(i, j)$ throughout this appendix:
repeated indices $i$ and indices $j$ indicate summation. In addition to the
weighted ellipticity, photon noise also biases the estimation of its shear
response due to the nonlinearity in the shear response. We correct for noise
biases in the shear response, namely $\langle \tilde{w}\,
\tilde{e}_{\alpha;\alpha} \rangle + \langle \tilde{w}_{;\alpha}\,
\tilde{e}_{\alpha} \rangle$ to accurately measure the shear response. These
corrections are composed of
\begin{equation}
\label{eq:app_noicor_d1}
\begin{split}
\Delta \langle \tilde{w} \tilde{e}_{\alpha;\alpha} \rangle_{\text{noise}}
&=\frac{1}{2}\left\langle
\tilde{w}\frac{\partial^2 \tilde{e}_{\alpha;\alpha}}{\partial v_i \partial v_j} K_{v_i}^{v_j}
+2\frac{\partial{\tilde{w}}}{\partial{v_i}}\frac{\partial \tilde{e}_{\alpha;\alpha}}{\partial{v_j}} K_{v_i}^{v_j}
\right\rangle\\
&+\frac{1}{2}\left\langle\tilde{e}_{\alpha;\alpha}\frac{\partial^2 \tilde{w}}{\partial v_i \partial v_j} K_{v_i}^{v_j}
\right\rangle\,
\end{split}
\end{equation}
and
\begin{equation}
\label{eq:app_noicor_d2}
\begin{split}
\Delta \langle \tilde{w}_{;\alpha} \tilde{e}_\alpha \rangle_{\text{noise}}
&=\frac{1}{2}\left\langle
\frac{\partial{\tilde{w}}}{\partial{v_k}}
\frac{\partial^2 \left(\tilde{v}_{k;\alpha}\tilde{e}_\alpha\right)}{\partial v_i \partial v_j}
K_{v_i}^{v_j} 
+2\frac{\partial^2{\tilde{w}}}{\partial{v_k}\partial{v_i}}
\frac{\partial \left(\tilde{v}_{k;\alpha}\tilde{e}_\alpha\right)}{\partial{v_j}}
K_{v_i}^{v_j}\right\rangle \\
&+\frac{1}{2}\left\langle \tilde{v}_{k;\alpha}\tilde{e}_\alpha
\frac{\partial^3 \tilde{w}}{\partial{v_k} \partial{v_i} \partial{v_j}} K_{v_i}^{v_j}
\right\rangle\,. 
\end{split}
\end{equation}
Finally, the shear transform equation of the expectation value of the
\textit{noisy} weighted ellipticity is
\begin{equation}
\label{eq:app_noisy_shear_transform}
\begin{split}
\langle \tilde{w} \tilde{e}_\alpha \rangle-\Delta \langle \tilde{w} \tilde{e}_\alpha \rangle_{\text{noise}}
&= \gamma_\alpha \left(
\langle \tilde{w}_{;\alpha} \tilde{e}_\alpha \rangle
+\langle \tilde{w} \tilde{e}_{\alpha;\alpha} \rangle
\right )\\
&-\gamma_\alpha \left(
\Delta \langle \tilde{w}_{;\alpha} \tilde{e}_\alpha \rangle_{\text{noise}}
+\Delta \langle \tilde{w} \tilde{e}_{\alpha;\alpha} \rangle_{\text{noise}}
\right)\\
&+ \mathcal{O}\left(\gamma_{\alpha}^3\right)+ \mathcal{O}\left((\delta{v})^4\right)
\end{split}
\end{equation}

Here we separate the noise bias correction terms defined in
equations~\eqref{eq:app_noicor_n}--\eqref{eq:app_noicor_d2} into terms that
only include zeroth, first and higher-order derivatives of the selection weight
function with respect to the basis modes.

\subsection{Zeroth-order terms}
\label{app:noirev_0}
The correction terms only including the zeroth-order derivative of
selection weight function are
$$
\tilde{w}\left\{\frac{\partial^2 \tilde{e}_\alpha}
{2\,\partial v_i \partial v_j} K_{v_i}^{v_j}\right\}
\qquad\text{and}\qquad
\tilde{w} \left\{\frac{\partial^2 \tilde{e}_{\alpha;\alpha}}
{2\,\partial v_i \partial v_j} K_{v_i}^{v_j}\right\}\,.
$$
The terms in the bracket are the second-order corrections for the noise biases
in ellipticity and its shear response. These noise bias corrections have been
presented in equations~(21), (A1) and (A2) of \citet{FPFS_Li2022}.

It is worth pointing out that the notation in \citet{FPFS_Li2022} is slightly
different from the above, and we translate these equations here. We can derive
the correction term from equation~(21) of
\citet{FPFS_Li2022}:
\begin{equation}
    \begin{split}
    \frac{\partial^2 \tilde{e}_1}{2\,\partial v_i \partial v_j} K_{v_i}^{v_j}
    =& e_1 \frac{K^{M_{00}}_{M_{00}}}{(M_{00}+C)^2}
    -\frac{K^{M_{00}}_{M_{22c}}}{(M_{00}+C)^2} ,\\
    \frac{\partial^2 \tilde{e}_2}{2\,\partial v_i \partial v_j} K_{v_i}^{v_j}
    =& e_2\frac{K^{M_{00}}_{M_{00}}}{(M_{00}+C)^2}
    -\frac{K^{M_{00}}_{M_{22s}}}{(M_{00}+C)^2}
    \end{split}
\end{equation}

The shear response is composed of the following terms: $s_0$, $s_4$ and
$(e_\alpha)^2$. We can derive the second-order noise bias correction terms for
them, which is similar to equations~(A1) and (A2) of \citet{FPFS_Li2022}:
\begin{equation}
    \begin{split}
    \frac{\partial^2 \tilde{s}_0}{2\,\partial v_i \partial v_j} K_{v_i}^{v_j}
    &= s_0 \frac{K^{M_{00}}_{M_{00}}}{(M_{00}+C)^2}
    -\frac{K^{M_{00}}_{M_{00}}}{(M_{00}+C)^2},\\
    \frac{\partial^2 \tilde{s}_4}{2\,\partial v_i \partial v_j} K_{v_i}^{v_j}
    &= s_4 \frac{K^{M_{00}}_{M_{00}}}{(M_{00}+C)^2}
    -\frac{K^{M_{40}}_{M_{00}}}{(M_{00}+C)^2},\\
    \frac{\partial^2 (\tilde{e}_1)^2}{2\,\partial v_i \partial v_j} K_{v_i}^{v_j}
    &= 3 \frac{K^{M_{00}}_{M_{00}}}{(M_{00}+C)^2}
    + \frac{K^{M_{22c}}_{M_{22c}}}{(M_{00}+C)^2}
    - 4e_1\frac{K^{M_{00}}_{M_{22c}}}{(M_{00}+C)^2},\\
    \frac{\partial^2 (\tilde{e}_2)^2}{2\,\partial v_i \partial v_j} K_{v_i}^{v_j}
    &= 3 \frac{K^{M_{00}}_{M_{00}}}{(M_{00}+C)^2}
    + \frac{K^{M_{22s}}_{M_{22s}}}{(M_{00}+C)^2}
    - 4e_2\frac{K^{M_{00}}_{M_{22s}}}{(M_{00}+C)^2}\,.
    \end{split}
\end{equation}

\subsection{First-order terms}
\label{app:noirev_1}
Three terms include the first-order derivative of the selection
weight with respect to the basis modes:
\begin{equation}
\label{eq:app_noicor_term10}
\frac{\partial{\tilde{w}}}{\partial{v_k}}\left \{
\frac{\partial^2{(\tilde{v}_{k;\alpha}\tilde{e}_\alpha)}}
{2\,\partial v_i \partial v_j} K_{v_i}^{v_j}
\right\}\,,
\end{equation}
\begin{equation}
\frac{\partial{\tilde{w}}}{\partial{v_i}}\frac{\partial
\tilde{e}_\alpha}{\partial{v_j}} K_{v_i}^{v_j}\,.
\qquad \text{and} \qquad
\label{eq:app_B7}
\frac{\partial{\tilde{w}}}{\partial{v_i}}\frac{\partial
\tilde{e}_{\alpha;\alpha}}{\partial{v_j}} K_{v_i}^{v_j}
\end{equation}

The term in the bracket of equation~\eqref{eq:app_noicor_term10} is the
second-order noise bias correction for $\tilde{v}_{k;\alpha}
\tilde{e}_\alpha$\,. Specifically for our shear estimation, we are using peak
modes ($q_k$) for detection and shapelets modes ($M_{00}$ and $M_{20}$) for
galaxy selection. The shear responses of peak modes are $q_{k;1}$ and
$q_{k;2}$; and the shear responses of shapelet modes, $M_{00}$ ($M_{20}$) are
composed of $M_{22c}$ and $M_{22s}$ ($M_{42c}$ and $M_{42s}$), as shown in
equations~\eqref{eq:shear_response_m00} (\eqref{eq:shear_response_m20}).
Keeping the second-order terms in the noise bias contributions, the
relationships between the expectation values of the noisy observables (e.g.,
$\tilde{e}_1\tilde{q}_{k;1}$, $\tilde{e}_1\tilde{M}_{22c}$,
$\tilde{e}_1\tilde{M}_{42c}$ ) and the noiseless correspondents are derived
using the Hessian matrix and the covariance matrix in
equation~\eqref{eq:app_noicor_term10}:
\begin{equation}
\begin{split}
\langle\tilde{e}_1\tilde{q}_{k;1}\rangle=&
\left \langle e_1 q_{k;1}\left(1+\frac{K_{M_{00}}^{M_{00}}}{D^2}\right) \right\rangle
-\left\langle \frac{q_{k;1}K_{M_{00}}^{M_{22c}}}{D^2}\right\rangle
- \left\langle \frac{e_1K_{M_{00}}^{q_{k;1}}}{D}\right\rangle\\
&+\left\langle \frac{K_{M_{22c}}^{q_{k;1}}}{D}\right\rangle\,,\\
\langle\tilde{e}_1\tilde{M}_{22c}\rangle=&
\left \langle e_1 M_{22c}\left(1+\frac{K_{M_{00}}^{M_{00}}}{D^2}\right)-
2\,\frac{e_1K_{M_{00}}^{M_{22c}}}{D}\right\rangle
+\left\langle \frac{K_{M_{22c}}^{M_{22c}}}{D}\right\rangle\,,\\
\langle\tilde{e}_1\tilde{M}_{42c}\rangle=&
\left \langle e_1 M_{42c}\left(1+\frac{K_{M_{00}}^{M_{00}}}{D^2}\right) \right\rangle
-\left\langle \frac{\epsilon_1 K_{M_{00}}^{M_{22c}}}{D}\right\rangle
- \left\langle \frac{e_1K_{M_{00}}^{M_{42c}}}{D}\right\rangle\\
&+\left\langle \frac{K_{M_{22c}}^{M_{42c}}}{D}\right\rangle\,,
\end{split}
\end{equation}
where $D=M_{00}+C$. We only take the terms related to the first component of
ellipticity as an example, and the terms for the second component have a
similar form.

Next we provide the noise bias correction for the first term in
equation~\eqref{eq:app_B7} using the derivatives of ellipticity and selection
weight:
\begin{equation}
\label{eq:app_noicor_term11}
\begin{split}
\left\langle
\frac{\partial{\tilde{w}}}{\partial{v_i}}
\frac{\partial \tilde{e}_1}{\partial{v_j}} K_{v_i}^{v_j}
\right\rangle&=
\left\langle \frac{\partial{\tilde{w}}}{\partial{v_i}}\frac{K^{M_{22c}}_{v_i}}{D}
- \frac{\partial{\tilde{w}}}{\partial{v_i}}\frac{e_1 K^{M_{00}}_{v_i}}{D}
\right\rangle\,,\\
\left\langle\frac{\partial{\tilde{w}}}{\partial{v_i}}
\frac{\partial \tilde{e}_2}{\partial{v_j}} K_{v_i}^{v_j}\right\rangle&=
\left\langle\frac{\partial{\tilde{w}}}{\partial{v_i}}
\frac{K^{M_{22s}}_{v_i}}{D}-\frac{\partial{\tilde{w}}}{\partial{v_i}}
\frac{e_2 K^{M_{00}}_{v_i}}{D}\right\rangle\,.
\end{split}
\end{equation}
Also, the noise bias correction for the second term in
equation~\eqref{eq:app_B7}, using the derivatives of shear response and
selection weight, is as follows:
\begin{equation}
\label{eq:app_noicor_term12}
\begin{split}
\left\langle\frac{\partial{\tilde{w}}}{\partial{v_i}}
\frac{\partial \tilde{e}_{1;1}}{\partial{v_j}} K_{v_i}^{v_j}\right\rangle&=
\frac{1}{\sqrt{2}}\left\langle \frac{\partial{\tilde{w}}}{\partial{v_i}}\frac{K^{M_{00}}_{v_i}}{D}
\left(\frac{C}{D}+s_4 - 4 e_1^2 \right)\right\rangle\\
&+\frac{4}{\sqrt{2}}\left\langle
\frac{\partial{\tilde{w}}}{\partial{v_i}}\frac{e_1 K^{M_{22c}}_{v_i}}{D}
\right\rangle
-\frac{1}{\sqrt{2}}\left\langle
\frac{\partial{\tilde{w}}}{\partial{v_i}}\frac{K^{M_{40}}_{v_i}}{D}
\right\rangle, \\
\left\langle\frac{\partial{\tilde{w}}}{\partial{v_i}}
\frac{\partial \tilde{e}_{2;2}}{\partial{v_j}} K_{v_i}^{v_j}\right\rangle&=
\frac{1}{\sqrt{2}}\left\langle \frac{\partial{\tilde{w}}}{\partial{v_i}} \frac{K^{M_{00}}_{v_i}}{D}
\left(\frac{C}{D}+s_4 - 4 e_2^2 \right)\right\rangle\\
&+\frac{4}{\sqrt{2}}\left\langle
\frac{\partial{\tilde{w}}}{\partial{v_i}}\frac{e_2 K^{M_{22s}}_{v_i}}{D}
\right\rangle
-\frac{1}{\sqrt{2}}\left\langle
\frac{\partial{\tilde{w}}}{\partial{v_i}}\frac{K^{M_{40}}_{v_i}}{D}
\right\rangle\,.
\end{split}
\end{equation}
In equations~\eqref{eq:app_noicor_term11} and \eqref{eq:app_noicor_term12},
$v_i$ can be any basis modes used for detection and selection (e.g., $M_{00}$,
$M_{40}$, $q_{0\dots7}$).

\subsection{Other terms}

We neglect the second-order noise bias terms other than those in
Appendices~\ref{app:noirev_0} and \ref{app:noirev_1}. In \citet{imPT_Li2023},
we develop a pipeline to automatically derive the full second-order noise bias
correction using auto-differentiation in \texttt{jax}. As shown in
Figures~\ref{fig:res_hsc_magcut} and \ref{fig:res_hsc_r2cut}, the
multiplicative shear bias from neglecting the other terms is about $0.2\%$\,.

\section{Inhomogeneous Noise}
\label{app:inhomogeneous_noise}

In this appendix, we focus on noise on single exposures of space-based
observations (e.g., Euclid and Roman): photon noise and read noise are not
correlated over pixels on single exposures, but space-based images are
dominated by photon noise from galaxies, which is \textit{not homogeneous} at
galaxy scale.

Since image noise on single exposures is not correlated between pixels, the
off-diagonal terms of the covariance matrix are zero. The covariance matrix is
\begin{equation}
    \langle n_{\vx} n_{\vx'}^{*} \rangle=
    \left(f_\vx+B_\vx\right) \delta_\mathrm{D}(\vx-\vx')\,,
\end{equation}
where $(f_\vx+B_\vx)$ is the spectrum of the noise covariance matrix in
configuration space; $f_\vx$ is the inhomogeneous surface density field of the
galaxies; and $B_\vx$ is the homogeneous noise variance from sky background and
read noise. Since the covariance of the pixel noise in configuration space is
not homogeneous, the covariance of noise on Fourier wave numbers is correlated.
Therefore, we carry out our derivation in configuration space.

Taking a shapelet mode $M_{nm}$ as an example, the measurement error on the
shapelet mode is (Plancherel theorem)
\begin{equation}
    \delta{M}_{nm}= \iint \dd[2]{x} \,
    \left(g_{nm}(\vx)\right)^{*} n_\vx\,,
\end{equation}
where $g_{nm}(\vx)$ is the inverse Fourier transform of the shapelet basis
function deconvolved by the PSF:
\begin{equation}
    g_{nm}(\vx) = \left( \frac{1}{2\pi}\right)^2 \iint \dd[2]{k}
    \frac{\left(\tilde{\chi}_{nm}(\vk)\right)}{p_{\vk}}
    e^{i\vk \cdot \vx}\,.
\end{equation}
The covariance between $\delta{M}_{nm}$ and $\delta{M}_{n'm'}$ is
\begin{equation}
K_{M_{nm}}^{M_{n'm'}}
= \iint \dd[2]{x} \left(g_{nm}(\vx)\right)^{*} g_{n'm'}(\vx) \left( f_\vx+ B_\vx \right)\,.
\end{equation}
The covariances between measurement errors can be estimated in real observations
as long as we can estimate $f_\vx + B_\vx$ from single exposure images before
subtracting the sky background.

\section{Tests for the heavily blended case}
\label{app:test_blend2}

In this appendix, we show the multiplicative bias and additive bias for the
blended image simulation introduced in Section~\ref{sec:3_sim_blended} with
input galaxy number density of $170$~arcmin$^{-2}$\, (double our fiducial
density) in Figures~\ref{fig:res_blend2_magcut} and \ref{fig:res_blend2_r2cut}.
Similar to the tests in Section~\ref{sec:4_blend_bias}, we use two different
smoothing scales, $\sigma_h= 0\farcs59$ and $\sigma_h= 0\farcs45$\,, to detect,
select and measure shear from the simulated images. These are stress tests of
the algorithm under extreme conditions (e.g., near the center of a galaxy
cluster).

For each one of the smoothing scales, the result is consistent with the one
using the same smoothing scale in Section~\ref{sec:4_blend_bias} at the
$2\sigma$ level.

\label{lastpage}
\end{document}